\documentstyle[aps,rmp,epsfig]{revtex}

\newcommand{\dd}{{\rm d}}
\newcommand{\bx}{{\bf x}}
\newcommand{\aem}{\alpha_{_{\rm EM}}}
\newcommand{\ag}{\alpha_{_{\rm G}}}
\newcommand{\as}{\alpha_{_{\rm S}}}
\newcommand{\aw}{\alpha_{_{\rm W}}}
\newcommand{\gfermi}{G_{_{\rm F}}}
\def\ga{\mathrel{\raise.3ex\hbox{$>$\kern-.75em\lower1ex\hbox{$\sim$}}}}
\def\la{\mathrel{\raise.3ex\hbox{$<$\kern-.75em\lower1ex\hbox{$\sim$}}}}
\newcommand{\dslash}{D\!\!\!\!/}

\begin{document}
\twocolumn \title{The fundamental constants and their variation:
observational status and theoretical motivations}
\author{Jean-Philippe Uzan\footnote{e-mail: {\tt uzan@iap.fr}.}}

\address{Institut d'Astrophysique de
Paris, GReCO, CNRS-FRE 2435, 98 bis, Bd Arago, 75014 Paris, France.}

\address{Laboratoire de Physique Th\'eorique, CNRS-UMR 8627,
Universit\'e Paris Sud, b\^atiment 210, F-91405 Orsay cedex, France.}

\maketitle

\begin{abstract}
This article describes the various experimental bounds on the
variation of the fundamental constants of nature. After a discussion
on the role of fundamental constants, of their definition and link
with metrology, the various constraints on the variation of the fine
structure constant, the gravitational, weak and strong interactions
couplings and the electron to proton mass ratio are reviewed. This
review aims (1) to provide the basics of each measurement, (2) to show
as clearly as possible why it constrains a given constant and (3) to
point out the underlying hypotheses. Such an investigation is of
importance to compare the different results, particularly in view of
understanding the recent claims of the detections of a variation of
the fine structure constant and of the electron to proton mass ratio
in quasar absorption spectra. The theoretical models leading to the
prediction of such variation are also reviewed, including Kaluza-Klein
theories, string theories and other alternative theories and
cosmological implications of these results are discussed. The links
with the tests of general relativity are emphasized.
\end{abstract}
\tableofcontents
\section{Introduction}\label{sec_1}

The development of physics relied considerably on the Copernician
principle, which states that we are not living in a particular place
in the universe and stating that the laws of physics do not differ
from one point in spacetime to another. This contrasts with the
Aristotelian point of view in which the laws on Earth and in Heavens
differ. It is however natural to question this assumption. Indeed, it
is difficult to imagine a change of the form of physical laws (e.g. a
Newtonian gravitation force behaving as the inverse of the square of
the distance on Earth and as another power somewhere else) but a
smooth change in the physical constants is much easier to conceive.

Comparing and reproducing experiments is also a root of the scientific
approach which makes sense only if the laws of nature does not depend
on time and space.  This hypothesis of constancy of the constants
plays an important role in particular in astronomy and cosmology where
the redshift measures the look-back time. Ignoring the possibility of
varying constants could lead to a distorted view of our universe and
if such a variation is established corrections would have to be
applied. It is thus of importance to investigate this possibility
especially as the measurements become more and more
precise. Obviously, the constants have not undergone huge variations
on Solar system scales and geological time scales and one is looking
for tiny effects. Besides, the question of the values of the constants
is central to physics and one can hope to explain them dynamically as
predicted by some high-energy theories. Testing for the constancy of
the constants is thus part of the tests of general relativity. Let us
emphasize that this latter step is analogous to the transition from
the Newtonian description of mechanics in which space and time were
just a static background in which matter was evolving to the
relativistic description where spacetime becomes a dynamical quantity
determined by the Einstein equations (Damour, 2001).

Before discussing the properties of the constants of nature, we must
have an idea of which constants to consider. First, all constants of
physics do not play the same role, and some have a much deeper one
than others. Following Levy-Leblond (1979), we can define three
classes of fundamental constants, {\em class A} being the class of the
constants characteristic of particular objects, {\em class B} being
the class of constants characteristic of a class of physical
phenomena, and {\em class C} being the class of universal
constants. Indeed, the status of a constant can change with time. For
instance, the velocity of light was a initially a type A constant
(describing a property of light) which then became a type B constant
when it was realized that it was related to electro-magnetic phenomena
and, to finish, it ended as a type C constant (it enters many laws of
physics from electromagnetism to relativity including the notion of
causality...). It has even become a much more fundamental constant
since it has been chosen as the definition of the meter (Petley,
1983). A more conservative definition of a fundamental constant would
thus be to state that it is {\it any parameter} that can not be
calculated with our present knowledge of physics, i.e. a free
parameter of our theory at hand. Each free parameter of a theory is
in fact a challenge for future theories to explain its value.

How many fundamental constants should we consider? The set of
constants which are conventionally considered as fundamental (Flowers
and Petley, 2001) consists of the electron charge $e$, the electron
mass $m_{\rm e}$, the proton mass $m_{\rm p}$, the reduced Planck constant
$\hbar$, the velocity of light in vacuum $c$, the Avogadro constant
$N_{_{\rm A}}$, the Boltzmann constant $k_{_{\rm B}}$, the Newton
constant $G$, the permeability and permittivity of space,
$\varepsilon_0$ and $\mu_0$. The latter has a fixed value in the SI
system of unit ($\mu_0=4\pi\times10^{-7}\,{\rm H}\,{\rm m}^{-1}$) which
is implicit in the definition of the Ampere; $\varepsilon_0$ is then
fixed by the relation $\varepsilon_0\mu_0=c^{-2}$. The inclusion of
$N_{_{\rm A}}$ in the former list has been debated a lot (see e.g. Birge,
1929).  To compare with, the minimal standard model of particle
physics plus gravitation that describes the four known interactions
depends on 20 free parameters (Cahn, 1996; Hogan, 2000): the Yukawa
coefficients determining the masses of the six quark $(u,d,c,s,t,b)$
and three lepton $(e,\mu,\tau)$ flavors, the Higgs mass and vacuum
expectation value, three angles and a phase of the
Cabibbo-Kobayashi-Maskawa matrix, a phase for the QCD vacuum and three
coupling constants $g_{_{\rm S}}, g_{_{\rm W}}, g_1$ for the gauge
group $SU(3)\times SU(2)\times U(1)$ respectively. Below the $Z$ mass,
$g_1$ and $g_{_{\rm W}}$ combine to form the electro-magnetic coupling
constant
\begin{equation}
g_{_{\rm EM}}^{-2}=\frac{5}{3}g_1^{-2}+g_{_{\rm
W}}^{-2}.
\end{equation}

\begin{figure}
\begin{center}
\psfig{file=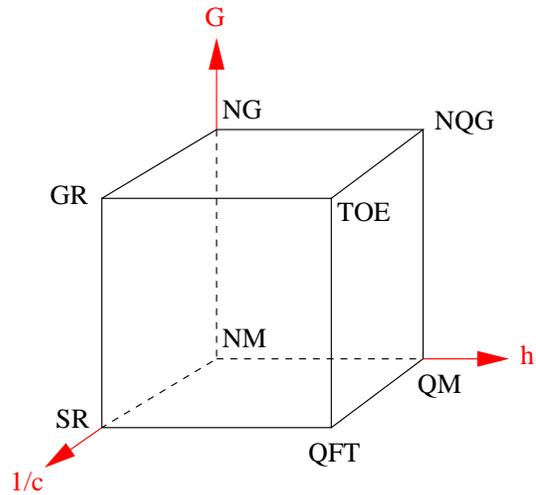,width=7cm} \vskip3mm \caption{The
cube of physical theories as presented by Okun (1991). At the
origin stands the part of Newtonian mechanics (NM) that does not
take gravity into account. NG, QM and SR then stand for Newtonian
gravity, quantum mechanics and special relativity which
respectively introduce the effect of one of the constants. Special
relativity `merges' respectively with quantum mechanics and
Newtonian gravity to give quantum field theory (QFT) and general
relativity (GR). Bringing quantum mechanics and Newtonian gravity
together leads to non-relativistic quantum gravity and all theories
together give the theory of everything (TOE). [From Okun (1991)].}
\label{fig0}
\end{center}
\end{figure}

The number of free parameters indeed depends on the physical model at
hand (see Weinberg, 1983). This issue has to be disconnected from the
number of required fundamental dimensionful constants.  Duff, Okun and
Veneziano (2002) recently debated this question, respectively arguing
for none, three and two (see also Wignall, 2000). Arguing for no
fundamental constant leads to consider them simply as conversion
parameters. Some of them are, like the Boltzmann constant, but some
others play a deeper role in the sense that when a physical quantity
becomes of the same order of this constant new phenomena appear, this
is the case e.g. of $\hbar$ and $c$ which are associated respectively
to quantum and relativistic effects. Okun (1991) considered that only
three fundamental constants are necessary, the underlying reason being
that in the international system of units which has 7 base units an 17
derived units, four of the seven base units are in fact derived
(Ampere, Kelvin, mole and candela). The three remaining base units
(meter, second and kilogram) are then associated to three fundamental
constants ($c$, $\hbar$ and $G$).  They can be seen as limiting
quantities: $c$ is associated to the maximum velocity and $\hbar$ to
the unit quantum of angular momentum and sets a minimum of uncertainty
whereas $G$ is not directly associated to any physical quantity (see
Martins 2002 who argues that $G$ is the limiting potential for a mass
that does not form a black hole). In the framework of quantum field
theory + general relativity, it seems that this set of three constants
has to be considered and it allows to classify the physical theories
(see figure~\ref{fig0}). However, Veneziano (1986) argued that in the
framework of string theory one requires only two dimensionful
fundamental constants, $c$ and the string length $\lambda_s$. The use
of $\hbar$ seems unnecessary since it combines with the string tension
to give $\lambda_s$. In the case of the Goto-Nambu action
$S/\hbar=(T/\hbar)\int\dd(Area)\equiv \lambda_s^{-2}\int\dd(Area)$ and
the Planck constant is just given by $\lambda_s^{-2}$. In this view,
$\hbar$ has not disappeared but has been promoted to the role of a UV
cut-off that removes both the infinities of quantum field theory and
singularities of general relativity. This situation is analogous to
pure quantum gravity (Novikov and Zel'dovich, 1982) where $\hbar$ and
$G$ never appear separately but only in the combination $\ell_{_{\rm
Pl}}=\sqrt{G\hbar/c^{3}}$ so that only $c$ and $\ell_{_{\rm Pl}}$ are
needed. Volovik (2002) considered the analogy with quantum
liquids. There, an observer knows both the effective and microscopic
physics so that he can judge whether the fundamental constants of the
effective theory remain fundamental constants of the microscopic
theory. The status of a constant depends on the considered theory
(effective or microscopic) and, more interestingly, on the observer
measuring them, i.e. on whether this observer belongs to the world of
low-energy quasi-particles or to the microscopic world.

Resolving this issue is indeed far beyond the scope of this article
and can probably be considered more as an epistemological question
than a physical one. But, as the discussion above tends to show, the
answer depends on the theoretical framework considered [see also
Cohen-Tannoudji (1985) for arguments to consider the Boltzmann
constant as a fundamental constant]. A more pragmatic approach is then
to choose a theoretical framework, so that the set of undetermined
fixed parameters is fully known and then to wonder why they have the
values they have and if they are constant.

We review in this article both the status of the experimental
constraints on the variation of fundamental constants and the
theoretical motivations for considering such variations. In
section~\ref{sec_2}, we recall Dirac's argument that initiated the
consideration of time varying constants and we briefly discuss how it
is linked to anthropic arguments.  Then, since the fundamental
constants are entangled with the theory of measurement, we make some
very general comments on the consequences of metrology. In
Sections~\ref{sec_4} and~\ref{sec_3}, we review the observational
constraints respectively on the variation of the fine structure and of
gravitational constants. Indeed, we have to keep in mind that the
obtained constraints depend on underlying assumptions on a certain set
of other constants. We summarize more briefly in Section~\ref{sec_5},
the constraints on other constants and we give, in
Section~\ref{sec_7}, some hints of the theoretical motivations arising
mainly from grand unified theories, Kaluza-Klein and string theories.
We also discuss a number of cosmological models taking these
variations into account.  For recent shorter reviews, see Varshalovich
{\em et al.} (2000a), Chiba (2001), Uzan (2002) and Martins (2002).

\noindent{\bf Notations:} In this work, we use SI units and the
following values of the fundamental constants today\footnote{see
{\tt http://physics.nist.gov/cuu/Constants/} for an up to date
list of the recommended values of the constants of nature.}
\begin{eqnarray}
c&=&299,792,458\,{\rm m\cdot s}^{-1}\\
\hbar&=&1.054 571 596(82)\times 10^{-34}\,{\rm J\cdot s} \\
G&=&  6.673(10)\times10^{-11}\,{\rm m}^3\cdot{\rm kg}^{-1}\cdot{\rm s}^{-2}\\
m_{\rm e}&=&9.109 381 88(72)\times10^{-31}\,{\rm kg}\\
m_{\rm p}&=&1.672 621 58(13)\times10^{-27}\,{\rm kg}\\
m_{\rm n}&=&1.674 927 16(13)\times10^{-27}\,{\rm kg}\\
e&=&1.602 176 462(63)\times10^{-29}\,{\rm C}
\end{eqnarray}
for the velocity of light, the reduced Planck constant, the Newton
constant, the masses of the electron, proton and neutron, and the
charge of the electron. We also define
\begin{equation}
q^2\equiv\frac{e^2}{4\pi\varepsilon_0}\,
\end{equation}
and the following dimensionless ratios
\begin{eqnarray}
\aem&\equiv&\frac{q^2}{\hbar c}\sim 1/137.035 999 76(50) \\
\aw&\equiv&\frac{\gfermi m_{\rm p}^2 c}{\hbar^3}\sim1.03\times10^{-5}\\
\as(E)&\equiv&\frac{g_{_{\rm s}}^2(E)}{\hbar c}\label{fort}\\
\ag&\equiv&\frac{Gm_{\rm p}^2}{\hbar c}\sim 5\times10^{-39}\\
\mu&\equiv&\frac{m_{\rm e}}{m_{\rm p}}\sim5.44617\times10^{-4}\\
x&\equiv&g_{\rm p}\aem^2\mu\sim 1.62\times10^{-7}\\
y&\equiv&g_{\rm p}\aem^2\sim2.977\times10^{-4}
\end{eqnarray}
which characterize respectively the strength of the electro-magnetic,
weak, strong and gravitational forces and the electron-proton mass
ratio, $g_{\rm p}\simeq5.585$ is the proton gyro-magnetic factor. Note
that the relation (\ref{fort}) between two quantities that depend
strongly on energy; this will be discussed in
more details in Section~\ref{sec_5}. We introduce the notations
\begin{eqnarray}
a_0&=&\frac{\hbar}{m_{\rm e}c\aem}=0.5291771~{\rm \AA}\label{Bohr}\\
-E_I&=&\frac{1}{2}m_{\rm e}c^2\aem^2=13.60580\,{\rm eV}\label{Eion}\\
R_\infty&=&-\frac{E_I}{hc}=1.0 973 731568 549(83)\times10^7\,{\rm
m}^{-1}\label{Rydberg}
\end{eqnarray}
respectively for the Bohr radius, the hydrogen ionization energy and
the Rydberg constant.

While working in cosmology, we assume that the universe is
described by a Friedmann-Lema\^{\i}tre spacetime
\begin{equation}
\dd s^2=-\dd t^2+ a^2(t)\gamma_{ij}\dd x^i\dd x^j,
\end{equation}
where $t$ is the cosmic time, $a$ the scale factor and
$\gamma_{ij}$ the metric of the spatial sections. We define the
redshift as
\begin{equation}
1+z\equiv\frac{a_0}{a}=\frac{\nu_e}{\nu_0}
\end{equation}
where $a_0$ is the value of the scale factor today while $\nu_e$
and $\nu_0$ are respectively the frequencies at emission and
today. We decompose the Hubble constant today as
\begin{equation}
H_0^{-1}=9.7776\times10^9\,h^{-1}\,{\rm yr}
\end{equation}
where $h=0.68\pm0.15$ is a dimensionless number, and the density
of the universe today is given by
\begin{equation}
\rho_0=1.879\times10^{-26}\Omega h^2\,{\rm kg\cdot m}^{-3}.
\end{equation}

\section{Generalities}\label{sec_2}
\subsection{From Dirac numerological principle to anthropic arguments}
\label{subsec_2.1}

The question of the constancy of the constants of physics was probably
first addressed by Dirac (1937, 1938, 1979) who expressed, in his
``Large Numbers hypothesis'', the opinion that very large (or small)
dimensionless universal constants cannot be pure mathematical numbers
and must not occur in the basic laws of physics. He suggested, on the
basis of this numerological principle, that these large numbers should
rather be considered as variable parameters characterizing the state
of the universe.  Dirac formed the five dimensionless ratios $\aem$,
$\aw$, $\ag$, $\delta\equiv H_0\hbar/m_{\rm p}c^2\sim
2h\times10^{-42}$ and $\epsilon\equiv G\rho_0/H_0^2\sim 5h^{-2}\times
10^{-4}$ and asked the question of which of these ratio is constant as
the universe evolves. Usually, only $\delta$ and $\epsilon$ vary as
the inverse of the cosmic time (note that with the value of the
density chosen by Dirac, the universe is not flat so that $a\propto t$
and $\rho\propto t^{-3}$). Dirac then noticed that $\ag\mu/\aem$,
representing the relative magnitude of electrostatic and gravitational
forces between a proton and an electron, was of the same order as
$H_0e^2/m_{\rm e}c^2=\delta\aem/\mu$ representing the age of the
universe in atomic time so that the five previous numbers can be
``harmonized'' if one assumes that $\ag$ and $\delta$ vary with time
and scale as the inverse of the cosmic time\footnote{The ratio
$\delta\aem/\mu$ represents roughly the inverse of the number of times
an electron orbits around a proton during the age of the
universe. Already, this suggested a link between micro-physics and
cosmological scales.}. This implies that the intensity of all
gravitational effects decrease with a rate of about $10^{-10}\,{\rm
yr}^{-1}$ and that $\rho\propto t^{-2}$ (since $\epsilon$ is constant)
which corresponds to a flat universe.  Kothari (1938) and
Chandrasekhar (1939) were the first to point out that some
astronomical consequences of this statement may be detectable. Similar
ideas were expressed by Milne (1937).

Dicke (1961) pointed out that in fact the density of the universe is
determined by its age, this age being related to the time needed to
form galaxies, stars, heavy nuclei... This led him to formulate that
the presence of an observer in the universe places constraints on the
physical laws that can be observed. In fact, what is meant by observer
is the existence of (highly?) organized systems and the anthropic
principle can be seen as a rephrasing of the question ``why is the
universe the way it is?'' (Hogan, 2000). Carter (1974, 1976, 1983),
who actually coined the term ``anthropic principle'', showed that the
numerological coincidence found by Dirac can be derived from physical
models of stars and the competition between the weakness of gravity
with respect to nuclear fusion. Carr and Rees (1979) then showed how
one can scale up from atomic to cosmological scales only by using
combinations of $\aem$, $\ag$ and $m_{\rm e}/m_{\rm p}$.

Dicke (1961, 1962b) brought Mach's principle into the discussion and
proposed (Brans and Dicke, 1961) a theory of gravitation based on this
principle. In this theory the gravitational constant is replaced by a
scalar field which can vary both in space and time.  It follows that,
for cosmological solutions, $G\propto t^{-n}$, $H\propto t^{-1}$ and
$\rho\propto t^{n-2}$ where $n$ is expressible in terms of an
arbitrary parameter $\omega_{_{\rm BD}}$ as $n^{-1}=2+3\omega_{_{\rm
BD}}/2$. Einstein gravity is recovered when $\omega_{_{\rm
BD}}\rightarrow\infty$. This predicts that $\ag\propto t^{-n}$ and
$\delta\propto t^{-1}$ whereas $\aem$, $\aw$ and $\epsilon$ are kept
constant. This kind of theory was further generalized to obtain
various functional dependences for $G$ in the formalization of
scalar-tensor theories of gravitation (see e.g. Damour and
Esposito-Far\`ese, 1992).

The first extension of Dirac's idea to non-gravitational forces was
proposed by Jordan (1937, 1939) who still considered that the weak
interaction and the proton to electron mass ratio were constant. He
realized that the constants has to become dynamical fields and used
the action
\begin{eqnarray}
S=\int\sqrt{-g}\dd^4\bx\phi^\eta\left[R-\xi\left(\frac{\nabla\phi}{\phi}
\right)^2-\frac{\phi}{2}F^2 \right],
\end{eqnarray}
$\eta$ and $\xi$ being two parameters.  Fierz (1956) realized that
with such a Lagrangian, atomic spectra will be space-dependent.
But, Dirac's idea was revived after Teller (1948) argued that the
decrease of $G$ contradicts paleontological evidences [see also
Pochoda and Schwarzschild (1964) and Gamow (1967c) for evidences
based on the nuclear resources of the Sun]. Gamow (1967a, 1967b)
proposed that $\aem$ might vary as $t$ in order to save the,
according to him, ``elegant'' idea of Dirac (see also
Stanyukovich, 1962).  In both Gamow (1967a, 1967b) and Dirac
(1937) theories the ratio $\ag/\aem$ decreases as $t^{-1}$. Teller
(1948) remarked that $\aem^{-1}\sim -\ln H_0t_{_{\rm Pl}}$ so that
$\aem^{-1}$ would become the logarithm of a large number.  Landau
(1955), de Witt (1964) and Isham {\em et al.} (1971) advocated
that such a dependence may arise if the Planck length provides a
cut-off to the logarithmic divergences of quantum electrodynamics.
In this latter class of models $\aem\propto1/\ln t$, $\ag\propto
t^{-1}$, $\delta\propto t^{-1}$ and $\aw$ and $\epsilon$ remain
constant. Dyson (1967), Peres (1967) and then Davies (1972)
showed, using respectively geological data of the abundance of
rhenium and osmium and the stability of heavy nuclei, that these
two hypothesis were ruled out observationally (see
Section~\ref{sec_4} for details on the experimental results).
Modern theories of high energy physics offer new arguments to
reconsider the variation of the fundamental constants (see
Section~\ref{sec_7}). The most important outcome of Dirac's
proposal and of the following assimilated theories [among which a
later version of Dirac (1974) theory in which there is matter
creation either where old matter was present or uniformly
throughout the universe] is that the hypothesis of the constancy
of the fundamental constants can and must be checked
experimentally.

A way to reconcile some of the large numbers is to consider the energy
dependence of the couplings as determined by the renormalization group
(see e.g.  Itzkyson and Zuber, 1980). For instance, concerning the
fine structure constant, the energy-dependence arises from vacuum
polarization that tends to screen the charge. This screening is less
important at small distance and the charge appears bigger so that the
effective coupling constant grows with energy. It follows from this
approach that the three gauge groups get unified into a larger grand
unification group so that the three couplings $\aem$, $\aw$ and $\as$
stem from the same dimensionless number $\alpha_{_{\rm GUT}}$. This
might explain some large numbers and answer some of Dirac concerns
(Hogan, 2000) but indeed, it does not explain the weakness of gravity
which has become known as the hierarchy problem.

Let us come back briefly to the anthropic considerations and show that
they allow to set an interval of admissible values for some
constants. Indeed, the anthropic principle does not tell whether the
constants are varying or not but it gives an insight on how special
our universe is. In such an approach, one studies the effect of small
variations of a constant around its observed value and tries to find a
phenomenon highly dependent on this constant.  This does not ensure
that there is no other set of constants (very different of the one
observed today) for which an organized universe may exist. It just
tells about the stability in a neighborhood of the location of our
universe in the parameter space of physical constants.  Rozental
(1988) argued that requiring that the lifetime of the proton
$\tau_p\sim\aem^{-2}(\hbar/m_{\rm p}c^2)\exp(1/\aem)\sim10^{32}$~yr is
larger than the age of the universe $t_{\rm u}\sim c/H_0\sim10^{17}$~s
implies that $\aem<1/80$. On the other side, if we believe in a grand
unified theory, this unification has to take place below the Planck
scale implying that $\aem>1/170$, this bound depending on assumptions
on the particle content. Similarly requiring that the electromagnetic
repulsion is much smaller than the attraction by strong interaction in
nuclei (which is necessary to have nuclei) leads to $\aem<1/20$. The
thermonuclear reactions in stars are efficient if $k_{_{\rm
B}}T\sim\aem m_{\rm p}c^2$ and the temperature of a star of radius
$R_{_{\rm S}}$ and mass $M_{_{\rm S}}$ can roughly be estimated as
$k_{_{\rm B}}T\sim GM_sm_{\rm p}/R_{_{\rm S}}$, which leads to the
estimate $\aem\sim10^{-3}$.  One can indeed think of many other
examples to put such bounds.  From the previous considerations, we
retain that the most stringent is
\begin{equation}
{1}/{170}<\aem<{1}/{80}.
\end{equation}
It is difficult to believe that these arguments can lead to
much sharper constraints. They are illustrative and give a hint that
the constants may not be ``random'' parameters without giving any
explanation for their values.

Rozental (1988) also argued that the existence of hydrogen and the
formation of complex elements in stars (mainly the possibility of the
reaction $3\alpha\rightarrow {}^{12}{\rm C}$) set constraints on the
values of the strong coupling constant. The production of ${}^{12}{\rm
C}$ in stars requires a triple tuning: (i) the decay lifetime of
${}^8{\rm Be}$, of order $10^{-6}$~s, is four orders of magnitude
longer than the time for two $\alpha$ particles to scatter, (ii) an
excited state of the carbon lies just above the energy of ${}^8{\rm
Be}+\alpha$ and finally (iii) the energy level of ${}^{16}{\rm O}$ at
7.1197~MeV is non resonant and below the energy of ${}^{12}{\rm
C}+\alpha$, of order 7.1616~MeV, which ensures that most of the carbon
synthetized is not destroyed by the capture of an $\alpha$-particle
(see Livio {\em et al.}, 2000). Oberhummer {\em et al.} (2000) showed
that outside a window of respectively 0.5\% and 4\% of the values of
the strong and electromagnetic forces, the stellar production of
carbon or oxygen will be reduced by a factor 30 to 1000 (see also
Pochet {\em et al}, 1991). Concerning the gravitational constant,
galaxy formation require $\ag<10^4$. Other such constraints on the
other parameters listed in the previous section can be obtained.

\subsection{Metrology}
\label{subsec_2.2}

The introduction of constants in physical law is closely related
to the existence of systems of units. For instance, Newton's law
states that the gravitational force between two masses is
proportional to each mass and inversely proportional to their
separation. To transform the proportionality to an equality one
requires the use of a quantity with dimension of ${\rm
m}^3\cdot{\rm kg}^{-1}\cdot s^{-2}$ independent of the separation
between the two bodies, of their mass, of their composition
(equivalence principle) and on the position (local position
invariance). With an other system of units this constant could have
simply been anything.

The determination of the laboratory value of constants relies
mainly on the measurements of lengths, frequencies, times,... (see
Petley, 1985 for a treatise on the measurement of constants and
Flowers and Petley, 2001, for a recent review). Hence, any
question on the variation of constants is linked to the definition
of the system of units and to the theory of measurement. The
choice of a base units affects the possible time variation of
constants.

The behavior of atomic matter is mainly determined by the value of the
electron mass and of the fine structure constant. The Rydberg energy
sets the (non-relativistic) atomic levels, the hyperfine structure
involves higher powers of the fine structure constant, and molecular
modes (including vibrational, rotational...modes) depend on the ratio
$m_{\rm e}/m_{\rm p}$. As a consequence, if the fine structure
constant is spacetime dependent, the comparison between several
devices such as clocks and rulers will also be spacetime
dependent. This dependence will also differ from one clock to another
so that {\it metrology becomes both device and spacetime dependent}.

Besides this first metrologic problem, the choice of units has
implications on the permissible variations of certain dimensionful
constant. As an illustration, we follow Petley (1983) who
discusses the implication of the definition of the meter. The
definition of the meter via a prototype platinum-iridium bar
depends on the interatomic spacing in the material used in the
construction of the bar. Atkinson (1968) argued that, at first
order, it mainly depends on the Bohr radius of the atom so that
this definition of the meter fixes the combination (\ref{Bohr}) as
constant. Another definition was based on the wavelength of the
orange radiation from krypton-86 atoms. It is likely that this
wavelength depends on the Rydberg constant and on the reduced mass
of the atom so that it ensures that $m_{\rm e}c^2\aem^2/2\hbar$ is
constant. The more recent definition of the meter as the length of
the path traveled by light in vacuum during a time of
$1/299792458$ of a second imposes the constancy of the speed of
light\footnote{Note that the velocity of light is not assigned a
fixed value {\it directly}, but rather the value is fixed as a
consequence of the definition of the meter.} $c$. Identically, the
definitions of the second as the duration of 9,192,631,770 periods
of the transition between two hyperfine levels of the ground state
of cesium-133 or of the kilogram via an international prototype
respectively impose that $m_{\rm e}^2c^2\aem^4/\hbar$ and $m_{\rm
p}$ are fixed.

Since the definition of a system of units and the value of the
fundamental constants (and thus the status of their constancy) are
entangled, and since the measurement of any dimensionful quantity is in
fact the measurements of a ratio to standards chosen as units, {\it it
only makes sense to consider the variation of dimensionless ratios}.

In theoretical physics, we often use the fundamental constants as
units (see McWeeny, 1973 for the relation between natural units
and SI units). The international system of units (SI) is more
appropriate to human size measurements whereas natural systems of
units are more appropriate to the physical systems they refer to.
For instance $\hbar$, $c$ and $G$ allows to construct the Planck
mass, time and length which are of great use as units while
studying high-energy physics and the same can be done from
$\hbar$, $e$, $m_{\rm e}$ and $\varepsilon_0$ to construct a unit
mass ($m_{\rm e}$), length ($4\pi\varepsilon_0 h^2/m_{\rm e} e^2$)
and time ($2\varepsilon_0h^3/\pi m_{\rm e} e^4$). A physical
quantity can always be decomposed as the product of a label
representing a standard quantity of reference and a numerical
value representing the number of times the standard has to be
taken to build the required quantity. It follows that a given
quantity $X$ that can be expressed as
$X=k_1F_1(\rm{m},\rm{kg},\rm{s},\ldots)$ with $k_1$ a
dimensionless quantity and $F_1$ a function of the base units
(here SI) to some power.  Let us decompose $X$ as
$X=k_2F_2(\hbar,e,c,\ldots)$ where $k_2$ is another dimensionless
constant and $F_2$ a function of a sufficient number of
fundamental constants to be consistent with the initial base
units. The time variation of $X$ is given by
$$
\frac{\dd\ln X}{\dd t}=\frac{\dd\ln k_1}{\dd t}+
\frac{\dd\ln F_1}{\dd t}=\frac{\dd\ln k_2}{\dd t}+
\frac{\dd\ln F_2}{\dd t}.
$$
Since only ${\dd k_1}/{\dd t}$ or ${\dd k_2}/{\dd t}$ can be
measured, it is necessary to have chosen a system of units, the
constancy of which is assumed (i.e. that either $\dd F_1/\dd t=0$
or $\dd F_2/\dd t=0$) to draw any conclusion concerning the time
variation of $X$, in the same way as the description of a motion
needs to specify a reference frame.

\begin{figure}
\begin{center}
\psfig{file=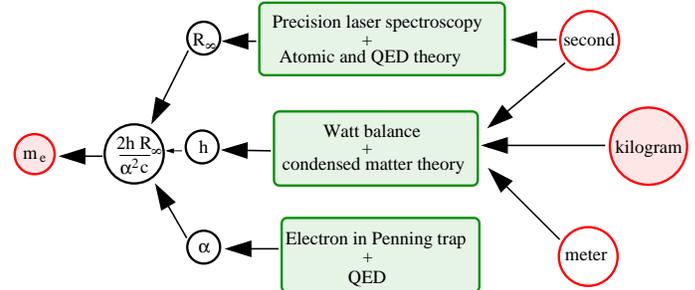,width=9cm}
\caption{ Sketch of the experimental and theoretical chain leading to
the determination of the electron mass. Note that, as expected, the
determination of $\aem$ requires no dimensional input. [From Mohr and
Taylor (2001)].}
\label{figme}
\end{center}
\end{figure}

To illustrate the importance of the choice of units and the
entanglement between experiment and theory while measuring a
fundamental constant, let us sketch how one determines $m_{\rm e}$
in the SI system (following Mohr and Taylor, 2001), i.e. in
kilogram (see figure~\ref{figme}). The kilogram is defined from a
platinum-iridium bar to which we have to compare the mass of the
electron. The key to this measurement is to express the electron
mass as $m_{\rm e}=2hR_\infty/\aem^2c$. From the definition of the
second, $R_\infty$ is determined by precision laser-spectroscopy
on hydrogen and deuterium and the theoretical expression for the
$1s$-$2s$ hydrogen transition as $\nu=(3/4)R_\infty c[1-\mu
+11\aem^2/48+(56\aem^3)/(9\pi)\ln\aem+\ldots]$ arising from QED.
The fine structure constant is determined by comparing theory and
experiment for the anomalous magnetic moment of the electron
(involving again QED). Finally, the Planck constant is determined
by a Watt balance comparing a Watt electrical power to a Watt
mechanical power (involving classical mechanics and classical
electromagnetism only: $h$ enters through the current and voltage
calibration based on two condensed-matter phenomena: Josephson and
quantum Hall effects so that it involves the theories of these two
effects).

As a conclusion, let us recall that (i) in general, the values of the
constants are not determined by a direct measurement but by a chain
involving both theoretical and experimental steps, (ii) they depend on
our theoretical understanding, (iii) the determination of a
self-consistent set of values of the fundamental constants results
from an adjustment to achieve the best match between theory and a
defined set of experiments (see e.g., Birge, 1929) (iv) that the
system of units plays a crucial role in the measurement chain, since
for instance in atomic units, the mass of the electron could have been
obtained directly from a mass ratio measurement (even more precise!)
and (v) fortunately the test of the variability of the constants does
not require {\it a priori} to have a high-precision value of the
considered constant.

In the following, we will thus focus on the variation of
dimensionless ratios which, for instance, characterize the
relative magnitude of two forces, and are independent of the
choice of the system of units and of the choice of standard rulers
or clocks. Let us note that some (hopeless) attempts to constraint
the time variation of dimensionful constants have been tried and
will be briefly discussed in Section~\ref{sec_5.5}. This does not
however mean that a physical theory cannot have dimensionful
varying constants. For instance, a theory of varying fine
structure constant can be implemented either as a theory with
varying electric charge or varying speed of light.

\subsection{Overview of the methods}
\label{subsec_2.3}

Before going into the details of the constraints, it is worth
taking some time to discuss the kind of experiments or
observations that we need to consider and what we can hope to
infer from them.

As emphasized in the previous section, we can only measure the
variation of dimensionless quantities (such as the ratio of two
wavelengths, two decay rates, two cross sections ...) and the idea is
to pick up a physical system which depends strongly on the value of a
set of constants so that a small variation will have dramatic
effects. The general strategy is thus to constrain the spacetime
variation of an observable quantity as precisely as possible and then
to relate it to a set of fundamental constants.

Basically, we can split all the methods into three classes: (i) {\it
atomic methods} including atomic clocks, quasar absorption spectra and
observation of the cosmic microwave background radiation (CMBR) where
one compares ratios of atomic transition frequencies. The CMB
observation depends on the dependence of the recombination process on
$\aem$; (ii) {\it nuclear methods} including nucleosynthesis,
$\alpha$- and $\beta$-decay, Oklo reactor for which the observables
are respectively abundances, lifetimes and cross sections; and (iii)
{\it gravitational methods} including the test of the violation of the
universality of free fall where one constrains the relative
acceleration of two bodies, stellar evolution\ldots

These methods are either {\it experimental} (e.g. atomic clocks) for
which one can have a better control of the systematics, {\it
observational} (e.g. geochemical, astrophysical and cosmological
observations) or {\it mixed} ($\alpha$- and $\beta$-decay,
universality of free fall).  This sets the time scales on which a
possible variation can be measured. For instance, in the case of the
fine structure constant (see Section~\ref{sec_4}), one expects to be
able to constrain a relative variation of $\aem$ of order $10^{-8}$
[geochemical (Oklo)], $10^{-5}$ [astrophysical (quasars)],
$10^{-3}-10^{-2}$ [cosmological methods], $10^{-13}-10^{-14}$
[laboratory methods] respectively on time scales of order $10^9$~yr,
$10^9-10^{10}$~yr, $10^{10}$~yr and $1-12$ months. This brings up the
question of the comparison and of the compatibility of the different
measurements since one will have to take into account e.g. the rate of
change of $\aem$ which is often assumed to be constant. In general,
this requires to specify a model both to determine the law of
evolution and the links between the constants. Long time scale
experiments allow to test a slow drift evolution while short time
scale experiments enable to test the possibility of a rapidly varying
constant.

The next step is to convert the bound on the variation of some
measured physical quantities (decay rate, cross section,...) into
a bound on some constants. It is clear that in general (for atomic
and nuclear methods at least) it is impossible to consider the
electromagnetic, weak and strong effects independently so that
this latter step involves some assumptions.

{\it Atomic methods} are mainly based on the comparison of the
wavelengths of different transitions. The non relativistic spectrum
depends mainly on $R_\infty$ and $\mu$, the fine structure on
$R_\infty\aem^2$ and the hyperfine structure on $g_{\rm
p}R_\infty\aem^2$. Extending to molecular spectra to include
rotational and vibrational transitions allows to have access to
$\mu$. It follows that we can hope to disentangle the observations of
the comparisons of different transitions to constrain on the
variation of $(\aem,\mu,g_{\rm p})$. The exception is CMB which
involves a dependence on $\aem$ and $m_{\rm e}$ mainly due to the
Thomson scattering cross section and the ionization fraction.
Unfortunately the effect of these parameters have to be disentangled
from the dependence on the usual cosmological parameters which render
the interpretation more difficult.

The internal structure and mass of the proton and neutron are
completely determined by strong gauge fields and quarks
interacting together. Provided we can ignore the quark masses and
electromagnetic effects, the whole structure is only dependent on
an energy scale $\Lambda_{_{\rm QCD}}$. It follows that the
stability of the proton greatly depends on the electromagnetic
effects and the masses $m_{\rm u}$ and $m_{\rm d}$ of the up and
down quarks.  In nuclei, the interaction of hadrons can be thought
to be mediated by pions of mass $m_\pi^2\sim m_{\rm p}(m_{\rm
u}+m_{\rm d})$. Since the stability of the nucleus mainly results
from the balance between this attractive nuclear force, the
nucleon degeneracy pressure and the Coulomb repulsion, it will
mainly involve $m_{\rm u}$, $m_{\rm d}$, $\aem$.

{\it Big bang nucleosynthesis} depends on $G$ (expansion rate),
$\gfermi$ (weak interaction rates), $\as$ (binding of light elements),
$\aem$ (via the electromagnetic contribution to $m_{\rm n}-m_{\rm p}$
but one will also have to take into account the contribution of a
possible variation of the mass of the quarks, $m_{\rm u}$ and $m_{\rm
d}$). Besides, if $m_{\rm n}-m_{\rm p}$ falls below $m_{\rm e}$ the
$\beta$-decay of the neutron is no longer energetically possible. The
abundance of helium is mainly sensitive to the freeze-out temperature
and the neutron lifetime and heavier element abundances to the nuclear
rates.

All {\it nuclear methods} involve a dependence on the mass of the
nuclei of charge $Z$ and atomic number $A$
$$
m(A,Z)=Zm_{\rm p}+(A-Z)m_{\rm n}+E_{_{\rm S}}+E_{_{\rm EM}},
$$
where $E_{_{\rm S}}$ and $E_{_{\rm EM}}$ are respectively the
strong and electromagnetic contributions to the binding energy.
The Bethe-Weiz\"acker formula gives that
\begin{equation}\label{bethe}
E_{_{\rm EM}}=98.25\frac{Z(Z-1)}{A^{1/3}}\aem\,{\rm MeV}.
\end{equation}
If we decompose $m_{\rm p}$ and $m_{\rm n}$ as (see Gasser and
Leutwyler, 1982) $m_{({\rm p,n})}=u_3 +b_{({\rm u,d})}m_{\rm
u}+b_{({\rm d,u})}m_{\rm d}+B_{({\rm p,n})}\aem$ where $u_3$ is the
pure QCD approximation of the nucleon mass ($b_{\rm u}$, $b_{\rm d}$
and $B_{({\rm n,p})}/u_3$ being pure numbers), it reduces to
\begin{eqnarray}\label{mass}
m(A,Z)&=&\left(Au_3+E_{_{\rm S}}\right)\\ &+&
(Zb_{\rm u}+Nb_{\rm d})m_{\rm u}+(Zb_{\rm d}+Nb_{\rm u})m_{\rm d}\nonumber\\
&+&\left(ZB_{\rm p}+NB_{\rm n}+98.25\frac{Z(Z-1)}{A^{1/3}}\,{\rm
MeV}\right)\aem,\nonumber
\end{eqnarray}
with $N=A-Z$, the neutron number. This depends on our understanding of
the description of the nucleus and can be more sophisticated. For an
atom, one would have to add the contribution of the electrons,
$Zm_{\rm e}$. The form (\ref{mass}) depends on strong, weak and
electromagnetic quantities. The numerical coefficients $B_{({\rm
n,p})}$ are given explicitly by (Gasser and Leutwiller, 1982)
\begin{equation}\label{gl}
B_{\rm p}\aem=0.63\,{\rm MeV}\quad
B_{\rm n}\aem=-0.13\,{\rm MeV}.
\end{equation}

It follows that it is in general difficult to disentangle the
effect of each parameter and compare the different methods. For
instance comparing the constraint on $\mu$ obtained from
electromagnetic methods to the constraints on $\as$ and $\gfermi$
from nuclear methods requires to have some theoretical input such
as a theory to explain the fermion masses. Moreover, most of the
theoretical models predict a variation of the coupling constants
from which one has to infer the variation of $\mu$ etc...

For macroscopic bodies, the mass has also a negative
contribution
\begin{equation}\label{llr8}
  \Delta m(G)=-\frac{G}{2c^2}\int\frac{\rho(\vec r)\rho(\vec r')}{|\vec r-\vec r'|}
\dd^3\vec r\dd^3\vec r'
\end{equation}
from the gravitational binding energy. As a conclusion, from
(\ref{mass}) and (\ref{llr8}), we expect the mass to depend on all the
coupling constant, $m(\aem,\aw,\as,\ag,...)$.

This has a profound consequence concerning the motion of any
body. Let $\alpha$ be any fundamental constant, assumed to be a
scalar function and having a time variation of cosmological origin
so that in the privileged cosmological rest-frame it is given by
$\alpha(t)$. A body of mass $m$ moving at velocity $\vec v$ will
experience an anomalous acceleration
\begin{eqnarray}\label{hh}
\delta\vec a&\equiv&\frac{1}{m}\frac{\dd m\vec v}{\dd t}-
            \frac{\dd \vec v}{\dd t}
            =\frac{\partial\ln m}{\partial\alpha}\dot\alpha\vec v.
\end{eqnarray}
Now, in the rest-frame the body, $\alpha$ has a spatial dependence
$\alpha[(t'+\vec v.\vec r'/c^2)/\sqrt{1-v^2/c^2}]$ so that, as long as
$v\ll c$, $\nabla\alpha=({\dot \alpha}/{c^2})\vec v$.  The anomalous
acceleration can thus be rewritten as
\begin{equation}
\delta\vec a=-\left(\frac{\alpha}{m}\frac{\delta
mc^2}{\delta\alpha}\right)\nabla\ln\alpha.
\end{equation}
In the most general case, for non-relativistically moving body,
\begin{equation}
\delta\vec a=-\left(\frac{\alpha}{m}\frac{\delta
mc^2}{\delta\alpha}\right)\left(\frac{\nabla\alpha}{\alpha}
+\frac{\dot \alpha}{\alpha}\frac{\vec v}{c^2}\right).
\end{equation}
It reduces to Eq.~(\ref{hh}) in the appropriate limit and the
additional gradient term will be produced by local matter sources.
This anomalous acceleration is generated by the change in the
(electromagnetic, gravitational,...) binding energy (Dicke, 1964;
Dicke, 1969; Haugan, 1979; Eardley, 1979; Nordtvedt, 1990).  Besides,
the $\alpha$-dependence is a priori composition-dependent (see
e.g. Eq.~\ref{mass}). As a consequence, any variation of the
fundamental constants will entail a violation of the universality of
free fall: the total mass of the body being space dependent, an
anomalous force appears if energy is to be conserved. The variation of
the constants, deviation from general relativity and violation of the
weak equivalence principle are in general expected together, e.g. if
there exists a new interaction mediated by a massless scalar field.

Gravitational methods include the constraints that can be derived
from the test of the theory of gravity such as the test of the
universality of free fall, the motion of the planets in the Solar
system, stellar and galactic evolutions. They are based on the
comparison of two time scales, the first (gravitational time)
dictated by gravity (ephemeris, stellar ages,\ldots) and the
second (atomic time) is determined by any system not determined by
gravity (e.g. atomic clocks,\ldots) (Canuto and Goldman, 1982).
For instance planet ranging, neutron star binaries observations,
primordial nucleosynthesis and paleontological data allow to
constraint the relative variation of $G$ respectively to a level
of $10^{-12}-10^{-11}$, $10^{-13}-10^{-12}$, $10^{-12}$,
$10^{-10}$ per year.

Attacking the full general problem is a hazardous and dangerous
task so that we will first describe the constraints obtained in
the literature by focusing on the fine structure constant and the
gravitational constant and we will then extend to some other (less
studied) combinations of the constants. Another and complementary
approach is to predict the mutual variations of different
constants in a given theoretical model (see Section~\ref{sec_7}).

\section{Fine structure constant}\label{sec_4}
\subsection{Geological constraints}
\label{subsec_4.1}

\subsubsection{The Oklo phenomenon}\label{oklo}

Oklo is a prehistoric natural fission reactor that operated about
$2\times10^{9}$~yr ago during $(2.3\pm0.7)\times10^5$~yr in the Oklo
uranium mine in Gabon. This phenomenon was discovered by the French
Commissariat \`a l'\'Energie Atomique in 1972 (see Naudet, 1974,
Maurette, 1976 and Petrov, 1977 for early studies and Naudet, 2000 for
a general review) while monitoring for uranium ores.  Two billion
years ago, uranium was naturally enriched (due to the difference of
decay rate between ${}^{235}{\rm U}$ and ${}^{238}{\rm U}$) and
${}^{235}{\rm U}$ represented about 3.68\% of the total uranium
(compared with 0.72\% today). Besides, in Oklo the concentration of
neutron absorbers which prevent the neutrons from being available for
the chain fission was low; water played the role of moderator and
slowed down fast neutrons so that they can interact with other
${}^{235}{\rm U}$ and the reactor was large enough so that the
neutrons did not escape faster than they were produced.

From isotopic abundances of the yields, one can extract informations
about the nuclear reactions at the time the reactor was operational
and reconstruct the reaction rates at that time.  One of the key
quantity measured is the ratio ${}^{149}_{62}{\rm
Sm}/{}^{147}_{62}{\rm Sm}$ of two light isotopes of samarium which are
not fission products. This ratio of order of 0.9 in normal samarium,
is about 0.02 in Oklo ores. This low value is interpreted by the
depletion of ${}^{149}_{62}{\rm Sm}$ by thermal neutrons to which it
was exposed while the reactor was active.

Shlyakhter (1976) pointed out that the capture cross section of
thermal neutron by ${}^{149}_{62}{\rm Sm}$
\begin{equation}\label{oklo1}
n+{}^{149}_{62}{\rm Sm}\longrightarrow {}^{150}_{62}{\rm Sm}+\gamma
\end{equation}
is dominated by a capture resonance of a neutron of energy of about
0.1 eV. The existence of this resonance is a consequence of an almost
cancellation between the electromagnetic repulsive force and the
strong interaction.

To obtain a constraint, one first needs to measure the neutron capture
cross section of ${}^{149}_{62}{\rm Sm}$ at the time of the reaction
and to relate it to the energy of the resonance. One has finally to
translate the constraint on the variation of this energy on a
constraint on the time variation of the considered constant.

The cross section of the neutron capture (\ref{oklo1}) is strongly
dependent on the energy of a resonance at $E_{r}=97.3$~meV and is
well described by the Breit-Wigner formula
\begin{equation}\label{oklo2}
\sigma_{(n,\gamma)}(E)=\frac{g_0\pi}{2}\frac{\hbar^2}{m_{\rm n}E}
\frac{\Gamma_{\rm n}\Gamma_\gamma}{(E-E_r)^2+\Gamma^2/4}
\end{equation}
where $g_0\equiv(2J+1)(2s+1)^{-1}(2I+1)^{-1}$ is a statistical factor
which depends on the spin of the incident neutron $s=1/2$, of the
target nucleus $I$ and of the compound nucleus $J$; for the reaction
(\ref{oklo1}), we have $g_0=9/16$. The total width
$\Gamma\equiv\Gamma_{\rm n}+\Gamma_\gamma$ is the sum of the neutron partial
width $\Gamma_{\rm n}=0.533$~meV (at $E_r$) and of the radiative partial
width $\Gamma_\gamma=60.5$~meV.

The effective absorption cross section is defined by
\begin{equation}\label{hatsig}
\hat\sigma(E_r,T)=\frac{1}{v_0}\frac{2}{\sqrt{\pi}}
\int\sigma_{(n,\gamma)}(E)\sqrt{\frac{2E}{m_{\rm n}}}
\frac{\hbox{e}^{-E/k_{_{\rm B}}T}}{(k_{_{\rm B}}T)^{3/2}}\sqrt{E}\dd E
\end{equation}
where the velocity $v_0=2200\,{\rm m\cdot s}^{-1}$ corresponds to
an energy $E_0=25.3$~meV and the effective neutron flux is
similarly given by
\begin{equation}
\hat\phi=v_0\frac{2}{\sqrt{\pi}}
\int\sqrt{\frac{2E}{m_{\rm n}}}
\frac{\hbox{e}^{-E/k_{_{\rm B}}T}}{(k_{_{\rm B}}T)^{3/2}}\sqrt{E}\dd E.
\end{equation}

The samples of the Oklo reactors were exposed (Naudet, 1974) to an
integrated effective fluence $\int\hat\phi\dd t$ of about
$10^{21}$~neutron$\cdot{\rm cm}^{-2}=1~{\rm kb}^{-1}$. It implies
that any process with a cross section smaller than 1~kb can be
neglected in the computation of the abundances; this includes
neutron capture by ${}^{144}_{62}{\rm Sm}$ and ${}^{148}_{62}{\rm
Sm}$. On the other hand, the fission of ${}^{235}_{92}{\rm U}$,
the capture of neutron by ${}^{143}_{60}{\rm Nd}$ and by
${}^{149}_{62}{\rm Sm}$ with respective cross sections
$\sigma_{5}\simeq0.6$~kb, $\sigma_{143}\sim0.3$~kb and
$\sigma_{149}\geq70$~kb are the dominant processes. It follows
that the equations of evolution for the number densities
$N_{147}$, $N_{148}$, $N_{149}$ and $N_{235}$ of
${}^{147}_{62}{\rm Sm}$, ${}^{148}_{62}{\rm Sm}$,
${}^{149}_{62}{\rm Sm}$ and ${}^{235}_{92}{\rm U}$ are (Damour and
Dyson, 1996; Fujii {\em et al.}, 2000)
\begin{eqnarray}
\frac{\dd N_{147}}{\dd t}&=&-\hat\sigma_{147}\hat\phi N_{147}+
       \hat\sigma_{f235}\hat\phi N_{235}\\
\frac{\dd N_{148}}{\dd t}&=&\hat\sigma_{147}\hat\phi N_{147}\\
\frac{\dd N_{149}}{\dd t}&=&-\hat\sigma_{149}\hat\phi N_{149}+
       \hat\sigma_{f235}\hat\phi N_{235}\\
\frac{\dd N_{235}}{\dd t}&=&-\sigma^*_5N_{235}
\end{eqnarray}
where the system has to be closed by using a modified absorption cross
section $\sigma^*_5=\sigma_5(1-C)$ (see references in Damour and Dyson,
1996). This system can be integrated under the assumption that the cross
sections are constant and the result compared with the natural
abundances of the samarium to extract the value of $\hat\sigma_{149}$ at
the time of the reaction.  Shlyakhter (1976) first claimed that
$\hat\sigma_{149}=55\pm8$~kb (at cited by Dyson, 1978).  Damour and
Dyson (1996) re-analized this result and found that $57~{\rm
kb}\leq\hat\sigma_{149}\leq93\,{\rm kb}$. Fujii {\em et al.} (2000)
found that $\hat\sigma_{149}=91\pm6$~kb.

By comparing this measurements to the current value of the cross
section and using (\ref{hatsig}) one can transform it into a
constraint on the variation of the resonance energy. This step
requires to estimate the neutron temperature. It can be obtained
by using informations from the abundances of other isotopes such
as lutetium and gadolinium.  Shlyakhter (1976) deduced that
$|\Delta E_r|<20\,\hbox{meV}$ but assumed the much too low
temperature of $T=20^{\rm o}$~C. Dyson and Damour (1996) allowed
the temperature to vary between $180^{\rm o}$~C and $700^{\rm
o}$~C and deduced the conservative bound $-120\,\hbox{meV}<\Delta
E_r<90\,\hbox{meV}$ and Fujii {\em et al.}  (2000) obtained two
branches, the first compatible with a null variation $\Delta
E_r=9\pm11$~meV and the second indicating a non-zero effect
$\Delta E_r=-97\pm8$~meV both for $T=200-400^{\rm o}$~C and argued
that the first branch was favored.

Damour and Dyson (1996) related the variation of $E_r$ to the fine
structure constant by taking into account that the radiative capture
of the neutron by ${}^{149}_{62}{\rm Sm}$ corresponds to the existence
of an excited quantum state ${}^{150}_{62}{\rm Sm}$ (so that
$E_r=E_{150}^*-E_{149}-m_{\rm n}$) and by assuming that the nuclear
energy is independent of $\aem$. It follows that the variation of
$\aem$ can be related to the difference of the Coulomb energy of these
two states. The computation of this latter quantity is difficult and
requires to be related to the mean-square radii of the protons in the
isotopes of samarium and Damour and Dyson (1996) showed that the
Bethe-Weiz\"acker formula (\ref{bethe}) overestimates by about a
factor the 2 the $\aem$-sensitivity to the resonance energy.  It
follows from this analysis that
\begin{equation}
\aem\frac{\Delta E_r}{\Delta\aem}\simeq-1.1\,{\rm MeV},
\end{equation}
which, once combined with the constraint on $\Delta E_r$, implies
\begin{equation}
-0.9\times10^{-7}<\Delta\aem/\aem<1.2\times10^{-7}
\end{equation}
corresponding to the range $-6.7\times10^{-17}\,{\rm
yr}^{-1}<\dot\aem/\aem<5.0\times10^{-17}\,{\rm yr}^{-1}$ if $\dot\aem$
is assumed constant. This tight constraint arises from the large
amplification between the resonance energy ($\sim0.1$~eV) and the
sensitivity ($\sim1$~MeV).  Fujii {\em et al.} (2000) re-analyzed the
data and included data concerning gadolinium and found the favored
result $\dot\aem/\aem=(-0.2\pm0.8)\times10^{-17}\,{\rm yr}^{-1}$ which
corresponds to
\begin{equation}
\Delta\aem/\aem=(-0.36\pm1.44)\times10^{-8}
\end{equation}
and another branch $\dot\aem/\aem=(4.9\pm0.4)\times10^{-17}\,{\rm
yr}^{-1}$. The first bound is favored given the constraint on the
temperature of the reactor. Nevertheless, the non-zero result
cannot be eliminated, even using results from gadolinium
abundances (Fujii, 2002). Note however that spliting the analysis
in two branches seems to be at odd with the aim of obtaining a
constraint. Olive {\em et al}. (2002) refined the analysis and
confirmed the previous results.

Earlier studies include the original work by Shlyakhter (1976) who
found that $|\dot\aem/\aem|<10^{-17}\,{\rm yr}^{-1}$ corresponding to
\begin{equation}
|\Delta\aem/\aem|<1.8\times10^{-8}.
\end{equation}
In fact he stated that the variation of the strong interaction
coupling constant was given by $\Delta g_{_{\rm S}}/g_{_{\rm S}}\sim
\Delta E_r/V_0$ where $V_0\simeq 50\,{\rm MeV}$ is the depth of a
square potential well.  Arguing that the Coulomb force increases the
average inter-nuclear distance by about 2.5\% for $A\sim150$, he
concluded that $\Delta\aem/\aem\sim20\Delta g_{_{\rm S}}/g_{_{\rm
S}}$, leading to $|\dot\aem/\aem|<10^{-17}\,{\rm yr}^{-1}$.  Irvine
(1983a,b) quoted the bound $|\dot\aem/\aem|<5\times10^{-17}\,{\rm
yr}^{-1}$. The analysis of Sisterna and Vucetich (1990) used,
according to Damour and Dyson (1996) an ill-motivated
finite-temperature description of the excited state of the compound
nucleus. Most of the studies focus on the effect of the fine structure
constant mainly because the effects of its variation can be well
controlled but, one would also have to take the effect of the
variation of the Fermi constant, or identically $\aw$, (see
Section~\ref{subsec_5.1}).  Horv\'ath and Vucetich (1988) interpreted
the results from Oklo in terms of null-redshift experiments.

\subsubsection{$\alpha$-decay}

The fact that $\alpha$-decay can be used to put constraints on the
time variation of the fine structure constant was pointed out by
Wilkinson (1958) and then revived by Dyson (1972, 1973). The main
idea is to extract the $\aem$-dependence of the decay rate and to
use geological samples to bound its time variation.

The decay rate, $\lambda$, of the $\alpha$-decay of a nucleus ${}^A_Z{\rm X}$
of charge $Z$ and atomic number $A$
\begin{equation}
{}_{Z+2}^{A+4}{\rm X}\longrightarrow {}_Z^A{\rm X}+ {}_2^4{\rm He}
\end{equation}
is governed by the penetration of the Coulomb barrier
described by the Gamow theory and well approximated by
\begin{equation}
\lambda\simeq\Lambda(\aem,v)\hbox{e}^{-4\pi Z\aem c/v}
\end{equation}
where $v$ is the escape velocity of the $\alpha$ particle and
where $\Lambda$ is a function that depends slowly on $\aem$ and
$v$. It follows that the variation of the decay rate with respect
to the fine structure constant is well approximated by
\begin{equation}
\frac{\dd\ln\lambda}{\dd\aem}\simeq-4\pi Z\frac{c}{v}\left(1-
\frac{1}{2}\frac{\dd\ln \Delta E}{\dd\ln\aem}\right)
\end{equation}
where $\Delta E\equiv 2mv^2$ is the decay energy. Considering that the
total energy is the sum of the nuclear energy $E_{_{\rm nuc}}$ and
of the Coulomb energy $E_{_{\rm EM}}/80~{\rm MeV}\simeq
Z(Z-1)A^{-1/3}\aem$ and that the former does not depend on $\aem$,
one deduces that
\begin{equation}\label{40}
\frac{\dd\ln \Delta E}{\dd\ln\aem}\simeq\left(
\frac{\Delta E}{0.6\,{\rm MeV}}\right)^{-1}f(A,Z)
\end{equation}
with $f(A,Z)\equiv\left[(Z+2)(Z+1)(A+4)^{-1/3}\right.$ $\left.-Z(Z-1)A^{-1/3}\right]$.  It
follows that the sensitivity of the decay rate on the fine structure
constant is given by
\begin{eqnarray}\label{s}
s&\equiv&\frac{\dd\ln\lambda}{\dd\ln\aem}\nonumber\\
&\simeq&
4\pi Z\frac{c}{v}\aem\left\lbrace
\left(\frac{0.3\,{\rm MeV}}{\Delta E}\right)f(A,Z)-1\right\rbrace.
\end{eqnarray}
This result can be qualitatively understood since an increase of
$\aem$ induces an increase in the height of the Coulomb barrier at
the nuclear surface while the depth of the nuclear potential below
the top remains the same. It follows that the $\alpha$ particle
escapes with greater energy but at the same energy below the top
of the barrier. Since the barrier becomes thinner at a given
energy below its top, the penetrability increases.  This
computation indeed neglects the effect of a variation of $\aem$ on
the nucleus that can be estimated to be dilated by about 1\% if
$\aem$ increases by 1\%.

Wilkinson (1958) considered the most favorable $\alpha$-decay reaction
which is the decay of $^{238}_{92}{\rm U}$
\begin{equation}
^{238}_{92}{\rm U}\rightarrow {}^{235}_{90}{\rm Th}+{}^4_2{\rm He}
\end{equation}
for which $\Delta E\simeq4.27\,$MeV ($s\simeq540$). By comparing the
geological dating of the Earth by different methods, he concluded
that the decay constant $\lambda$ of $^{238}{\rm U}$, $^{235}{\rm U}$
and ${}^{232}{\rm Th}$ have not changed by more than a factor 3 or 4
during the last $3-4\times10^{9}$~years from which it follows that
$|\dot\aem/\aem|<2\times10^{-12}\,{\rm yr}^{-1}$ and thus
\begin{equation}
\left|{\Delta\aem}/{\aem}\right|<8\times10^{-3}.
\end{equation}
This bound is very rough but it agrees with Oklo on comparable time
scale. This constraint was revised by Dyson (1972) who claimed that
the decay rate has not changed by more than 20\%, during the past
$2\times10^9$ years, which implies
\begin{equation}
\left|{\Delta\aem}/{\aem}\right|<4\times10^{-4}.
\end{equation}
These data were recently revisited by Olive {\em et al.} (2002). Using
laboratory and meteoric data for $^{147}{\rm Sm}$ ($\Delta
E\simeq2.31$~MeV, $s\simeq770$) for which $\Delta\lambda/\lambda$ was
estimated to be of order $7.5\times10^{-3}$ they concluded that
\begin{equation}
\left|{\Delta\aem}/{\aem}\right|<10^{-5}.
\end{equation}

\subsubsection{Spontaneous fission}

$\alpha$-emitting nuclei are classified into four generically
independent decay series (the thorium, neptunium, uranium and
actinium series). The uranium series is the longest known series.
It begins with $^{238}_{92}{\rm U}$, passes a second time through
$Z=92$ ($^{234}_{92}{\rm U}$) as a consequence of an
$\alpha$-$\beta$-decay and then passes by five $\alpha$-decays and
finishes by an $\alpha$-$\beta$-$\beta$-decay to end with
$^{206}_{82}{\rm Pb}$.  The longest lived member is
$^{238}_{92}{\rm U}$ with a half-life of $4.47\times10^9$~yr,
which four orders of magnitude larger than the second longest
lived elements.  $^{238}_{92}{\rm U}$ thus determines the time
scale of the whole series.

The expression of the lifetime in the case of spontaneous fission
can be obtained from Gamow theory of $\alpha$-decay by replacing
the charge $Z$ by the product of the charges of the two fission
products.

Gold (1968) studied the fission of $^{238}_{92}{\rm U}$ with a decay
time of $7\times10^{-17}\,{\rm yr}^{-1}$.  He obtained a sensitivity
(\ref{s}) of $s=120$. Ancient rock samples allow to conclude, after
comparison of rock samples dated by potassium-argon and
rubidium-strontium, that the decay time of $^{238}_{92}{\rm U}$ has
not varied by more than 10\% in the last $2\times10^9$~yr. Indeed, the
main uncertainty comes from the dating of the rock. Gold (1968)
concluded on that basis that
\begin{equation}
\left|{\Delta\aem}/{\aem}\right|<4.66\times10^{-4}
\end{equation}
which corresponds to
$\left|{\dot\aem}/{\aem}\right|<2.3\times10^{-13}\,{\rm
yr}^{-1}$ if one assumes that $\dot\aem$ is constant. This bound
is indeed comparable, in order of magnitude, to the one obtained
by $\alpha$-decay data.

Chitre and Pal (1968) compared the uranium-lead and potassium-argon
dating methods respectively governed by $\alpha$- and $\beta$-
decay to date stony meteoric samples. Both methods have different
$\aem$-dependence (see below) and they concluded that
\begin{equation}
\left|{\Delta\aem}/{\aem}\right|<(1-5)\times10^{-4}.
\end{equation}
Dyson (1972) argued on similar basis that the decay rate of
$^{238}_{92}{\rm U}$ has not varied by more than 10\% in the past
$2\times10^9$~yr so that
\begin{equation}
\left|{\Delta\aem}/{\aem}\right|<10^{-3}.
\end{equation}

\subsubsection{$\beta$-decay}

Dicke (1959) stressed that the comparison of the rubidium-strontium
and potassium-argon dating methods to uranium and thorium rates
constrains the variation of $\aem$. He concluded that there was no
evidence to rule out a time variation of the $\beta$-decay rate.

Peres (1968) discussed qualitatively the effect of a fine
structure constant increasing with time arguing that the nuclei
chart would have then been very different in the past since the
stable heavy element would have had $N/Z$ ratios much closer to
unity (because the deviation from unity is mainly due to the
electrostatic repulsion between protons). For instance $^{238}{\rm
U}$ would be unstable against double $\beta$-decay to $^{238}{\rm
Pu}$. One of its arguments to claim that $\aem$ has almost not
varied lies in the fact that $^{208}{\rm Pb}$ existed in the past
as $^{208}{\rm Rn}$, which is a gas, so that the lead ores on
Earth would be uniformly distributed.

As long as long-lived isotopes are concerned for which the decay
energy $\Delta E$ is small, we can use a non-relativistic
approximation for the decay rate
\begin{equation}
\lambda=\Lambda_\pm \left(\Delta E\right)^{p_\pm}
\end{equation}
respectively for $\beta^-$-decay and electron capture.
$\Lambda_\pm$ are functions that depend smoothly on $\aem$ and
which can thus be considered constant, $p_+=\ell+3$ and
$p_-=2\ell+2$ are the degrees of forbiddenness of the transition.
For high-$Z$ nuclei with small decay energy $\Delta E$, the
exponent $p$ becomes $p=2+\sqrt{1-\aem^2Z^2}$ and is independent
of $\ell$. It follows that the sensitivity (\ref{s}) becomes
\begin{equation}
s=p\frac{\dd\ln\Delta E}{\dd\ln\aem}.
\end{equation}
The second factor can be estimated exactly as in Eq.~(\ref{40})
for $\alpha$-decay but with $f(A,Z)=\pm(2Z+1)A^{-1/3}[0.6\,{\rm
MeV}/\Delta E]$, the $-$, $+$ signs corresponding respectively to
$\beta$-decay and electron capture.

The laboratory determined decay rates of rubidium to strontium by
$\beta$-decay
\begin{equation}
{}^{87}_{37}{\rm Rb}\longrightarrow{}^{87}_{38}{\rm Sr}+\bar\nu_e+e^-
\end{equation}
and to potassium to argon by electron capture
\begin{equation}
{}^{40}_{19}{\rm K}+e^-\longrightarrow{}^{40}_{18}{\rm Ar}+\nu_e
\end{equation}
are respectively $1.41\times10^{-11}\,{\rm yr}^{-1}$ and
$4.72\times10^{-10}$ yr$^{-1}$. The decay energies are respectively
$\Delta E=0.275$~MeV and $\Delta E=1.31$~MeV so that $s\simeq-180$ and
$s\simeq-30$. Peebles and Dicke (1962) compared these laboratories
determined values with their abundances in rock samples after dating
by uranium-lead method and with meteorite data (dated by uranium-lead
and lead-lead). They concluded that the variation of $\aem$ with $\ag$
cannot be ruled out by comparison to meteorite data.  Later, Yahil
(1975) used the concordance of the K-Ar and Rb-Sr geochemical ages to
put the limit
\begin{equation}
|\Delta\aem/\aem|<1.2
\end{equation}
over the past $10^{10}\,{\rm yr}$.

The case of the decay of osmium to rhenium by electron emission
\begin{equation}
{}^{187}_{75}{\rm Re}\longrightarrow{}^{187}_{76}{\rm Os}+\bar\nu_e+e^-
\end{equation}
was first considered by Peebles and Dicke (1962). They noted that
the very small value of its decay energy $\Delta E\simeq2.5$~keV
makes it a very sensitive indicator of the variation of $\aem$. In
that case $p\simeq 2.8$ so that $s\simeq-18000$. It follows that a
change of about $10^{-2}$\% of $\aem$ will induce a change in the
decay energy of order of the keV, that is of the order of the
decay energy itself. With  a time decreasing $\aem$, the decay
rate of rhenium will have slowed down and then osmium will have
become unstable.  Peebles and Dicke (1962) did not have reliable
laboratory determination of the decay rate to put any constraint.
Dyson (1967) compared the isotopic analysis of molybdenite ores,
the isotopic analysis of 14 iron meteorites and laboratory
measurements of the decay rate. Assuming that the variation of the
decay energy comes entirely from the variation of $\aem$, he
concluded that
\begin{equation}
\left|{\Delta\aem}/{\aem}\right|<9\times10^{-4}
\end{equation}
during the past $3\times10^9$ years. In a re-analysis (Dyson,
1972) he concluded that the rhenium decay-rate did not change by
more than 10\% in the past $10^9$ years so that
\begin{equation}
\left|{\Delta\aem}/{\aem}\right|<5\times10^{-6}.
\end{equation}
Using a better determination of the decay rate of ${}^{187}_{75}{\rm
Re}$ based on the growth of ${}^{187}_{}{\rm Os}$ over a 4-year period
into a large source of osmium free rhenium, Lindner {\em et al.} (1986)
deduced that
\begin{equation}
{\Delta\aem}/{\aem}=(-4.5\pm9)\times10^{-4}
\end{equation}
over a $4.5\times10^9$~yr period. This was recenlty updated (Olive
{\it et al.}, 2002) to take into account the improvements in the
analysis of the meteorite data which now show that the half-life has
not varied by more than $0.5\%$ in the past 4.6~Gyr (i.e. a redshift
of about 0.45). This implies that
\begin{equation}\label{meteorite}
\left|{\Delta\aem}/{\aem}\right|<3\times10^{-7}.
\end{equation}

We just reported the values of the decay rates as used at the time of
the studies. One could want to update these constraints by using new
results on the measurements on the decay rate,\ldots.  Even though,
they will not, in general, be competitive with the bounds obtained by
other methods. These results can also be altered if the neutrinos are
massive.

\subsubsection{Conclusion}

All the geological studies are on time scales of order of the age
of the Earth (typically $z\sim0.1-0.15$ depending on the values of
the cosmological parameters).

The Oklo data are probably the most powerful geochemical data to study
the variation of the fine structure constant but one has to understand
and to model carefully the correlations of the variation of $\aw$ and
$g_{_{\rm S}}$ as well as the effect of $\mu$ (see the recent study by
Olive {\em et al.}, 2002). This difficult but necessary task remains
to be done.

The $\beta$-decay results depend on the combination $\aem^s\aw^2$
and have the advantage not to depend on $G$. They may be
considered more as historical investigations than as competitive
methods to constraint the variation of the fine structure
constant, especially in view of the Oklo results. The dependence
and use of this method on $\as$ was studied by Broulik and Trefil
(1971) and Davies (1972) (see section~\ref{subsec_5.15}).

\subsection{Atomic spectra}
\label{subsec_4.2}

The previous bounds on the fine structure constant assume that
other constants like the Fermi constant do not vary. The use of
atomic spectra may offer cleaner tests since we expect them to
depend mainly on combinations of $\aem$, $\mu$ and $g_{\rm p}$.

We start by recalling some basics concerning atomic spectra in order
to desribe the modelling of the spectra of many-electron systems which
is of great use while studying quasar absorption spectra. We then
focus on laboratory experiments and the results from quasar
absorption spectra.

\subsubsection{$\aem$-dependence of atomic spectra}

As an example, let us briefly recall the spectrum of the hydrogen atom
(see e.g. Cohen-Tannoudji {\em et al.}, 1977). As long as we neglect
the effect of the spins and we work in the non-relativistic
approximation, the spectrum is simply obtained by solving the
Schr\"odinger equation with Hamiltonian
\begin{equation}
H_0=\frac{{\bf P}^2}{2m_{\rm e}}-\frac{e^2}{4\pi\varepsilon_0 r}
\end{equation}
the eigenfunctions of which is of the form
$\psi_{nlm}=R_n(r)Y_{lm}(\theta,\phi)$ where $n$ is the principal
quantum number. This solution has an energy
\begin{equation}\label{51}
E_n=-\frac{E_I}{n^2}\left(1-\frac{m_{\rm e}}{m_{\rm p}}\right)
\end{equation}
independently of the quantum numbers $l$ and $m$ satisfying $0\leq
l<n$, $|m|\leq l$. It follows that there are $n^2$ states with the
same energy. The spectroscopic nomenclature refers to a given
energy level by the principal quantum number and a letter
designing the quantum number $l$ ($s,p,d,f,g\ldots$ respectively
for $l=0,1,2,3,4\ldots$).

This analysis neglects relativistic effects which are expected to
be typically of order $\aem^4$ (since in the Bohr model,
$v/c=\aem$ for the orbit $n=1$), to give the {\it fine
structure} of the spectrum. The derivation of this fine structure
spectrum requires to solve the Dirac equation for a particle in a
potential $-q^2/r$ and then to develop the solution in the
non-relativist limit. Here, we simply use a perturbative approach
in which the Hamiltonian of the system is expanded in $v/c$ as
\begin{equation}
H=H_0+W
\end{equation}
where the corrective term $W$ has different contributions. The
spin-orbit interaction is described by
\begin{equation}
W_{_{\rm S.O.}}=\frac{\aem}{2m_{\rm e}c^2}\frac{\hbar c}{r^3}{\bf L}.{\bf S}.
\end{equation}
Since $r$ is of order of the Bohr radius, it follows that
$W_{_{\rm S.O.}}\sim\aem^2H_0$. The splitting is indeed small: for
instance, it is of order $4\times10^{-5}$~eV between the levels
$2p_{3/2}$ and $2p_{1/2}$, where we have added in indices the
total electron angular moment quantum number $J$.  The second
correction arises from the $(v/c)^2$-relativistic terms and is of
the form
\begin{equation}
W_{_{\rm rel}}=-\frac{{\bf P}^4}{8m_{\rm e}^3c^2}
\end{equation}
and it is easy to see that its amplitude is also of order
$W_{_{\rm rel}}\sim\aem^2H_0$. The third and last correction,
known as the Darwin term, arises from the fact that in the Dirac
equation the interaction between the electron and the Coulomb
field is local. But, the non-relativist approximation leads to a
non-local equation for the electron spinor that is sensitive to
the field on a zone of order of the Compton wavelength centered in
$\bf r$. It follows that
\begin{equation}
W_{_{\rm D}}=\frac{\pi\hbar^2 q^2}{m_{\rm e}^2c^2}\delta(\bf r).
\end{equation}
The average in an atomic state is of order $\left<W_{_{\rm
D}}\right>=\pi\hbar^2q^2/(2m_{\rm e}^2c^2)|\psi({\bf 0})|^2\sim
m_{\rm e}c^2\aem^4\sim\aem^2H_0$. In conclusion all the relativistic
corrections are of order $\aem^2\sim(v/c)^2$. The energy of a fine
structure level is
$$
E_{nlJ}=m_{\rm e}c^2-\frac{E_I}{n^2}
-\frac{m_{\rm e}c^2}{2n^4}\left(\frac{n}{J+1/2}-\frac{3}{4}\right)\aem^4
+\ldots
$$
and is independent\footnote{This is valid to
all order in $\aem$ and the Dirac equation directly gives $E_{nlJ}=
m_{\rm e}c^2\left[1+\aem^2\left(n-J-1/2+\sqrt{(J+1/2)^2-\aem^2}\right)^{-2}
\right]^{-1/2}$.} of the quantum number $l$.

A much finer effect, referred to as {\it hyperfine structure},
arises from the interaction between the spins of the electron,
${\bf S}$, and the proton, ${\bf I}$. They are respectively
associated to the magnetic moments
\begin{equation}
{\bf M}_S=\frac{q\hbar}{2m_{\rm e}}\frac{{\bf
S}}{\hbar}
\,,\qquad
{\bf M}_I=-g_{\rm p}\frac{q\hbar}{2m_{\rm p}}\frac{{\bf
I}}{\hbar}.
\end{equation}
Note that at this stage, the spectrum becomes dependent on the
strong interaction via $g_{\rm p}$ (and via $g_I$ in more general
cases).  This effect can be taken into account by adding the
Hamiltonian
\begin{eqnarray}\label{hfH}
W_{_{\rm hf}}&=& -\frac{\mu_0}{4\pi}
\left\lbrace\frac{q}{r^3}{\bf L}.{\bf M}_I+
\frac{8\pi}{3}{\bf M}_I.{\bf M}_S\delta({\bf r})
\right.\nonumber\\
&&\left.+\quad
\frac{1}{r^3}\left[ 3({\bf M}_S.{\bf n})({\bf M}_I.{\bf n})
-{\bf M}_I.{\bf M}_S\right]\right\rbrace
\end{eqnarray}
where ${\bf n}$ is the unit vector pointing from the proton to the
electron.  The order of magnitude of this effect is typically
$e^2\hbar^2/(m_{\rm e}m_{\rm p}c^2r^3)$ hence roughly 2000 times
smaller than the effect of the spin-orbit coupling. It splits each
fine level in a series of hyperfine levels labelled by
$F\in[|J-I|,I+J]$. For instance for the level $2s_{1/2}$ and
$2p_{1/2}$, we have $J=1/2$ and $F$ can take the two values 0 and
1, for the level $2p_{3/2}$, $J=3/2$ and $F=1$ or $F=2$ etc\ldots
(see figure~\ref{fig1} for an example). This description neglects
the quantum aspect of the electromagnetic field; one effects of
the coupling of the atom to this field is to lift the degeneracy
between the levels $2s_{1/2}$ and $2p_{1/2}$. This is called the
Lamb effect.

In more complex situations, the computation of the spectrum of a given
atom has to take all these effects into account but the solution of
the Schr\"odinger equation depends on the charge distribution and has
to be performed numerically.

The easiest generalization concerns hydrogen-like atoms of charge
$Z$ for which the spectrum can be obtained by replacing $e^2$ by
$(Ze)^2$ and $m_{\rm p}$ by $Am_{\rm p}$. For an external electron in a
many-electron atoms, the electron density near the nucleus is
given (see e.g. Dzuba {\em et al.}, 1999a) by $Z_a^2Z/(n_* a_0)^3$
where $Z_a$ is the effective charge felt by the external electron
outside the atom, $n_*$ an effective principal quantum number
defined by $E_{n_*}=-E_IZ_a^2/n_*^2$. It follows that the
relativistic corrections to the energy level are given by
$$
\Delta E_{n_*,l,J}=\frac{E_I}{n_*^4}Z_a^2Z^2\aem^2
\left[\frac{n_*}{J+1/2}-\frac{Z_a}{Z}\left(1-\frac{Z_a}{4Z}\right)\right].
$$
Such a formula does not take into account many-body effects and
one expects in general a formula of the form $\Delta
E_{n_*,l,J}=E_{n*}Z^2\aem^2[1/J+1/2-C(Z,J,l)]/n_*$.  Dzuba {\em et
al.} (1999b) developed a method to compute the atomic spectra of
many-electrons atoms including relativistic effects. It is based
on many-body perturbation theory (Dzuba {\em et al.}, 1996)
including electron-electron correlations and use a
correlation-potential method for the atom (Dzuba {\em et al.},
1983).

Laboratory measurements can provide these spectra but only for
$\aem=\aem^{(0)}$. In order to detect a variation of $\aem$, one
needs to compute them for different values of $\aem$. Dzuba {\em
et al.} (1999a) describe the energy levels within one
fine-structure multiplet as
\begin{eqnarray}
E&=&E_0+Q_1\left[\left(\frac{\aem}{\aem^{(0)}}\right)^2-1\right]+Q_2
\left[\left(\frac{\aem}{\aem^{(0)}}\right)^4-1\right]\nonumber\\
&+&K_1{\bf L}.{\bf
S}\left(\frac{\aem}{\aem^{(0)}}\right)^2+K_2({\bf L}.{\bf
S})^2\left(\frac{\aem}{\aem^{(0)}}\right)^4
\end{eqnarray}
where $E_0$, $Q_1$ and $Q_2$ describe the configuration center. The
terms in ${\bf L}.{\bf S}$ induce the spin-orbit coupling, second order
spin-orbit interaction and the first order of the Breit
interaction. Experimental data can be fitted to get $K_1$ and
$K_2$ and then numerical simulations determine $Q_1$ and $Q_2$. The
result is conveniently written as
\begin{equation}\label{dzubpara}
\omega=\omega_0+q_1x+q_2y
\end{equation}
with $x\equiv [\aem/\aem^{(0)}]^2-1$ and $y\equiv
[\aem/\aem^{(0)}]^4-1$. As an example, let us cite the result of
Dzuba {\em et al.} (1999b) for Fe~II
\begin{eqnarray}
6d\quad&J=9/2\quad&\omega=38458.9871+1394x+38y \nonumber\\
       &J=7/2\quad&\omega=38660.0494+1632x+0y \nonumber\\
6f\quad&J=11/2\quad&\omega=41968.0642+1622x+3y \nonumber\\
       &J=9/2\quad&\omega=42114.8329+1772x+0y \nonumber\\
       &J=7/2\quad&\omega=42237.0500+1894x+0y \nonumber\\
6p\quad&J=7/2\quad&\omega=42658.2404+1398x-13y
\end{eqnarray}
with the frequency in cm$^{-1}$ for transitions from the ground-state. An
interesting case is Ni~II (Dzuba {\em et al.}, 2001) which has large
relativistic effects of opposite signs
\begin{eqnarray}
2f\quad&J=7/2\quad&\omega= 57080.373-300x\nonumber\\
6d\quad&J=5/2\quad&\omega= 57420.013-700x\nonumber\\
6f\quad&J=5/2\quad&\omega= 58493.071+800x.
\end{eqnarray}
Such results are particularly useful to compare with spectra
obtained from quasar absorption systems as e.g. in the analysis
by Murphy {\em et al.} (2001c).

In conclusion, the key point is that the spectra of atoms depend mainly
on $\mu$, $\aem$ and $g_{\rm p}$ and contain terms both in $\aem^2$ and
$\aem^4$ and that typically
\begin{equation}
H=\aem^2\widetilde H_0+\aem^4\widetilde W_{_{\rm fine}}+
g_{\rm p}\mu^2\aem^4\widetilde W_{_{\rm hyperfine}}\,,
\end{equation}
so that by comparing different kind of transitions in different atoms
there is hope to measure these constants despite the fact that $\as$
plays a role via the nuclear magnetic moment. We describe in the next
section the laboratory experiments and then turn to the measurement of
quasar absorption spectra.

\subsubsection{Laboratory experiments}

Laboratory experiments are based on the comparison either of different
atomic clocks or of atomic clock with ultra-stable oscillators. They
are thus based only on the quantum mechanical theory of the atomic
spectra. They also have the advantage to be more reliable and
reproducible, thus allowing a better control of the systematics and a
better statistics. Their evident drawback is their short time scales,
fixed by the fractional stability of the least precise standards. This
time scale is of order of a month to a year so that the obtained
constraints are restricted to the instantaneous variation today, but
it can be compensated by the extreme sensibility. They involve the
comparison of either ultra-stable oscillators to different composition
or of atomic clocks with different species. Solid resonators,
electronic, fine structure and hyperfine structure transitions
respectively give access to $R_\infty/\aem$, $R_\infty$,
$R_\infty\aem^2$ and $g_{\rm p}\mu R_\infty\aem^2$.

Turneaure and Stein (1974) compared cesium atomic clocks with
superconducting microwave cavities oscillator. The frequency of the
cavity-controlled oscillators was compared during 10 days that one of
a cesium beam. The relative drift rate was
$(-0.4\pm3.4)\times10^{-14}\,{\rm day}^{-1}$.  The dimensions of the
cavity depends on the Bohr radius of the atom while the cesium clock
frequency depends on $g_{\rm p}\mu\aem^2$ (hyperfine transition). It
follows that ${\nu_{_{\rm Ce}}}/{\nu_{_{\rm cavity}}}\propto g_{\rm
p}\mu\aem^3$ so that
\begin{equation}
\frac{\dd}{\dd t}\ln\left(g_{\rm p}\mu\aem^3\right) <4.1\times
10^{-12}\,{\rm yr}^{-1}.
\end{equation}

Godone {\em et al.} (1993) compared the frequencies of cesium and
magnesium atomic beams. The cesium clock, used to define the
second in the SI system of units, is based on the {\it hyperfine
transition} $F=3,\, m_{\rm F}=0 \rightarrow F=4,\, m_{\rm F}=0$ in
the ground-state $6^2s_{1/2}$ of ${}^{133}{\rm Ce}$ with frequency
given, at lowest order and neglecting relativistic and quantum
electrodynamic corrections, by
\begin{equation}
\nu_{_{\rm Ce}}=\frac{32cR_\infty
Z_s^3\aem^2}{3n^3}g_I\mu\sim9.2\,{\rm GHz},
\end{equation}
where $Z_s$ the effective nuclear charge and $g_I$ the cesium nucleus
gyromagnetic ratio. The magnesium clock is based on the frequency of the
{\it fine structure} transition $3p_1\rightarrow 3p_0,\, \Delta m_j=0$
in the meta-stable triplet of $^{24}{\rm Mg}$
\begin{equation}
\nu_{_{\rm Hg}}=\frac{cR_\infty Z_s^4\aem^2}{6n^3}\sim601\,{\rm GHz}.
\end{equation}
It follows that
\begin{equation}
\frac{\dd}{\dd t}\ln\frac{\nu_{_{\rm Ce}}}{\nu_{_{\rm Hg}}}=
\left[\frac{\dd}{\dd t}\ln\left(g_I\mu\right)\right]
\times\left(1\pm10^{-2}\right).
\end{equation}
The experiment led to the bound
\begin{equation}
\left|\frac{\dd}{\dd t}\ln(g_{\rm p}\mu)\right|
<5.4\times10^{-13}\,{\rm yr}^{-1}
\end{equation}
after using the constraint $\dd\ln(g_{\rm p}/g_I)/\dd
t<5.5\times10^{-14}\,{\rm yr}^{-1}$ (Demidov {\em et al.}, 1992).  When
combined with the astrophysical result by Wolfe {\em et al.}  (1976) on
the constraint of $g_{\rm p}\mu\aem^2$ (see Section~\ref{subsec_5.3}) it is
deduced that
\begin{equation}
\left|{\dot\aem}/{\aem}\right|<2.7\times10^{-13}\,{\rm yr}^{-1}.
\end{equation}
We note that relativistic corrections were neglected.

Prestage {\em et al.} (1995) compared the rates of different atomic
clocks based on hyperfine transitions in alkali atoms with different
atomic numbers. The frequency of the hyperfine transition between
$I\pm1/2$ states is given by (see e.g. Vanier and Audoin, 1989)
\begin{eqnarray}
\nu_{_{\rm alkali}}&=&\frac{8}{3}\left(I+\frac{1}{2}\right)\aem^2g_IZ
\frac{z^2}{n_*^3}\left(1-\frac{\dd\Delta_n}{\dd n}\right)
F_{\rm rel}(\aem Z)\nonumber\\
&&(1-\delta)(1-\epsilon)\mu R_\infty c,
\end{eqnarray}
where $z$ is the charge of the remaining ion once the valence
electron has been removed and $\Delta_n=n-n_*$. The term
$(1-\delta)$ is the correction to the potential with respect to
the Coulomb potential and $(1-\epsilon)$ a correction for the
finite size of the nuclear magnetic dipole moment. It is estimated
that $\delta\simeq4\%-12\%$ and $\epsilon\simeq0.5\%$. $F_{\rm
rel}(\aem Z)$ is the Casimir relativistic contribution to the
hyperfine structure and one takes advantage of the increasing
importance of $F_{\rm rel}$ as the atomic number increases (see
figure~\ref{figfrel}). It follows that
\begin{equation}
\frac{\dd}{\dd t}\ln\frac{\nu_{_{\rm alkali}}}{\nu_{_{\rm H}}}
=\frac{\dot\aem}{\aem}\frac{\dd\ln F_{\rm rel}(\aem
Z)}{\dd\ln\aem},
\end{equation}
where $\nu_{_{\rm H}}$ is the frequency of a H maser and when comparing
two alkali atoms
\begin{equation}
\frac{\dd}{\dd t}\ln\frac{\nu_{_{\rm alkali1}}}{\nu_{_{\rm
alkali2}}} =\frac{\dot\aem}{\aem}\left(\left.\frac{\dd\ln F_{\rm
rel}}{\dd\ln\aem} \right|_1-\left.\frac{\dd\ln F_{\rm
rel}}{\dd\ln\aem} \right|_2\right).
\end{equation}
The comparison of different alkali clocks was performed and the
comparison of ${\rm Hg}^+$ ions with a cavity tuned H maser over a
period of 140 days led to the conclusion that
\begin{equation}
\left|{\dot\aem}/{\aem}\right|<3.7\times10^{-14}\,{\rm yr}^{-1}.
\end{equation}
This method constrains in fact the variation of the quantity
$\aem g_{\rm p}/g_I$. One delicate point is the evaluation of the
correction function and the form used by Prestage {\em et al.}
(1995) [$F_{\rm rel}\sim1+11(Z\aem)^2/6+\ldots$] differs with the
$1s$ [$F_{\rm rel}\sim1+3(Z\aem)^2/2+\ldots$] and $2s$ [$F_{\rm
rel}\sim1+17(Z\aem)^2/8+\ldots$] results for hydrogen like atoms
(Breit, 1930).

Sortais {\em et al.} (2001) compared a rubidium to a cesium clock over
a period of 24 months and deduced that $\dd\ln(\nu_{_{\rm
Rb}}/\nu_{_{\rm Cs}})/\dd t=(1.9\pm3.1)\times10^{-15}\,{\rm yr}^{-1}$,
hence improving the uncertainty by a factor 20 relatively to Prestage
{\em et al.} (1995). Assuming $g_{\rm p}$ constant, they deduced
\begin{equation}
{\dot\aem}/{\aem}=(4.2\pm6.9)\times10^{-15}\,{\rm yr}^{-1}
\end{equation}
if all the drift can be attributed to the Casimir relativistic
correction $F_{\rm rel}$.

All the results and characteristics of these experiments are summed up
in table~\ref{table9}. Recently, Braxmaier {\em et al.}  (2001)
proposed a new method to test the variability of $\aem$ and $\mu$
using electromagnetic resonators filled with a dielectric.  The index
of the dielectric depending on both $\aem$ and $\mu$, the comparison
of two oscillators could lead to an accuracy of $4\times10^{-15}\,{\rm
yr}^{-1}$.  Torgerson (2000) proposed to compare atom-stabilized
optical frequency using an optical resonator. On an explicit example
using indium and thalium, it is argued that a precision of
$\dot\aem/\aem\sim10^{-18}/t$, $t$ being the time of the experiment,
can be reached.

Finally, let us note that similar techniques were used to test
local Lorentz invariance (Lamoreaux {\em et al.}, 1986, Chupp {\em
et al.}, 1989) and CPT symmetry (Bluhm {\em et al}., 2002). In the
former case, the breakdown of local Lorentz invariance would cause
shifts in the energy levels of atoms and nuclei that depend on the
orientation of the quantization axis of the state with respect to
a universal velocity vector, and thus on the quantum numbers of
the state.

\subsubsection{Astrophysical observations}\label{subsec_4.5}

The observation of spectra of distant astrophysical objects
encodes information about the atomic energy levels at the position
and time of emission. As long as one sticks to the
non-relativistic approximation, the atomic transition energies are
proportional to the Rydberg energy and all transitions have the
same $\aem$-dependence, so that the variation will affect all the
wavelengths by the same factor. Such a uniform shift of the
spectra can not be distinguished from a Doppler effect due to the
motion of the source or to the gravitational field where it sits.

The idea is to compare different absorption lines from different
species or equivalently the redshift associated with them.
According to the lines compared one can extract information about
different combinations of the constants at the time of emission
(see table~\ref{table0}).

While performing this kind of observations a number of problems
and systematic effects have to be taken into account and
controlled:
\begin{enumerate}
\item Errors in the determination of laboratory wavelengths to which the
observations are compared,
\item while comparing wavelengths from different atoms one has to take
into account that they may be located in different regions of the
cloud with different velocities and hence with different Doppler
redshift.
\item One has to ensure that there is no light blending.
\item The differential isotopic saturation has to be
controlled. Usually quasars absorption systems are expected to have
lower heavy element abundances (Prochoska and Wolfe, 1996, 1997,
2000). The spatial inhomogeneity of these abundances may also play a
role.
\item Hyperfine splitting can induce a saturation similar to isotopic
abundances.
\item The variation of the velocity of the Earth during the integration
of a quasar spectrum can induce differential Doppler shift,
\item Atmospheric dispersion across the spectral direction of the
spectrograph slit can stretch the spectrum. It was shown that this
can only mimic a negative $\Delta\aem/\aem$ (Murphy {\em et al.},
2001b).
\item The presence of a magnetic field will shift the energy levels by
Zeeman effect.
\item Temperature variations during the observation will change the
air refractive index in the spectrograph.
\item Instrumental effects such as variations of the intrinsic
instrument profile have to be controlled.
\end{enumerate}
The effect of these possible systematic errors are discussed by
Murphy {\em et al.} (2001b). In the particular case of the
comparison of hydrogen and molecular lines, Wiklind and Combes
(1997) argued that the detection of the variation of $\mu$ was
limited to $\Delta\mu/\mu\simeq10^{-5}$.  A possibility to reduce
the systematics is to look at atoms having relativistic
corrections of different signs (see Section~\ref{subsec_4.2})
since the systematics are not expected, a priori, to simulate the
correlation of the shift of different lines of a multiplet (see
e.g. the example of Ni~II Dzuba {\em et al.}, 2001).  Besides the
systematics, statistical errors were important in early studies
but have now enormously decreased.

An efficient method is to observe fine-structure doublets for which
\begin{equation}
\Delta\nu=\frac{\aem^2Z^4R_\infty}{2n^3}\,{\rm cm}^{-1},
\end{equation}
$\Delta\nu$ being the frequency splitting between the two lines of
the doublet and $\bar\nu$ the mean frequency (Bethe and Salpeter,
1977). It follows that $\Delta\nu/\bar\nu\propto\aem^2$ and thus
$\Delta\ln\lambda\vert_z/\Delta\ln\lambda\vert_0=[1+\Delta\aem/\aem]^2$.
It can be inverted to give $\Delta\aem/\aem$ as a function of
$\Delta\lambda$ and $\bar\lambda$ as
\begin{equation}\label{doublet}
  \left(\frac{\Delta\aem}{\aem}\right)(z)=\frac{1}{2}
\left[\left(\frac{\Delta\lambda}{\bar\lambda}\right)_z/\left(
\frac{\Delta\lambda}{\bar\lambda}\right)_0 -1\right].
\end{equation}
As an example, it takes the following form for Si~IV (Varshalovich
{\it et al.}, 1996a)
\begin{equation}\label{doubletsi}
  \left(\frac{\Delta\aem}{\aem}\right)(z)=77.55
\left(\frac{\Delta\lambda}{\bar\lambda}\right)_z-0.5.
\end{equation}
Since the observed wavelengths are redshifted as $\lambda_{\rm
obs}=\lambda_{em}(1+z)$ it reduces to
\begin{equation}\label{doubletsi2}
  \left(\frac{\Delta\aem}{\aem}\right)(z)=77.55
\frac{\Delta z}{1+\bar z}.
\end{equation}
As a conclusion, by measuring the two wavelengths of the doublet
and comparing to laboratory values, one can measure the time
variation of the fine structure constant. This method has been
applied to different systems and is the only one that gives a
direct measurement of $\aem$.

Savedoff (1956) was the first to realize that the fine and
hyperfine structures can help to disentangle the redshift effect
from a possible variation of $\aem$ and Wilkinson (1958) pointed
out that ``the interpretation of redshift of spectral lines
probably implies that atomic constants have not changed by less
than $10^{-9}$ parts per year''.

Savedoff (1956) used the data by Minkowski and Wolson (1956) of
the spectral lines of H, N~II, O~I, O~II, Ne~III and N~V for the
radio source Cygnus A of redshift $z\sim 0.057$.  Using the data
for the fine-structure doublet of N~II and Ne~III and assuming
that the splitting was proportional to $\aem^2(1+z)$ led to
\begin{equation}
{\Delta\aem}/{\aem}=\left(1.8\pm1.6\right)\times10^{-3}.
\end{equation}
Bahcall and Salpeter (1965) used the fine structure splitting of
the O~III and Ne~III emission lines in the spectra of the
quasi-stellar radio sources 3C~47 and 3C~147.  Bahcall {\em et al.}
(1967) used the observed fine structure of Si~II and Si~IV in the
quasi-stellar radio sources 3C~191 to deduce that
\begin{equation}
\Delta\aem/\aem=(-2\pm5)\times10^{-2}
\end{equation}
at a redshift $z=1.95$.  Gamow (1967) criticized this measurements
and suggested that the observed absorption lines were not
associated with the quasi-stellar source but were instead produced
in the intervening galaxies. But Bahcall {\em et al.} (1967)
showed on the particular example of 3C~191 that the excited fine
structure states of Si~II were seen to be populated in the
spectrum of this object and that the photon fluxes required to
populate these states were orders of magnitude too high to be
obtained in intervening galaxies.

Bahcall and Schmidt (1967) then used the absorption lines of the O~III
multiplet of the spectra of five radio galaxies with redshift of order
$z\sim0.2$ to improve the former bound to
\begin{equation}
\Delta\aem/\aem=(1\pm2)\times10^{-3},
\end{equation}
considering only statistical errors.

Wolfe {\em et al.} (1976) studied the spectrum of AO~0235+164, a
BL Lac object with redshift $z\sim0.5$. From the comparison of the
hydrogen hyperfine frequency with the resonance line for ${\rm
Mg}^+$, they obtained a constraint on $g_{\rm p}\mu\aem^2$ (see
Section~\ref{subsec_5.3}). From the comparison with the ${\rm
Mg}^+$ fine structure separations they constrained $g_{\rm
p}\mu\aem$, and the ${\rm Mg}^+$ fine structure doublet splitting
gave
\begin{equation}
|\Delta\aem/\aem|<3\times10^{-2}.
\end{equation}
Potekhin and Varshalovich (1994) extended this method based on the
absorption lines of alkali-like atoms and compared the wavelengths
of a catalog of transitions $2s_{1/2}-2p_{3/2}$ and
$2s_{1/2}-2p_{1/2}$ for a set of five elements. The advantages of
such a method are that (1) it is based on the measurement of the
difference of wavelengths which can be measured much more
accurately than (broader) emission lines and (2) these transitions
correspond to transitions from a single level and are thus not
affected by differences in the radial velocity distributions of
different ions.  They used data on 1414 absorption doublets of
C~IV, N~V, O~VI, Mg~II, Al~III and Si~IV and obtained
\begin{equation}
\Delta\aem/\aem=(2.1\pm2.3)\times10^{-3}
\end{equation}
at $z\sim3.2$ and $|\dd\ln\aem/\dd z|<5.6\times10^{-4}$ between
$z=0.2$ and $z=3.7$ at $2\sigma$ level. In these measurements
Si~IV, the most widely spaced doublet, is the most sensitive to a
change in $\aem$.  The use of a large number of systems allows to
reduce the statistical errors and to obtain a redshift dependence
after averaging over the celestial sphere. Note however that
averaging on shells of constant redshift implies that we average
over a priori non-causally connected regions in which the value of
the fine structure constant may a priori be different. This result
was further constrained by Varshalovitch and Potekhin (1994) who
extended the catalog to 1487 pairs of lines and got
\begin{equation}
|\Delta\aem/\aem|<1.5\times10^{-3}
\end{equation}
at $z\sim3.2$. It was also shown that the fine structure splitting was
the same in eight causally disconnected regions at $z=2.2$ at a
$3\sigma$ level.

Cowie and Songaila (1995) improved the previous analysis to get
\begin{equation}
\Delta\aem/\aem=(-0.3\pm1.9)\times10^{-4}
\end{equation}
for quasars between
$z=2.785$ and $z=3.191$. Varshalovich {\em et al.} (1996a) used
the fine-structure doublet of Si~IV to get
\begin{equation}
\Delta\aem/\aem=(2\pm7)\times10^{-5}
\end{equation}
at $2\sigma$ for quasars between $z=2.8$ and $z=3.1$ (see also
Varshalovich {\em et al.}, 1996b).

Varshalovich {\em et al.} (2000a) studied the doublet lines of
Si~IV, C~IV and Ng~II and focused on the fine-structure doublet of
Si~IV to get
\begin{equation}
\Delta\aem/\aem=(-4.5\pm4.3[\hbox{stat}]\pm1.4[\hbox{syst}])\times10^{-5}
\end{equation}
for $z=2-4$. An update of this analysis (Ivanchik {\em et al.},
1999) with 20 absorption systems between $z=2$ and $z=3.2$  gave
\begin{equation}
\Delta\aem/\aem=(-3.3\pm6.5[\hbox{stat}]\pm8[\hbox{syst}])\times10^{-5}.
\end{equation}
Murphy {\em et al.} (2001d) used the same method
with 21 Si~IV absorption system toward 8 quasars with redshift
$z\sim 2-3$ to get
\begin{equation}
\Delta\aem/\aem=(-0.5\pm1.3)\times10^{-5}
\end{equation}
hence improving the previous constraint by a factor 3.\\

Recently Dzuba {\em et al.} (1999a,b) and Webb {\em et al.} (1999)
introduced a new method referred to as the {\it many multiplet}
method in which one correlates the shift of the absorption lines
of a set of multiplets of different ions. It is based on the
parametrization (\ref{dzubpara}) of the computation of atomic
spectra. One advantage is that the correlation between different
lines allows to reduce the systematics. An improvement is that one
can compare the transitions from different ground-states and using
ions with very different atomic mass also increases the
sensitivity because the difference between ground-states
relativistic corrections can be very large and even of opposite
sign (see the example of Ni~II by Dzuba {\em et al.}, 2001).

Webb {\em et al.} (1999) analyzed one transition of the Mg~II
doublet and five Fe~II transitions from three multiplets. The
limit of accuracy of the method is set by the frequency interval
between Mg~II~2796 and Fe~II~2383 which induces a fractional
change of $\Delta\aem/\aem\sim10^{-5}$. Using the simulations by
Dzuba {\em et al.} (1999a,b) it can be deduced that a change in
$\aem$ induces a large change in the spectrum of Fe~II and a small
one for Mg~II (the magnitude of the effect being mainly related to
the atomic charge). The method is then to measure the shift of the
Fe~II spectrum with respect to the one of Mg~II. This comparison
increases the sensitivity compared with methods using only alkali
doublets. Using 30 absorption systems toward 17 quasars they
obtained
\begin{eqnarray}
\Delta\aem/\aem=(-0.17\pm0.39)\times 10^{-5}&&\\
\Delta\aem/\aem=(-1.88\pm0.53)\times 10^{-5}&&
\end{eqnarray}
respectively for $0.6<z<1$ and $1<z<1.6$.  There is no signal of a
variation of $\aem$ for redshift smaller than 1 but a 3.5$\sigma$
deviation for redshifts larger than 1 and particularly in the
range $z\sim0.9-1.2$. The summary of these measurements are
depicted on figure~\ref{figwebb}. A possible explanation is a
variation of the isotopic ratio but the change of $^{26}{\rm
Mg}/{}^{24}{\rm Mg}$ would need to be substantial to explain the
result (Murphy {\em et al.}, 2001b). Calibration effects can also
be important since Fe~II and Mg~II lines are situated in different
order of magnitude of the spectra.

Murphy {\em et al.} (2001a) extended this technique of fitting of
the absorption lines to the species Mg~I, Mg~II, Al~II, Al~III,
Si~II, Cr~II, Fe~II, Ni~II and Zn~II for 49 absorption systems
towards 28 quasars with redhsift $z\sim0.5-3.5$ and got
\begin{eqnarray}
\Delta\aem/\aem=(-0.2\pm0.3)\times 10^{-5}&&\\
\Delta\aem/\aem=(-1.2\pm0.3)\times 10^{-5}&&
\end{eqnarray}
respectively for $0.5<z<1$ and $1<z<1.8$ at $4.1\sigma$.  The low
redshift part is a re-analysis of the data by Webb {\em et al.}
(1999). Over the whole sample ($z=0.5-1.8$) it gives the constraint
\begin{equation}
\Delta\aem/\aem=(-0.7\pm0.23)\times 10^{-5}.
\end{equation}

Webb {\em et al.} (2001) re-analyzed their initial sample and included
new optical QSO data to have 28 absorption systems with redshift
$z=0.5-1.8$ plus 18 damped Lyman-$\alpha$ absorption systems towards
13 QSO plus 21 Si~IV absorption systems toward 13 QSO . The analysis
used mainly the multiplets of Ni~II, Cr~II and Zn~II and Mg~I, Mg~II,
Al~II, Al~III and Fe~II were also included.  One improvement compared
with the analysis by Webb {\em et al.}  (1999) is that the ``$q$''
coefficient of Ni~II, Cr~II and Zn~II in Eq.~(\ref{dzubpara}) vary
both in magnitude and sign so that lines shift in opposite
directions. The data were reduced to get 72 individual estimates of
$\Delta\aem/\aem$ spanning a large range of redshift. From the Fe~II
and Mg~II sample they obtained
\begin{equation}
\Delta\aem/\aem=(-0.7\pm0.23)\times10^{-5}
\end{equation}
for $z=0.5-1.8$ and from
the Ni~II, Cr~II and Zn~II they got
\begin{equation}
\Delta\aem/\aem=(-0.76\pm0.28)\times10^{-5}
\end{equation}
for $z=1.8-3.5$ at a $4\sigma$ level. The fine-structure of Si~IV gave
\begin{equation}
\Delta\aem/\aem=(-0.5\pm1.3)\times10^{-5}
\end{equation}
for $z=2-3$.

This series of results is of great importance since all other
constraints are just upper bounds. Note that they are incompatible
with both Oklo ($z\sim0.1$) and meteorites data ($z\sim0.45$) if the
variation is linear with time. Such a non-zero detection, if
confirmed, will have tremendous implications concerning our
understanding of physics. Among the first questions that arise, it is
interesting to test whether this variation is compatible with other
bounds (e.g. test of the universality of free fall), to study the
level of detection needed by the other experiments knowing the level
of variation by Webb {\em et al.} (2001), to sort out the amplitude of
the variation of the other constants and to be ensure that no
systematic effects has been forgotten. For instance, the fact that
Mg~II and Fe~II are a priori not in the same region of the cloud was
not modelled; this could increase the errors even if it is difficult
to think that it can mimic the observed variation of $\aem$. If one
forgets the two points arising from HI~21~cm and molecular absorption
systems (hollow squares in figure~\ref{figwebb}), the best fit of the
data of figure~\ref{figwebb} does not seem to favor today's value of
the fine structure constant. This could indicate an unknown systematic
effect. Besides, if the variation of $\aem$ is monotonic then these
observations seem to be incompatible with the Oklo results.

\subsection{Cosmological constraints}\label{subsec_4.3}

\subsubsection{Cosmic microwave background}

The Cosmic Microwave Background Radiation (CMBR) is composed of
the photons emitted at the time of the recombination of hydrogen
and helium when the universe was about 300,000 years old [see e.g.
Hu and Dodelson (2002) or Durrer (2002) for recent reviews on CMBR
physics]. This radiation is observed to be a black body with a
temperature $T=2.723$~K with small anisotropies of order of the
$\mu$K. The temperature fluctuation in a direction
$(\vartheta,\varphi)$ is usually decomposed on a basis of
spherical harmonics as
\begin{equation}\label{cmb1}
\frac{\delta
T}{T}(\vartheta,\varphi)=\sum_{\ell}\sum_{m=-\ell}^{m=+\ell}a_{\ell
m}Y_{\ell m}(\vartheta,\varphi).
\end{equation}
The angular power spectrum miltipole $C_\ell=\langle \vert
a_{lm}\vert^2 \rangle$ is the coefficient of the decomposition of
the angular correlation function on Legendre polynomials. Given a
model of structure formation and a set of cosmological parameters,
this angular power spectrum can be computed and compared to
observational data in order to constraint this set of parameters.

Prior to recombination, the photons are tightly coupled to the
electrons, after recombination they can be considered mainly as
free particles. Changing the fine structure constant modifies the
strength of the electromagnetic interaction and thus the only
effect on CMB anisotropies arises from the change in the
differential optical depth of photons due to the Thomson
scattering
\begin{equation}\label{cmb2}
\dot\tau=x_{\rm e}n_{\rm e}c\sigma_{\rm T}
\end{equation}
which enters in the collision term of the Boltzmann equation
describing the evolution of the photon distribution function and where
$x_{\rm e}$ is the ionization fraction (i.e. the number density of
free electrons with respect to their total number density $n_{\rm
e}$).  The first dependence of the optical depth on the fine structure
constant arises from the Thomson scattering cross-section given by
\begin{equation}\label{cmb3}
\sigma_{\rm T}=\frac{8\pi}{3}\frac{\hbar^2}{m_{\rm e}^2c^2}\aem^2
\end{equation}
and the scattering by free protons can be neglected since $m_{\rm
e}/m_{\rm p}\sim5\times10^{-4}$. The second, and more subtle
dependence, comes from the ionization fraction. Recombination
proceeds via 2-photon emission from the $2s$ level or via the
Ly-$\alpha$ photons which are redshifted out of the resonance line
(Peebles, 1968) because recombination to the ground state can be
neglected since it leads to immediate reionization of another
hydrogen atom by the emission of a Ly-$\alpha$ photon . Following
Ma and Bertschinger (1995) and Peebles (1968) and taking into
account only the recombination of hydrogen, the equation of
evolution of the ionization fraction takes the form
$$
\frac{\dd x_{\rm e}}{\dd t}={\cal
C}\left[\beta\left(1-x_{\rm e}\right)\hbox{exp}
\left(-\frac{B_1-B_2}{k_{_{\rm B}}T}\right)
-{\cal R}n_{\rm p}x_{\rm e}^2\right].
$$
$B_n=-E_I/n^2$ is the energy of the $n$th hydrogen
atomic level, $\beta$ is the
ionization coefficient, ${\cal R}$ the recombination coefficient,
${\cal C }$ the correction constant due to the redshift of
Ly-$\alpha$ photons and to 2-photon decay and $n_p=n_e$ is the
number of proton. $\beta$ is related to ${\cal R}$ by the
principle of detailed balance so that
\begin{equation}\label{cmb5}
\beta={\cal R}\left(\frac{2\pi m_{\rm e}
k_{_{\rm B}}T}{h^2}\right)\hbox{exp}\left(-\frac{B_2}{k_{_{\rm B}}T}\right).
\end{equation}
The recombination rate to all other excited levels is
$$
{\cal R}=\frac{8\pi}{c^2}\left(\frac{k_{_{\rm B}}T}{2\pi m_{\rm e}}\right)^{3/2}
\sum_{n,l}^*(2l+1)\hbox{e}^{B_n/k_{_{\rm B}}T}\int_{B_n/k_{_{\rm B}}T}^\infty
\sigma_{nl}\frac{y^2\dd y}{\hbox{e}^y-1}
$$
where $\sigma_{nl}$ is the ionization cross section for the
$(n,l)$ excited level of hydrogen. The star indicates that the sum
needs to be regularized and the $\aem$-, $m_{\rm e}$-dependence of the
ionization cross section is complicated to extract. It can however
be shown to behave as
$\sigma_{nl}\propto\aem^{-1}m_{\rm e}^{-2}f(h\nu/B_1)$.

Finally, the factor ${\cal C}$ is given by
\begin{equation}\label{cmb7}
{\cal
C}=\frac{1+K\Lambda_{2s}(1-x_e)}{1+K(\beta+\Lambda_{2s})(1-x_e)}
\end{equation}
where $\Lambda_{2s}$ is the rate of decay of the $2s$ excited
level to the ground state via 2 photons; it scales as
$m_{\rm e}\aem^8$. The constant $K$ is given in terms of the
Ly-$\alpha$ photon $\lambda_{\alpha}=16\pi\hbar/(3m_{\rm e}\aem^2c)$
by $K=n_p\lambda_\alpha^3/(8\pi H)$ and scales as
$m_{\rm e}^{-3}\aem^{-6}$.

Changing $\aem$ will thus have two effects: first it changes the
temperature at which the last scattering happens and secondly it
changes the residual ionization after recombination. Both effects
influence the CMB temperature anisotropies [see Kaplinghat {\em et
al.} (1999) and Battye {\em et al.} (2001) for discussions]. The
last scattering can roughly be determined by the maximum of the
visibility function $g=\dot\tau\exp(-\tau)$ which measures the
differential probability for a photon to be scattered at a given
redshift. Increasing $\aem$ shifts $g$ to higher redshift at which
the expansion rate is faster so that the temperature and $x_e$
decrease more rapidly, resulting in a narrower $g$. This induces a
shift of the $C_\ell$ spectrum to higher multipoles and an
increase of the values of the $C_\ell$. The first effect can be
understood by the fact that pushing the last scattering surface to
a higher redshift leads to a smaller sound horizon at decoupling.
The second effect results from a smaller Silk damping.

Hannestad (1999) and then Kaplinghat {\em et al.} (1999) implemented
these equations in a Boltzmann code, taking into account only the
recombination of hydrogen and neglecting the one of helium, and showed
that coming satellite experiments such as MAP\footnote{{\tt
http://map.gsfc.nasa.gov/}} and Planck\footnote{{\tt
http://astro.estec.esa.nl/SA-general/Projects/Planck/}} should provide
a constraint on $\aem$ at recombination with a precision
$\vert\dot\aem/\aem\vert\leq 7\times 10^{-13}\,{\rm yr}^{-1}$, which
corresponds to a sensitivity
$\vert\Delta\aem/\aem\vert\sim10^{-2}-10^{-3}$ at a redshift of about
$z\sim1,000$. Avelino {\em et al.}  (2000) studied the dependence of
the position of the first acoustic peak on $\aem$. Hannestad (1999)
chose the underlying $\Lambda$CDM model $(\Omega,\Omega_{\rm
b},\Lambda,h,n,N_\nu, \tau,\aem)=(1,0.08,0,0.5,1,3,0,\aem^{(0)})$ and
performed a 8 parameters fit to determine to which precision the
parameters can be extracted. Kaplinghat {\em et al.} (1999) worked
with the parameters $(h,\Omega_{\rm b},\Lambda,N_\nu,Y_{\rm
p},\aem)$. They showed that the precision on $\Delta\aem/\aem$ varies
from $10^{-2}$ if the maximum observed CMB multipole is of order
500-1000 to $10^{-3}$ if one observes multipoles higher than 1500.

Avelino {\em et al.} (2000) claim that BOOMERanG and MAXIMA data
favor a value of $\aem$ smaller by a few percents in the past (see
also Martins {\em et al.}, 2002) and Battye {\em et al.} (2001)
showed that the fit to current CMB data are improved by allowing
$\Delta\aem\not=0$ and pointed out that the evidence of a
variation of the fine structure constant can be thought of as
favoring a delayed recombination model (assuming $\Omega=1$ and
$n=1$). Avelino {\em et al.} (2001) then performed a joint
analysis of nucleosynthesis and CMB data and did not find any
evidence for a variation of $\aem$ at one-sigma level at either
epoch. They consider $\Omega_{\rm b}$ and $\Delta\aem$ as independent
and the marginalization over one of the two parameters lead to
\begin{equation}
-0.09<\Delta\aem<0.02
\end{equation}
at 68\% confidence level. Martins {\em et al.} (2002) concluded that
MAP and Planck will allow to set respectively a 2.2\% and 0.4\%
constraint at 1$\sigma$ if all other parameters are
marginalized. Landau {\em et al.} (2001) concluded from the study of
BOOMERanG, MAXIMA and COBE data in spatially flat models with adiabatic
primordial fluctuations that, at $2\sigma$ level,
\begin{equation}
-0.14<\Delta\aem<0.03.
\end{equation}

All these works assume that only $\aem$ is varying but, as can been
seen from Eqs. (\ref{cmb1}-\ref{cmb7}), one has to assume the
constancy of the electron mass. Battye {\em et al.} (2001) show that
the change in the fine structure constant and in the mass of the
electron are degenerate according to $\Delta\aem\approx0.39\Delta m_{\rm e}$
but that this degeneracy was broken for multipoles higher than 1500.
The variation of the gravitational constant can also have similar
effects on the CMB (Riazuelo and Uzan, 2002).  All the works also
assume the $\aem$-dependence of ${\cal R}$ to be negligible and Battye
{\em et al.} (2001) checked that the helium recombination was
negligible in the range of $\Delta\aem$ considered.

In conclusion, strong constraints on the variation of $\aem$ can
be obtained from the CMB only if the cosmological parameters are
independently known. This method is thus non competitive unless
one has strong bounds on $\Omega_{\rm b}$ and $h$ (and the result
will always be conditional to the model of structure formation)
and assumptions about the variation of other constants such as the
electron mass, gravitational constant are made.

\subsubsection{Nucleosynthesis}\label{subsec_4.4}

The amount of $^{4}{\rm He}$ produced during the big bang
nucleosynthesis is mainly determined by the neutron to proton
ratio at the freeze-out of the weak interactions that interconvert
neutrons and protons. The result of Big Bang nucleosynthesis (BBN)
thus depends on $G$, $\aw$, $\aem$ and $\as$ respectively through
the expansion rate, the neutron to proton ratio, the
neutron-proton mass difference and the nuclear reaction rates,
besides the standard parameters such as e.g. the number of
neutrino families.  The standard BBN scenario (see e.g. Malaney,
1993, Reeves, 1994) proceeds in three main steps:
\begin{enumerate}
\item for $T>1$~MeV, ($t<1$~s) a first stage during which the
neutrons, protons, electrons, positrons an neutrinos are kept in
statistical equilibrium by the (rapid) weak interaction
\begin{eqnarray}\label{bbn0}
&&n\longleftrightarrow p+e^-+\bar\nu_e,\quad
n+\nu_e\longleftrightarrow  p+e^-,\nonumber\\
&&n+e^+\longleftrightarrow p+\bar\nu_e.
\end{eqnarray}
As long as statistical equilibrium holds, the neutron to proton ratio
is
\begin{equation}
(n/p)=\hbox{e}^{-Q/k_{_{\rm B}}T}
\end{equation}
where $Q\equiv (m_{\rm n}-m_{\rm p})c^2=1.29$~MeV. The abundance of the other
light elements is given by (Kolb and Turner, 1993)
\begin{eqnarray}
Y_A&=&g_A\left(\frac{\zeta(3)}{\sqrt{\pi}}\right)^{A-1}2^{(3A-5)/2}A^{5/2}
     \nonumber\\
   &&\left[\frac{k_{_{\rm B}}T}{m_{\rm N}c^2}\right]^{3(A-1)/2}
      \eta^{A-1}Y_{\rm p}^ZY_{\rm n}^{A-Z}\hbox{e}^{B_A/k_{_{\rm B}}T},
\end{eqnarray}
where $g_A$ is the number of degrees of freedom of the nucleus
$_Z^A{\rm X}$, $m_{\rm N}$ is the nucleon mass, $\eta$ the
baryon-photon ratio and $B_A\equiv(Zm_{\rm p}+(A-Z)m_{\rm n}-m_A)c^2$
the binding energy.
\item Around $T\sim0.8$~MeV ($t\sim2$~s), the weak interactions freeze
out at a temperature $T_{\rm f}$ determined by the competition between
the weak interaction rates and the expansion rate of the universe and
thus determined by $\Gamma_{_{\rm w}}(T_{\rm f})\sim H(T_{\rm f})$ that is
\begin{equation}
\gfermi^2(k_{_{\rm B}}T_{\rm f})^5\sim\sqrt{GN_*}(k_{_{\rm B}}T_{\rm f})^2
\end{equation}
where $\gfermi$ is the Fermi constant and $N_*$ the number of
relativistic degrees of freedom at $T_{\rm f}$. Below $T_{\rm f}$,
the number of neutrons and protons change only from the neutron
$\beta$-decay between $T_{\rm f}$ to $T_{\rm N}\sim0.1$~MeV when
$p+n$ reactions proceed faster than their inverse dissociation.
$T_{\rm N}$ is determined by demanding that the relative number of
photons with energy larger that the deuteron binding energy,
$E_{\rm D}$, is smaller than one, i.e. so that
$n_\gamma/n_p\sim\exp(E_{\rm D}/T_{\rm N})\sim1$.
\item For $0.05$~MeV$<T<0.6$~MeV ($3\,{\rm s}<t<6\,{\rm min}$), the
synthesis of light elements occurs only by two-body reactions.
This requires the deuteron to be synthetized ($p+n\rightarrow D$)
and the photon density must be low enough for the
photo-dissociation to be negligible. This happens roughly when
\begin{equation}\label{n0}
\frac{n_{\rm d}}{n_\gamma}\sim\eta^2\exp(-E_{\rm D}/T_{\rm N})\sim 1
\end{equation}
with $\eta\sim3\times10^{-10}$. The abundance of $^4{\rm He}$ by
mass, $Y_{\rm p}$, is then well estimated by
\begin{equation}\label{n1}
Y_{\rm p}\simeq2\frac{(n/p)_{\rm N}}{1+(n/p)_{\rm N}}
\end{equation}
with
\begin{equation}\label{n2}
(n/p)_{\rm N}=(n/p)_{\rm f}\exp(-t_{\rm N}/\tau_{\rm n})
\end{equation}
with $t_{\rm N}\propto G^{-1/2}T_{\rm N}^{-2}$ and $\tau_{\rm
n}^{-1}=1.636\gfermi^2(1+3g_A^2)m_{\rm e}^5/(2\pi^3)$, with
$g_A\simeq1.26$ being the axial/vector coupling of the
nucleon. Assuming that $E_{\rm D}\propto\as^2$, this gives a
dependence $t_{\rm N}/\tau_{\rm p}\propto G^{-1/2}\as^2\gfermi^2$ (see
Section~\ref{subsec_5.15}).

The helium abundance depends thus mainly on $Q$, $T_{\rm f}$ and
$T_{\rm N}$ (and hence mainly on the neutron lifetime, $\tau_{\rm n}$)
and the abundances of the other elements depends also on the nuclear
reaction rates.
\end{enumerate}

The light element abundances are thus sensible to the freeze-out
temperature, which depends on $\gfermi$, $G$, on the
proton-neutron mass difference $Q$, and on the values of the
binding energies $B_A$ so that they mainly depend $\aem$, $\aw$,
$\as$, $\ag$ and the mass of the quarks. An increase in $G$ or
$N_*$ results in a higher expansion rate and thus to an earlier
freeze-out, i.e. a higher $T_{\rm f}$. A decrease in $\gfermi$,
corresponding to a longer neutron lifetime, leads to a decrease of
the weak interaction rates and also results in a higher $T_{\rm
f}$. It inplies, assuming uncorrelated variations, that $|\Delta
G/G|<0.25$ (see Section~\ref{sec_3}) and
$|\Delta\gfermi/\gfermi|<6\times10^{-2}$ (see
Section~\ref{subsec_5.1}).

First, the radiative and Coulomb corrections for the weak reactions
(\ref{bbn0}) have been computed by Dicus {\em et al.} (1982) and shown
to have a very small influence on the abundances.

The constraints on the variation of these quantities were first
studied by Kolb {\em et al.} (1986) who calculated the dependence
of primordial ${}^4{\rm He}$ on $G$, $\gfermi$ and $Q$. They
studied the influence of independent changes of the former
parameters and showed that the helium abundance was mostly
sensitive in the change in $Q$.  Other abundances are less
sensitive to the value of $Q$, mainly because $^4{\rm He}$ has a
larger binding energy; its abundances is less sensitive to the
weak reaction rate and more to the parameters fixing the value of
$(n/p)$. To extract the constraint on the fine structure constant,
one needs a particular model for the $\aem$-dependence of $Q$.
Kolb {\em et al.} (1986) decomposed $Q$ as
\begin{equation}
Q=\aem Q_\alpha+\beta Q_\beta
\end{equation}
where the first part represents the electromagnetic contribution and
the second part corresponds to all non-electromagnetic
contributions. Assuming that $Q_\alpha$ and $Q_\beta$ are constant and
that the electromagnetic contribution is the dominant part of $Q$,
they deduce that $Q/Q_0\simeq\aem/\aem^{(0)}$ and thus that
$(n/p)\simeq (n/p)_0\left[1-q_0T_{\rm f}\aem/\aem^{(0)}\right]$. To consider
the effect of the dependent variation of $G$, $\gfermi$ and $\aem$,
the time variation of these constants was related to the time
variation of the volume of an internal space of characteristic size
$R$ for a 10-dimensional superstring model and Kaluza-Klein models
(see Section~\ref{sec_7} for details on these models)\footnote{Their
hypothesis on the variation of the Fermi constant are questionable,
see Section~\ref{subsec_5.1} for details.}. They concluded that
\begin{equation}
|\Delta\aem/\aem|<10^{-2}
\end{equation}
and showed that if one requires that the abundances of $^2{\rm H}$ and
$^3{\rm He}$ remains unchanged it is impossible to compensate the
change in $\aem$ by a change in the baryon-to-photon ratio.  Indeed,
the result depends strongly on the hypothesis of the functional
dependence. Khare (1986) then showed that the effect of the
extra-dimensions can be cancelled if the primordial neutrinos are
degenerate. This approach was generalized by Vayonakis (1988) who
considered the 10-dimensional limit of superstring and by Coley
(1990) for the case of 5-dimensional Kaluza-Klein theory.

Campbell and Olive (1995) kept track of the changes in $T_{\rm f}$ and
$Q$ separately and deduced that
\begin{equation}
\frac{\Delta Y_{\rm p}}{Y_{\rm p}}\simeq\frac{\Delta T_{\rm f}}{T_{\rm
f}}-\frac{\Delta Q}{Q}.
\end{equation}
They used this to study the constraints on $\gfermi$ (see
Section~\ref{subsec_5.1}).

Bergstr\"om {\em et al.} (1999) extended the original work by Kolb
{\em et al.} (1986) by considering other nuclei. They assumed the
dependence of $Q$ on $\aem$
\begin{equation}\label{qansatz}
Q\simeq\left(1.29-0.76{\Delta\aem}/{\aem}\right)~\hbox{MeV}
\end{equation}
that relies on a change of quark masses due to strong and
electromagnetic energy  binding. Since the abundances of other
nuclei depend mostly on the weak interaction rates, they studied
the dependence of the thermonuclear rates on $\aem$.  In the
non-relativistic limit, it is obtained as the thermal average of
the cross section times the relative velocity times the number
densities. The key point is that for charged particles the cross
section takes the form
\begin{equation}\label{bir}
\sigma(E)=\frac{S(E)}{E}\hbox{e}^{-2\pi\eta(E)}
\end{equation}
where $\eta(E)$ arises from the Coulomb barrier and is given in terms of
the charges and the reduced mass $\mu$ of the two particles as
\begin{equation}
\eta(E)=\aem Z_1Z_2\sqrt{\frac{\mu c^2}{2E}}.
\end{equation}
The factor $S(E)$ has to be extrapolated from experimental nuclear
data which allows Bergstr\"om {\em et al.} (1999) to determine the
$\aem$-dependence of all the relevant reaction rates. Let us note
that the $\aem$-dependence of the reduced mass $\mu$ and of $S(E)$
were neglected; the latter one is polynomial in $\aem$ (Fowler
{\em et al.}, 1975).  Keeping all other constants fixed, assuming
no exotic effects and taking a lifetime of 886.7~s for the
neutron, it was deduced that
\begin{equation}
\left|{\Delta\aem}/{\aem}\right|<2\times10^{-2}.
\end{equation}
In the low range of $\eta\sim1.8\times10^{-10}$ the ${}^7{\rm Li}$
abundance does not depend strongly on $\aem$ and the one of $^4{\rm
He}$ has to be used to constrain $\aem$. But it has to be noted that
the observational status of the abundance of $^4{\rm He}$ is still a
matter of debate and that the theoretical prediction of its variation
with $\aem$ depends on the model-dependent ansatz (\ref{qansatz}). For
the high range of $\eta\sim5\times10^{-10}$, the variation of
${}^7{\rm Li}$ with $\aem$ is rapid, due to the exponential Coulomb
barrier and limits the variation of $\aem$.

Nollet and Lopez (2002) pointed out that Eq. (\ref{bir}) does not
contain all the $\aem$-dependence. They argue that (i) the factor $S$
depends linearly on $\aem$, (ii) when a reaction produces two charged
particles there should be an extra $\aem$ contribution arising from
the fact that the particles need to escape the Coulomb potential,
(iii) the reaction energies depend on $\aem$ and (iv) radiative
captures matrix elements are proportional to $\aem$. The most secure
constraint arising from D/H measurements and combining with CMB data to
determine $\Omega_B$ gives
\begin{equation}
\Delta\aem/\aem=(3\pm7)\times 10^{-2}
\end{equation}
at $1\sigma$ level.

Ichikawa and Kawasaki (2002) included the effect of the quark mass and
by considering a joint variation of the different couplings as it
appears from a dilaton. $Q$ then takes the form
\begin{equation}
Q=a\aem\Lambda_{_{\rm QCD}}+b(y_{\rm d}-y_{\rm u})v
\end{equation}
where $a$ and $b$ are two parameters and $y_{\rm d}$, $y_{\rm u}$ the Yukawa
couplings. The neutron lifetime then behaves as
\begin{equation}
\tau_{\rm n}=(1/v y_{\rm e}^5)f^{-1}(Q/m_{\rm e}),
\end{equation}
with $f$ is a known function. Assuming that all the couplings vary
due to the effect of a dilaton, such that the Higgs vacuum
expectation value $v$ remains fixed, they constrained the
variation of this dilaton and deduced
\begin{equation}
\Delta\aem/\aem=(-2.24\pm3.75)\times10^{-4}.
\end{equation}

In all the studies, one either assumes all other constants fixed
or a functional dependence between them, as inspired from string
theory. The bounds are of the same order of magnitude that the
ones obtained from the CMB; they have the advantage to be at
higher redshift but suffer from the drawback to be
model-dependent.

\subsubsection{Conclusion}

Even if cosmological observations allow to test larger time scales, it
is difficult to extract tight constraints on the variation of the fine
structure constant from them.

The CMB seems clean at first glance since the effect of the fine
structure constant is well decoupled from the effect of the weak and
strong coupling constants. Still, it is entangled with assumption on
$G$. Besides, it was shown that degeneracy between some parameters
exists and mainly between the fine structure constant, the electron to
proton mass ratio, the baryonic density and the dark energy equation of
state (Huey {\em et al.}, 2001).

Nucleosynthesis is degenerate in the four fundamental coupling
constants. In some specific models where the variation of these
constants are linked it allows to constraint them and definitively the
helium abundance alone cannot constraint the fine structure constant.

\subsection{Equivalence Principle}\label{subsec_4.6}

The equivalence principle is closely related to the development of the
theory of gravity from Newton's theory to general relativity (see
Will, 1993 and Will, 2001 for reviews).  Its first aspect is the {\it
weak equivalence} principle stating that the weight of a body is
proportional to its mass or equivalently that the trajectory of any
freely falling body does not depend on its internal structure,
mass and composition.  Einstein formulated a stronger equivalence
principle usually referred to as {\it Einstein equivalence principle}
stating that (1) the weak equivalence principle holds, (2) any
non-gravitational experiment is independent of the velocity of the
laboratory rest-frame (local Lorentz invariance) and (3) of when an
where it is performed (local position invariance).

If the Einstein equivalence principle is valid then gravity can be
described as the consequence of a curved spacetime and is a metric
theory of gravity, an example of which are general relativity and the
Brans-Dicke (1961) theory. This statement is not a ``theorem'' but
there are a lot of indications to back it up (see Will, 1993,
2001). Note that superstring theory violates the Einstein equivalence
principle since it introduces additional fields (e.g. dilaton,
moduli...) that have gravitational-strength couplings which violates
of the weak equivalence principle.  A time variation of a fundamental
constant is in contradiction with Einstein equivalence principle since
it violates the local position invariance. Dicke (1957, 1964) was
probably the first to try to use the result of E\"otv\"os {\em et al.}
(1922) experiment to argue that the strong interaction constant was
approximatively position independent.  All new interactions that
appear in the extension of standard physics implies extra scalar or
vector fields and thus an expected violation of the weak equivalence
principle, the only exception being metric theories such as the class
of tensor-scalar theories of gravitation in which the dilaton couples
universally to all fields and in which one can have a time variation
of gravitational constant without a violation of the weak equivalence
principle (see e.g. Damour and Esposito-Far\`ese, 1992).

The difference in acceleration between two bodies of different
composition can be measured in E\"otv\"os-type experiments (E\"otv\"os
{\em et al.}, 1922) in which the acceleration of various pairs of
material in the Earth gravitational field are compared.  The results
of this kind of laboratory experiments are presented as bounds on the
parameter $\eta$
\begin{equation}
\eta\equiv2\frac{|\vec a_1-\vec a_2|}{|\vec a_1+\vec a_2|}.
\end{equation}
The most accurate constraints on $\eta$ are
$\eta=(-1.9\pm2.5)\times10^{-12}$ between beryllium and copper (Su
{\em et al.}, 1994) and $|\eta|<5.5\times10^{-13}$ between
Earth-core-like and Moon-mantle-like materials (Baessler {\em et al.},
1999).  The Lunar Laser Ranging (experiment) gives the bound
$\eta=(3.2\pm4.6)\times10^{-13}$ (Williams {\em et al.}, 1996) and
$\eta=(3.6\pm4)\times10^{-13}$ (M\"uller and Nordtvedt, 1998; M\"uller
{\em et al.}, 1999).  Note however that, as pointed by Nordtvedt
(1988, 2001a), the LLR measurement are ambiguous since the Earth and
the Moon have (i) a different fraction of gravitational self-energy
and (ii) a difference of composition (the core of the Earth having a
larger Fe/Ni ratio than the Moon). This makes this test sensititive
both to self-gravity and to non-gravitational forms of energy. The
experiment by Baessler {\em et al.}, (1999) lifts the degeneracy by
considering miniature ``Earth'' and ``Moon''.

As explained in Section~\ref{subsec_2.3}, if the self-energy depends
on position, the conservation of energy implies the existence of an
anomalous acceleration.  In the more general case where the long
range force is mediated by a scalar field $\phi$, one has to determine
the dependence $m_i(\phi)$ of the different particles. If it is
different for neutron and proton, then the force will be composition
dependent. At the Newtonian approximation, the interaction potential
between two particles is of the form (Damour and Esposito-Far\`ese,
1992)
\begin{equation}
V(r)=-G\left(1+\alpha_{12}\hbox{e}^{-r/\lambda}\right)\frac{m_1m_2}{r}
\end{equation}
with $\alpha_{12}\equiv f_1f_2$ and $f_i$ defined as
\begin{equation}
f_i\equiv M_4\frac{\partial\ln m_i(\phi)}{\partial\phi}
\end{equation}
where $M_4^{-2}\equiv 8\pi G/\hbar c$ is the four dimensional Planck mass.
The coefficient $\alpha_{12}$ is thus not a fundamental constant and
depends {\it a priori} on the chemical composition of the two test
masses.  It follows that
\begin{equation}
\eta_{12}=\frac{f_{\rm ext}|f_1-f_2|}{1+f_{\rm ext}(f_1+f_2)/2}
         \simeq M_4f_{\rm ext}\left|\partial_\phi\ln\frac{m_1}{m_2}\right|.
\end{equation}

To set any constraint, one has to determine the functions
$f_i(\phi)$, which can only be made in a model-dependent approach
[see e.g. Damour (1996) for a discussion of the information that
can be extracted in a model-independent way].  For instance, if
$\phi$ couples to a charge $Q$ the additional potential is
expected to be of the form
\begin{equation}
V(r)=-f_Q\frac{Q_1Q_2}{r}\hbox{e}^{-r/\lambda}
\end{equation}
with $f_Q$ being a fundamental constant ($f_Q>0$ for scalar exchange and
$f_Q<0$ for vector exchange). It follows that
$\alpha_{12}$ depends explicitly of the composition of the two bodies
as
\begin{equation}
\alpha_{12}=\xi_Q\frac{Q_1}{\mu_1}\frac{Q_2}{\mu_2}
\end{equation}
where $\mu_i\equiv m_i/m_{\rm H}$ and $\xi_Q={f_Q}/{Gm_{\rm
H}^2}$. Their relative acceleration in an external field $\vec
g_{\rm ext}$ is
\begin{equation}
\Delta\vec a_{12}=\xi_Q\left(\frac{Q}{\mu}\right)_{\rm ext}
\left[\frac{Q_1}{\mu_1}-\frac{Q_2}{\mu_2}\right]\vec g_{\rm ext}.
\end{equation}
For instance, in the case of a fifth force induced by a dilaton or
string moduli, Damour and Polyakov (1994a,b) showed that there are
three charges $B=N+Z$, $D=N-Z$ and $E=Z(Z-1)B^{1/3}$ representing
respectively the baryon number, the neutron excess and a term
proportional to the nuclear Coulomb energy.  The test of the
equivalence principle results in an exclusion plot in the plane
($\xi_Q,\lambda$) (see Figure~\ref{figwep}).

To illustrate the link between the variation of the constants and the
tests of relativity, let us considered the string-inspired model
developed by Damour and Polyakov (1994a, 1994b), in which the fine
structure constant is given in terms of a function of the four
dimensional dilaton as $\aem=B_F^{-1}(\phi)$. The QCD mass scale can
be expressed in terms of the string mass scale,
$M_s\sim3\times10^{17}$~GeV [see Section~\ref{subsec_7.1} for details
and Eq.~(\ref{qcd})]. In the chiral limit, the (Einstein-frame) hadron
mass is proportional to the QCD mass scale so that,
\begin{equation}\label{fhad}
f_{\rm hadron}\simeq-\left(\ln\frac{M_s}{m_{\rm hadron}}
+\frac{1}{2}\right)\frac{\partial\ln\aem}{\partial\phi}.
\end{equation}
With the expected form $\ln B_F(\phi)=-\kappa(\phi-\phi_m)^2/2$ (see
Section~\ref{subsec_7.1}), the factor of the r.h.s. of the previous
equation is of order $40\kappa(\phi-\phi_m)$. The exchange of the
scalar field excitation induces a deviation from general relativity
characterized, at post-Newtonian level, by the Eddington parameters
\begin{eqnarray}
&&1-\gamma_{_{\rm Edd}}\simeq2(40\kappa)^2(\phi_0-\phi_m)^2,\\
&&\beta_{_{\rm Edd}}-1\simeq(40\kappa)^2(\phi_0-\phi_m)^2/2.
\end{eqnarray}
Besides, the violation of the universality of free fall is given by
$\eta_{12}=\hat\delta_1-\hat\delta_2$ with
\begin{eqnarray}
\hat\delta_1&=&(1-\gamma_{_{\rm Edd}})\left[c_2\left(\frac{B}{\mu}\right)_1
+c_D\left(\frac{D}{\mu}\right)_1\right.\nonumber\\
&&\qquad\qquad\qquad\left.
+0.943\times10^{-5}
\left(\frac{E}{\mu}\right)_1\right]
\end{eqnarray}
obtained from the expression~(\ref{mass}) for the mass. In this
expression, the third term is expected to dominate. We see on this
example that the variation of the constants, the violation of the
equivalence principle and post-Newtonian deviation from general
relativity have to be considered together.

Similarly, in an effective 4-dimensional theory, the only consistent
approach to make a Lagrangian parameter time dependent is to consider
it as a field. The Klein-Gordon equation for this field
($\ddot\phi+3H\dot\phi+m^2\phi+\ldots=0$) implies that $\phi$ is
damped as $\dot\phi\propto a^{-3}$ if its mass is much smaller than
the Hubble scale. Thus, in order to be varying during the last Hubble
time, $\phi$ has to be very light with typical mass $m\sim
H_0\sim10^{-33}$~eV. This is analogous to the case of quintessence
models (see Section~\ref{sec_5.4} for details). As a consequence,
$\phi$ has to be very weakly coupled to the standard model fields.  To
illustrate this, Dvali and Zaldarriaga (2002) [followed by a
re-analysis by Chiba and Khori (2001), Wetterich (2002)] expanded
$\aem$ around it value today as
\begin{equation}
\aem=\aem(0)+\lambda\frac{\phi}{M_4}+{\cal
O}\left(\frac{\phi^2}{M_4^2} \right)
\end{equation}
from which it follows, from Webb {\em et al.} (2001) measure, that
$\lambda\Delta\phi/M_4\sim 10^{-7}$ during the last Hubble time.  The
change of the mass of the proton and of the neutron due to
electromagnetic effects was obtained from
Eqs.~(\ref{mass}-\ref{gl}) but with neglecting the last term. The
extra-Lagrangian for the field $\phi$ is thus
\begin{equation}
\delta L=\lambda\frac{\phi}{M_4}\left(B_{\rm p} p\bar p+ B_{\rm n}
n\bar n\right).
\end{equation}
A test body composed of $n_{\rm n}$ neutrons and $n_{\rm p}$ protons will
be characterized by a sensitivity
\begin{equation}
f_i=\frac{\lambda}{m_{\rm N}}(\nu_{\rm p}B_{\rm p}+\nu_{\rm n}B_{\rm n})
\end{equation}
where $\nu_{\rm n}$ (resp. $\nu_{\rm p}$) is the ratio of neutrons
(resp. protons) and where it has been assumed that $m_{\rm n}\sim
m_{\rm p}\sim m_{\rm N}$. Assuming\footnote{For copper $\nu_{\rm
p}=0.456$, for uranium $\nu_{\rm p}=0.385$ and for lead $\nu_{\rm
p}=0.397$.}  that $\nu_{{\rm n,p}}^{\rm Earth}\sim1/2$ and using that
the compactness of the Moon-Earth system $\partial\ln(m_{\rm
Earth}/m_{\rm Moon})/\partial\ln\aem\sim10^{-3}$, one gets
$\eta_{12}\sim10^{-3}\lambda^2$.  Dvali and Zaldarriaga (2002)
obtained the same result by considering that $\Delta\nu_{{\rm
n,p}}\sim6\times10^{-2}-10^{-1}$. This implies that $\lambda<10^{-5}$
which is compatible with the variation of $\aem$ if
$\Delta\phi/M_4>10^{-2}$ during the last Hubble period.

From cosmological investigations one can show that
$(\Delta\phi/M_4)^2\sim (\rho_\phi+P_\phi)/\rho_{\rm total}$. If
$\phi$ dominates the matter content of the universe, $\rho_{\rm
total}$, then $\Delta\phi\sim M_4$ so that $\lambda\sim 10^{-7}$
whereas if it is sub-dominant $\Delta\phi\ll M_4$ and $\lambda\gg
10^{-7}$. In conclusion
\begin{equation}
10^{-7}<\lambda<10^{-5}.
\end{equation}
This explicits the tuning on the parameter $\lambda$.

An underlying approximation is that the $\phi$-dependence arises only
from the electromagnetic self-energy. But, in general, one would
expect that the dominant contribution to the hadron mass, the QCD
contributions, also induces a $\phi$-dependence (as in the Damour and
Polyakov, 1994a,b approach).

In conclusion, the test of the equivalence principle offers a very
precise test of the variation of constants (Damour, 2001). The LLR
constrain $\eta\la10^{-13}$, i.e $|\vec a_{\rm Earth}-\vec a_{\rm
Moon}|\la 10^{-14}\,{\rm cm.s}^{-2}$, implies that on the size of the
Earth orbit $|\nabla\ln\aem|\la10^{-33}-10^{-32}\,{\rm
cm}^{-1}$. Extending this measurement to the Hubble size leads to the
estimate $\Delta\aem/\aem\la10^{-4}-10^{-5}$. This indicates that if
the claim by Webb {\em et al.} (2001) is correct then it should induce
a detectable violation of the equivalence principle by coming
experiments such as MICROSCOPE\footnote{{\tt
http://sci2.esa.int/Microscope/}} and STEP\footnote{{\tt
http://einstein.stanford.edu/STEP/}} will test it respectively at the
level $\eta\sim10^{-15}$ and $\eta\sim 10^{-18}$.  Indeed, this is a
rough estimate in which $\dot\aem$ is assumed to be constant, but this
is also the conclusion indicated by the result by Dvali and
Zaldarriaga (2002) and Bekenstein (1982).

Let us also note that this constraint has been discarded by a
some models (see Section~\ref{subsec_7.15}) and particularly while
claiming that a variation of $\aem$ of $10^{-5}$ was realistic
(Sandvik {\em et al.}, 2002; Barrow {\em et al.}, 2001) [see however
the recent discussion by Magueijo {\em et al.} (2002)].

\section{Gravitational constant}\label{sec_3}

As pointed by Dicke and Peebles (1965), the importance of gravitation
on large scales is due to the short range of the strong and weak
forces and to the fact that the electromagnetic force becomes weak
because of the global neutrality of the matter. As they provide tests
of the law of gravitation (planetary motions, light
deflection,\ldots), space science and cosmology also offer tests of
the constancy of the gravitational constant.

Contrary to most of the other fundamental constants, as the precision
of the measurements increased, the disparity between the measured
values of $G$ also increased. This led the CODATA\footnote{The CODATA
is the COmmittee on Data for Science and Technology, see {\tt
http://www.codata.org/}.} in 1998 to raise the relative uncertainty
for $G$ from 0.013\% to 0.15\% (Gundlach and Merkowitz, 2000).

\subsection{Paleontological and geophysical arguments}
\label{subsec_3.1}

Dicke (1964) stressed that the Earth is such a complex system that it
would be difficult to use it as a source of evidence for or against
the existence of a time variation of the gravitational constant. He
noted that among the direct effects, a weakening of the gravitational
constant induces a variation of the Earth surface temperature, an
expansion of the Earth radius and a variation of the length of the day
(Jordan, 1955, and then Murphy and Dicke, 1964; Hoyle, 1972).

\subsubsection{Earth surface temperature}

Teller (1948) first emphasized that Dirac hypothesis may be in
conflict with paleontological evidence. His argument is based on
the estimation of the temperature at the center of the Sun
$T_\odot\propto GM_\odot/R_\odot$ using the virial theorem. The
luminosity of the Sun is then proportional to the radiation energy
gradient times the mean free path of a photon times the surface of
the Sun, that is $L_\odot\propto T^7_\odot R_\odot^7M_\odot^{-2}$,
hence concluding that $L_\odot\propto T^7_\odot M_\odot^{5}$.
Computing the radius of the Earth orbit in Newtonian mechanics,
assuming the conservation of angular momentum (so that $GM_\odot
R_{\rm Earth}$ is constant) and stating that the Earth mean
temperature is proportional to the fourth root of the energy
received, he concluded that
\begin{equation}
  T_{\rm Earth}\propto G^{2.25}M_\odot^{1.75}.
\end{equation}
If $M_\odot$ is constant and $G$ was 10\% larger 300 million years
ago, the Earth surface temperature should have been 20\% higher,
that is close to the boiling temperature. This was in
contradiction with the existence of trilobites in the Cambrian.

Teller (1948) used a too low value for the age of the universe.
Gamow (1967a) actualized the numbers and showed that even if it
was safe at the Cambrian era, there was still a contradiction with
bacteria and alga estimated to have lived $4\times10^{9}$ years
ago. It follows that
\begin{equation}
\left|{\Delta G}/{G}\right|<0.1.
\end{equation}
Eichendorf and Reinhardt (1977) re-actualized Teller's argument in
light of a new estimate of the age of the universe and new
paleontological discoveries to get $|\dot G/G|<2.0\times10^{-11}\,{\rm
yr}^{-1}$ (cited by Petley, 1985).

When using such an argument, the heat balance of the atmosphere is
affected by many factors (water vapor content, carbon dioxide content,
circulatory patterns,\ldots) is completely neglected.  This renders
the extrapolation during several billion years very unreliable. For
instance, the rise of the temperature implies that the atmosphere is
at some stage mostly composed of water vapor so that its convective
mechanism is expected to change in such a way to increase the Earth
albedo and thus to decrease the temperature!

\subsubsection{Expanding Earth}

Egeyed (1961) first remarked that paleomagnetic data could be used to
calculate the Earth paleoradius for different geological epochs.
Under the hypothesis that the area of continental material has
remained constant while the bulk of the Earth has expanded, the
determination of the difference in paleolatitudes between two sites of
known separation give a measurement of the paleoradius.  Creer (1965)
showed that data older than $3\times10^8$ years form a coherent group
in $\dot r_{_{\rm Earth}}$ and Wesson (1973) concluded from a
compilation of data that the expansion was most probably of 0.66 mm
per year during the last $3\times10^9$ years.

Dicke (1962c, 1964) related the variation of the Earth
radius to a variation of the gravitational constant by
\begin{equation}
\Delta\ln r_{_{\rm Earth}}=-0.1\Delta\ln G.
\end{equation}
McElhinny {\em et al.} (1978) re-estimated the paleoradius of the Earth
and extended the analysis to the Moon, Mars and Mercury. Starting from
the hydrostatic equilibrium equation
\begin{equation}\label{radius1}
\frac{\dd P}{\dd r}=-G\frac{\rho(r)M(r)}{r^2},
\end{equation}
where $M(r)$ is the mass within radius $r$, they generalized Dicke's
result to get
\begin{equation}
\Delta\ln r_{_{\rm Earth}}=-\alpha\Delta\ln G
\end{equation}
where $\alpha$ depends on the equation of state $P(\rho)$,
e.g. $\alpha=1/(3n-4)$ for a polytropic gas, $P=C\rho^n$. In the case
of small planets, one can work in a small gravitational
self-compression limit and set $P=K_0(\rho/\rho_0-1)$.
Eq.~(\ref{radius1})then  gives $\alpha=({2}/{15})({\Delta\rho}/{\rho_0})$,
$\Delta\rho$ being the density difference between the center and
surface. This approximation is poor for the Earth and more
sophisticated model exist. They give $\alpha_{_{\rm
Earth}}=0.085\pm0.02$, $\alpha_{_{\rm Mars}}=0.032$, $\alpha_{_{\rm
Mercury}}=0.02\pm0.05$ and $\alpha_{_{\rm Moon}}=0.004\pm0.001$. Using
the observational fact that the Earth has not expanded by more than
$0.8\%$ over the past $4\times10^8$ years, the Moon of $0.06\%$ over
the past $4\times10^9$ years and Mars of $0.6\%$, they concluded that
\begin{equation}
-\dot G/G\la8\times10^{-12}\,{\rm yr}^{-1}.
\end{equation}
Despite any real evidence in favor of an expanding Earth, the rate of
expansion is also limited by another geophysical aspect, i.e. the
deceleration of the Earth rotation.

Dicke (1957) listed out some other possible consequences on the
scenario of the formation of the Moon and on the geomagnetic field but
none of them enable to put serious constraints. The paleontological
data give only poor limits on the variation of the gravitational
constant and even though the Earth kept a memory of the early
gravitational conditions, this memory is crude and geological data are
not easy to interpret.
\subsection{Planetary and stellar orbits}
\label{subsec_3.2}

Vinti (1974) studied the dynamics of two-body system in
Dirac cosmology. He showed that the equation of motion
\begin{equation}\label{llr2}
  \frac{\dd^2\vec r}{\dd t^2}= -G_0\frac{k+t_0}{k+t}m
  \frac{\vec r}{r^3}
\end{equation}
where $k$ is a constant and $G_0$ the gravitational constant
today, can be integrated. For bounded orbits, the solution
describes a growing ellipse with constant eccentricity, $e$,
pericenter argument, $\omega$ and a linearly growing semi-latus
rectum $p(t)=(l^2/G_0m)(k+t)/(k+t_0)$, where $l$ is the constant
angular momentum, of equation
\begin{equation}\label{llr3}
  r=\frac{p(t)}{1+e\cos(\theta-\omega)}.
\end{equation}
Similarly, Lynden-Bell (1982) showed that the equations of motion of the
$N$-body problem can be transformed to the standard equation if $G$
varies as $t^{-1}$.

It follows that in the Newtonian limit, the
orbital period of a two-body system is
\begin{equation}\label{llr4}
  P=\frac{2\pi l}{(Gm)^2}\frac{1}{(1-e^2)^{3/2}}\left[
  1+{\cal O}\left(\frac{G^2m^2}{c^2l^2}\right)\right]
\end{equation}
in which the correction terms represent the post-Newtonian corrections
to the Keplerian relationship. It is typically of order $10^{-7}$ and
$10^{-6}$ respectively for Solar system planetary orbits and for a
binary pulsar. It follows that
\begin{equation}\label{llr5}
  \frac{\dot P}{P}=3\frac{\dot l}{l}-2\frac{\dot G}{G}-2\frac{\dot
  m}{m}.
\end{equation}
Only for the orbits of bodies for which the gravitational self-energy
can be neglected does the previous equation reduce to
\begin{equation}\label{llr6}
  \frac{\dot P}{P}=-2\frac{\dot G}{G}.
\end{equation}
This leads to two observable effects in the Solar system (Shapiro,
1964; Counselman and Shapiro, 1968). First, the scale of the Solar
system changes and second, if $G$ evolves adiabatically as $G=G_0+\dot
G_0(t-t_0)$, there will be a quadratically growing increment in the
mean longitude of each body.

For a compact body, the mass depends on $G$ as well as other
post-Newtonian parameters. At first order in the post-Newtonian
expansion, there is a negative contribution (\ref{llr8}) to the mass
arising from the gravitational binding energy and one cannot neglect
$\dot m$ in Eq.~(\ref{llr5}). This is also the case if other constants
are varying.

\subsubsection{Early works}\label{earlywork}

Early works mainly focus on the Earth-Moon system and try to relate a
time variation of $G$ to a variation of the frequency or mean motion
($n=2\pi/P$) of the Moon around the Earth. Arguments on an expanding
Earth also raised interests in the determination of the Earth rotation
rate. One of the greatest problem is to evaluate and subtract the
contribution of the spin-down of the Earth arising from the friction
in the seas due to tides raised by the Moon [Van Flandern (1981)
estimated that $\dot n_{_{\rm tidal}}=(-28.8\pm1.5)''\,{\rm
century}^{-2}$] and a contribution from the Moon recession.

The determination of ancient rotation rates can rely on
paleontological data, ancient eclipse observations as well as
measurements of star declinations (Newton, 1970, 1974). It can be concluded
from these studies that there were about 400 days in a year during the
Devonian. Indeed, this studies are entailed by a lot of uncertainties,
for instance, Runcorn (1964) compared telescope observation from the
17th century to the ancient eclipse records and found a discrepancy of
a factor 2. As an example, Muller (1978) studied eclipses from 1374BC
to 1715AD to conclude that
\begin{equation}
\dot G/G=(2.6\pm15)\times10^{-11}\,{\rm yr}^{-1}
\end{equation}
and Morrison (1973) used ephemeris from 1663 to 1972 including 40,000
Lunar occultations from 1943 to 1972 to deduce that
\begin{equation}
|\dot G/G|<2\times10^{-11}\,{\rm yr}^{-1}.
\end{equation}

Paleontological data such as the growth rhythm found in fossil
bivalves and corals also enable to set constraint on the Earth
rotational history and the Moon orbit (Van Diggelen, 1976) [for
instance, in the study by Scrutton (1965) the fossils showed
marking so fine that the phases of the Moon were mirrored in the
coral growth]. Blake (1977b) related the variation of the number
of sidereal days in a sidereal year, $Y= n_E/n_S$, and in a
sidereal month, $M=n_E/n_M$, ($n_E$, $n_S$ and $n_M$ being
respectively the orbital frequencies of the motion of the Earth,
of the Moon around the Earth and of the Earth around the Sun) to
the variation of the Newton constant and the Earth momentum of
inertia $I$ as
\begin{equation}\label{fossil}
(\gamma-1)\frac{\Delta Y}{Y}-\gamma\frac{\Delta M}{M}=
\frac{\Delta I}{I}+2\frac{\Delta G}{G}
\end{equation}
with $\gamma=1.9856$ being a calculated constant. The fossil data
represent the number of Solar days in a tropical year and in a
synodic month which can be related to $Y$ and $M$ so that one
obtains a constraint on $\Delta I/I+2\Delta G/G$. Attributing the
variation of $I$ to the expansion of the Earth (Wesson, 1973), one
can argue that $\Delta I/I$ represents only 10-20\% of the r.h.s
of (\ref{fossil}). Blake (1977b) concluded that
\begin{equation}
\dot G/G=(-0.5\pm2)\times10^{-11}\,{\rm yr}^{-1}.
\end{equation}

Van Flandern (1971, 1975) studied the motion of the Moon from Lunar
occultation observations from 1955 to 1974 using atomic time which
differs from the ephemeris time relying on the motion of the Earth
around the Sun. He attributed the residual acceleration after
correction of tidal effect to a variation of $G$, $\dot n_{\rm
Moon}^G/2n_{\rm Moon}^G= (-8\pm5)\times10^{-9}\,{\rm
century}^{-2}$ to claim that
\begin{equation}
\dot G/G=(-8\pm5)\times10^{-11}\,{\rm yr}^{-1}.
\end{equation}
In a new analysis, Van Flandern (1981) concluded that $\dot n_{\rm
Moon}^G/ n_{\rm Moon}^G= (3.2\pm1.1)\times10^{-11}\,{\rm
yr}^{-1}$ hence that $G$ was increasing as
\begin{equation}
\dot G/G=(3.2\pm1.1)\times10^{-11}\,{\rm yr}^{-1}
\end{equation}
which has the opposite sign. In this comparison the time scale of the
atomic time is 20 years and the one of the ephemeris 200 years but is
less precise. It follows that the comparison is not obvious and that
these results are far from being convincing.  In this occultation
method, one has to be sure that the proper motions of the stars are
taken into account. One also has to assume that (1) the mass of the
planets are not varying (see Eq.~\ref{llr5}), which can happen if
e.g. the strong and fine structure constants are varying (2)
the fine structure constant is not varying while comparing with
atomic time and (3) the effect of the changing radius of the Earth was
not taken into account.

\subsubsection{Solar system}

Monitoring the separation of orbiting bodies offers a possibility to
constrain the time variation of $G$. This accounts for comparing a
gravitational time scale (set by the orbit) and an atomic time
scale and it is thus assumed that the variation of atomic
constants is negligible on the time of the experiment.

Shapiro {\em et al.} (1971) compared radar-echo time delays between
Earth, Venus and Mercury with a cesium atomic clock between 1964 and
1969. The data were fitted to the theoretical equation of motion for
the bodies in a Schwarzschild spacetime, taking into account the
perturbations from the Moon and other planets. They concluded that
\begin{equation}
|\dot G/G|<4\times10^{-10}\,{\rm yr}^{-1}.
\end{equation}
The data concerning Venus cannot be used due to
imprecision in the determination of the portion of the planet
reflecting the radar. This was improved to
\begin{equation}
|\dot G/G|<1.5\times10^{-10}\,{\rm yr}^{-1}
\end{equation}
by including Mariner 9 and Mars orbiter data (Reasenberg and Shapiro,
1976, 1978).  The analysis was further extended (Shapiro, 1990) to
give
\begin{equation}
\dot G/G=(-2\pm10)\times10^{-12}\,{\rm yr}^{-1}.
\end{equation}
The combination of Mariner 10 an Mercury and Venus ranging data
gives (Anderson {\em et al.}, 1991)
\begin{equation}
\dot G/G=(0.0\pm2.0)\times10^{-12}\,{\rm yr}^{-1}.
\end{equation}

The Lunar laser ranging (LLR) experiment has measured the position of
the Moon with an accuracy of about 1~cm for thirty years. This was
made possible by the American Appolo 11, 14 and 15 missions and
Soviet-French Lunakhod 1 and 4 which landed retro-reflectors on the
Moon that reflect laser pulse from the Earth (see Dickey {\em et al.}
(1994) for a complete description). Williams {\em et al.}  (1976)
deduced from the six first years of LLR data that $\omega_{_{\rm
BD}}>29$ so that
\begin{equation}
|\dot G/G|\la 3\times10^{-11}\,{\rm yr}^{-1}.
\end{equation}
M\"uller {\em et al.} (1991) used 20 years of data to improve this result to
\begin{equation}
|\dot G/G|<1.04\times10^{-11}\,{\rm yr}^{-1},
\end{equation}
the main error arising from the Lunar tidal acceleration. Dickey {\em
et al.} (1994) improved this constraint to
\begin{equation}
|\dot G/G|<6\times10^{-12}\,{\rm yr}^{-1}
\end{equation}
and Williams {\em et al.} (1996) with 24 years of data concluded that
\begin{equation}
|\dot G/G|<8\times10^{-12}\,{\rm yr}^{-1}
\end{equation}

Reasenberg {\em et al.} (1979) considered the 14 months data obtained
from the ranging of the Viking spacecraft and deduced that
$\omega_{_{\rm BD}}>500$ which implies
\begin{equation}
-\dot G/G<10^{-12}\,{\rm yr}^{-1}.
\end{equation}
Hellings {\em et al.} (1983)
using all available astrometric data and in particular the ranging
data from Viking landers on Mars deduced that
\begin{equation}
|\dot G/G|=(2\pm4)\times10^{-12}\,{\rm yr}^{-1}.
\end{equation}
The major contribution to the uncertainty is due to the modeling of
the dynamics of the asteroids on the Earth-Mars range.  Hellings {\em
et al.} (1983) also tried to attribute their result to a time
variation of the atomic constants.  Using the same data but a
different modeling of the asteroids, Reasenberg (1983) got
\begin{equation}
|\dot G/G|<3\times10^{-11}\,{\rm yr}^{-1}
\end{equation}
which was then improved by Chandler {\em et al.} (1993) to
\begin{equation}
|\dot G/G|<10^{-11}\,{\rm yr}^{-1}.
\end{equation}

All these measurements allow to test more than just the time
variation of the gravitational constant and offer a series of tests on
the theory of gravitation and constrain PPN parameters, geodetic
precession etc... (see Will, 1993).

\subsubsection{Pulsars}
\label{subsec_3.23}

Contrary to the Solar system case, the dependence of the gravitational
binding energy cannot be neglected while computing the time variation
of the period (Dicke, 1969; Eardley, 1975; Haugan, 1979). Here two
approaches can be followed; either one sticks to a model
(e.g. scalar-tensor gravity) and compute all the effects in this model
or one has a more phenomenological approach and tries to put some
model-independent bounds.

Eardley (1975) followed the first route and discussed the effects of a
time variation of the gravitational constant on binary pulsar in the
framework of the Brans-Dicke theory. In that case, both a dipole
gravitational radiation and the variation of $G$ induce a periodic
variation in the pulse period.  Nordtvedt (1990) showed that the
orbital period changes as
\begin{equation}
\frac{\dot P}{P}=-\left[2+\frac{2(m_1c_1+m_2c_2)+3(m_1c_2+m_2c_1)}{m_1+m_2}
\right]\frac{\dot G}{G}
\end{equation}
where $c_i\equiv\delta\ln m_i/\delta\ln G$. He concluded that for the pulsar
PSR~1913+16 ($m_1\simeq m_2$ and $c_1\simeq c_2$) one gets
\begin{equation}\label{pn}
\frac{\dot P}{P}=-\left[2+5c\right]\frac{\dot G}{G},
\end{equation}
the coefficient $c$ being model dependent.  As another
application, he estimated that $c_{_{\rm
Earth}}\sim-5\times10^{-10}$, $c_{_{\rm Moon}}\sim-10^{-8}$ and
$c_{_{\rm Sun}}\sim-4\times10^{-6}$ justifying the approximation
(\ref{llr6}) for the Solar system.

Damour {\em et al.} (1988) used the timing data of the binary pulsar
PSR~1913+16. They implemented the effect of the time variation of $G$
by considering the effect on $\dot P/P$ and making use of the
transformation suggested by Lynden-Bell (1982) to integrate the
orbit. They showed, in a theory-independent way, that $\dot
G/G=-0.5\delta\dot P/P$, where $\delta\dot P$ is the part of the
orbital period derivative that is not explained otherwise (by
gravitational waves radiation damping). It has to be contrasted with
the result (\ref{pn}) by Nordtvedt (1990). They got
\begin{equation}
\dot G/G=(1.0\pm2.3)\times10^{-11}\,{\rm yr}^{-1}.
\end{equation}
Damour and Taylor (1991) reexamined the data of PSR~1913+16 and the
upper bound
\begin{equation}
\dot G/G<(1.10\pm1.07)\times10^{-11}\,{\rm yr}^{-1}.
\end{equation}
Kaspi {\em et al.} (1994) used data from PSR~B1913+16 and PSR~B1855+09
respectively to get
\begin{equation}
\dot G/G=(4\pm5)\times10^{-12}\,{\rm yr}^{-1}
\end{equation}
and
\begin{equation}
\dot G/G=(-9\pm18)\times10^{-12}\,{\rm yr}^{-1},
\end{equation}
the latter case being more ``secure'' since the orbiting companion is
not a neutron star.

All the previous results concern binary pulsar but isolated can
also be used.  Heintzmann and Hillebrandt (1975) related the
spin-down of the pulsar JP1953 to a time variation of $G$. The
spin-down is a combined effect of electromagnetic losses, emission
of gravitational waves, possible spin-up due to matter accretion.
Assuming that the angular momentum is conserved so that
$I/P=$constant, one deduces that
\begin{equation}
\frac{\dot P}{P}_G=\left(\frac{\dd\ln I}{\dd\ln G}\right)
\frac{\dot G}{G}.
\end{equation}
The observational spin-down can be decomposed as
\begin{equation}
\frac{\dot P}{P}_{_{\rm obs}}=\frac{\dot P}{P}_{_{\rm mag}}
+\frac{\dot P}{P}_{_{\rm GW}}+\frac{\dot P}{P}_G.
\end{equation}
Since ${\dot P}/{P}_{_{\rm mag}}$ and ${\dot P}/{P}_{_{\rm GW}}$ are
positive definite, it follows that ${\dot P}/{P}_{_{\rm obs}}\geq{\dot
P}/{P}_G$ so that a bound on $\dot G$ can be inferred if the main
pulse period is the period of rotation.

Heintzmann and Hillebrandt (1975) modeled the pulsar by a polytropic
$(P\propto\rho^n$) white dwarf and deduced that ${\dd\ln
I}/{\dd\ln G}=2-3n/2$ so that
\begin{equation}
\vert\dot G/G\vert<10^{-10}\,{\rm yr}^{-1}.
\end{equation}
Mansfield (1976) assumed a relativistic degenerate, zero temperature
polytropic star and got
\begin{equation}\label{ll}
  -{\dot G}/{G}<5.8_{-1}^{+1}\times10^{-11}\,{\rm yr}^{-1}
\end{equation}
at a $2\sigma$ level. He also noted that a positive $\dot G$ induces a
spin-up counteracting the electromagnetic spin-down which can provide
another bound if an independent estimate of the pulsar magnetic field
can be obtained.  Goldman (1990), following Eardley (1975), used the
scaling relations $N\propto G^{-3/2}$ and $M\propto G^{-5/2}$ to
deduce that $2{\dd\ln I}/{\dd\ln G}=-5+3{\dd\ln I}/{\dd\ln N}$. He
used the data from the pulsar PSR~0655+64 to deduce
\begin{equation}\label{goldman}
  -\dot G/G<(2.2-5.5)\times10^{-11}\,{\rm yr}^{-1}.
\end{equation}

\subsection{Stellar constraints}
\label{subsec_3.4}

In early works, Pochoda and Schwarzschild (1964), Ezer and Cameron
(1966) and then Gamow (1967c) studied the Solar evolution in presence
of a time varying gravitational constant. They came to the conclusion
that under Dirac hypothesis, the original nuclear resources of the Sun
would have been burned by now. This results from the fact that an
increase of the gravitational constant is equivalent to an increase of
the star density (because of the Poisson equation).

A side effect of the change of luminosity is a change in the depth of
the convection zone. This induces a modification of the vibration
modes of the star and particularly to the acoustic waves, i.e
$p$-modes, (Demarque {\em et al.}, 1994). Demarque {\em et al.} (1994)
considered an ansatz in which $G\propto t^{-\beta}$ and showed that
$|\beta|<0.1$ over the last $4.5\times 10^9$ years, which corresponds
to
\begin{equation}\label{gsun1}
  \left\vert\dot G/G\right\vert<2\times10^{-11}\,{\rm yr}^{-1}.
\end{equation}
Guenther {\em et al.} (1995) also showed that $g$-modes could provide
even much tighter constraints but these modes are up to now very
difficult to observe. Nevertheless, they concluded, using the claim of
detection by Hill and Gu (1990), that
\begin{equation}\label{gsun2}
  \left\vert\dot G/G\right\vert<4.5\times10^{-12}\,{\rm yr}^{-1}.
\end{equation}
Guenther {\em et al.}  (1998) compared
the $p$-mode spectra predicted by different theories with varying
gravitational constant to the observed spectrum obtained by a network
of six telescopes and deduced that
\begin{equation}\label{gsun3}
  \left\vert\dot G/G\right\vert<1.6\times10^{-12}\,{\rm yr}^{-1}.
\end{equation}
The standard Solar model depends on few parameters and $G$ plays a
important role since stellar evolution is dictated by the balance
between gravitation and other interactions. Astronomical observations
determines very accurately $G M_\odot$ and a variation of $G$ with
$GM_\odot$ fixed induces a change of the pressue
($P=GM_\odot^2/R_\odot^2$) and density
($\rho=M_\odot/R_\odot^3$). Ricci and Villante (2002) studied the
effect of a variation of $G$ on the density and pressure profile of
the Sun and concluded that present data cannot constrain $G$ better
than $10{-2}\%$.

The late stages of stellar evolution are governed by the Chandrasekhar
mass $(\hbar c/G)^{3/2}m_{\rm n}^{-2}$ mainly determined by the
balance between the Fermi pressure of a degenerate electron gas and
gravity. Assuming that the mean neutron star mass is given by the
Chandrasekhar mass, one expects that $\dot G/G=-2\dot M_{_{\rm
NS}}/3M_{_{\rm NS}}$. Thorsett (1996) used the observations of five
neutron star binaries for which five Keplerian parameters can be
determined (the binary period $P_b$, the projection of the orbital
semi-major axis $a_1\sin i$, the eccentricity $e$, the time and
longitude of the periastron $T_0$ and $\omega$) as well as the
relativistic advance of the angle of the periastron $\dot \omega$.
Assuming that the neutron star masses vary slowly as $M_{_{\rm
NS}}=M_{_{\rm NS}}^{(0)}-\dot M_{_{\rm NS}} t_{_{\rm NS}}$, that their
age was determined by the rate at which $P_b$ is increasing (so that
$t_{NS}\simeq2P_b/\dot P_b$) and that the mass follows a normal
distribution, Thorsett (1996) deduced that, at $2\sigma$,
\begin{equation}\label{gsun4}
  \dot G/G=(-0.6\pm4.2)\times10^{-12}\,{\rm yr}^{-1}.
\end{equation}
Analogously, the Chandrasekhar mass sets the characteristic of the
light curves of supernovae (Riazuelo and Uzan, 2002).

Garcia-Berro {\em et al.} (1995) considered the effect of a variation
of the gravitational constant on the cooling of white dwarfs and on
their luminosity function. As first pointed out by Vila (1976), the
energy of white dwarfs is entirely of gravitational and thermal origin
so that a variation of $G$ will induce a modification of their energy
balance. Restricting to cold white dwarfs with luminosity smaller than
ten Solar luminosity, the luminosity can be related to the star
binding energy $B$ and gravitational energy, $E_{_{\rm grav}}$, as
\begin{equation}
L=-\frac{\dd B}{\dd t}+\frac{\dot G}{G}E_{_{\rm grav}}
\end{equation}
which simply results from the hydrostatic equilibrium. Again, the
variation of the gravitational constant intervenes via the Poisson
equation and the gravitational potential. The cooling process is
accelerated if $\dot G/G<0$ which then induces a shift in the position
of the cut-off in the luminosity function.
Garcia-Berro {\em et al.} (1995) concluded that
\begin{equation}\label{gsun}
  -\dot G/G<3^{+1}_{-3}\times10^{-11}\,{\rm yr}^{-1}.
\end{equation}
The result depends on the details of the cooling theory and on whether
the C/O white dwarf is stratified or not.

A time variation of $G$ also modifies the main sequence time of
globular clusters (Dicke 1962a; Roeder, 1967). Del'Innocenti {\em
et al.} (1996) calculated the evolution of low mass stars and
deduced the age of the isochrones. The principal effect is a
modification of the main sequence evolutionary time scale while
the appearance of the color-magnitude diagram remained undistorted
within the observational resolution and theoretical uncertainties.
Since the globular clusters must be younger than the universe, and
assuming that their age was between 8 and 20 Gyr, they concluded
\begin{equation}
\dot G/G=(-1.4\pm2.1)\times10^{-11}\,{\rm yr}^{-1}.
\end{equation}
This analysis was also applied to clusters of galaxies by Dearborn
and Schramm (1974). In that case a lower gravitational constant
allows the particle to escape from the cluster since the
gravitational binding energy also decreases. They deduced that the
decrease of $G$ that allows the existence of clusters at the
present epoch is
\begin{equation}
-\dot G/G<4\times10^{-11}\,{\rm yr}^{-1}.
\end{equation}

\subsection{Cosmological constraints}\label{subsec_3.3}

\subsubsection{CMB}

A time-dependent gravitational constant will have mainly three effects
on the CMB angular power spectrum (Riazuelo and Uzan, 2002):

(1) The variation of $G$ modifies the Friedmann equation and
therefore the age of the Universe (and, hence, the sound horizon). For
instance, if $G$ is larger at earlier time, the age of the Universe is
smaller at recombination, so that the peak structure is shifted
towards higher angular scales.

(2) The amplitude of the Silk damping is modified.  At small scales,
viscosity and heat conduction in the photon-baryon fluid produce a
damping of the photon perturbations (Silk, 1968). The damping scale
is determined by the photon diffusion length at recombination, and
therefore depends on the size of the horizon at this epoch, and hence,
depends on any variation of the Newton constant throughout the
history of the Universe.

(3) The thickness of the last scattering surface is modified. In the
same vein, the duration of recombination is modified by a variation of
the Newton constant as the expansion rate is different. It is well
known that CMB anisotropies are affected on small scales because the
last scattering ``surface'' has a finite thickness. The net effect is
to introduce an extra, roughly exponential, damping term, with the
cutoff length being determined by the thickness of the last scattering
surface. When translating redshift into time (or length), one has to
use the Friedmann equations, which are affected by a variation of the
Newton constant.  The relevant quantity to consider is the visibility
function $g$. In the limit of an infinitely thin last scattering
surface, $\tau$ goes from $\infty$ to $0$ at recombination epoch. For
standard cosmology, it drops from a large value to a much smaller one,
and hence, the visibility function still exhibits a peak, but is much
broader.

Chen and Kamionkowski (1999) studied the CMB spectrum in
Brans-Dicke theory and showed that CMB experiments such as MAP
will be able to constrain these theories for $\omega_{_{\rm
BD}}<100$ if all parameters are to be determined by the same CMB
experiment, $\omega_{_{\rm BD}}<500$ if all parameters are fixed
but the CMB normalization and $\omega_{_{\rm BD}}<800$ if one uses
the polarization. For the Planck mission these numbers are
respectively, 800, 2500 and 3200.

As far as we are aware, no complete study of the impact of the
variation of the gravitational constant (e.g. in scalar-tensor theory)
on the CMB has been performed yet. Note that, to compute the CMB
anisotropies, one needs not only the value of $G$ at the time of
decoupling but also its complete time evolution up to now, since it
will affect the integrated Sachs-Wolfe effect.

\subsubsection{Nucleosynthesis}

As explained in details in section~\ref{subsec_4.4}, changing the
value of the gravitational constant affects the freeze-out temperature
$T_{\rm f}$. A larger value of $G$ corresponds to a higher expansion rate.
This rate is determined by the combination $G\rho$ and
in the standard case the Friedmann equations imply that $G\rho t^2$ is
constant.  The density $\rho$ is determined by the number $N_*$ of
relativistic particles at the time of nucleosynthesis so that
nucleosynthesis allows to put a bound on the number of neutrinos
$N_\nu$. Equivalently, assuming the number of neutrinos to be three,
leads to the conclusion that $G$ has not varied from more than 20\%
since nucleosynthesis.  But, allowing for a change both in $G$ and
$N_\nu$ allows for a wider range of variation.  Contrary to the fine
structure constant the role of $G$ is less involved.

Steigmann (1976) used nucleosynthesis to put constraints on the Dirac
theory. Barrow (1978) assumed that $G\propto t^{-n}$ and obtained from
the helium abundances that $-5.9\times10^{-3}<n<7\times10^{-3}$ which
implies that
\begin{equation}
\left|{\dot G}/{G}\right|<(2\pm9.3)\,h\times 10^{-12}\,{\rm yr}^{-1},
\end{equation}
assuming a flat universe.  This corresponds in terms of the
Brans-Dicke parameter to $\omega_{_{\rm BD}}>25$, which is a much
smaller bounds that the ones obtained today.  Yang {\em et al.} (1979)
included the computation of the deuterium and lithium. They improved
the result by Barrow (1978) to $n<5\times10^{-3}$ which corresponds to
$\omega_{_{\rm BD}}>50$ and also pointed out that the constraint is
tighter if there are extra-neutrinos. It was further improved by
Rothman and Matzner (1982) to $|n|<3\times10^{-3}$ implying
\begin{equation}
\left|{\dot G}/{G}\right|<1.7\times 10^{-13}\,{\rm yr}^{-1}.
\end{equation}
Accetta {\em et al.}
(1990) studied the dependence of the abundances of D, $^3{\rm He}$,
$^4{\rm He}$ and $^7{\rm Li}$ upon the variation of $G$ and concluded
that
\begin{equation}
-0.3<{\Delta G}/{G}<0.4
\end{equation}
which roughly corresponds to $9\times10^{-3}<n<8\times10^{-3}$ and
to $|\dot G/G|<9\times10^{-13}\,{\rm yr}^{-1}$.

All previous investigations assumed that the other constants are kept
fixed and that physics is unchanged. Kolb {\em et al.} (1986) assumed
a correlated variation of $G$, $\aem$ and $\gfermi$ and got a bound on
the variation of the radius of the extra-dimensions.

The case of Brans-Dicke (1961) theory, in which only the gravitational
constant varies, was well studied. Casas {\em et al.} (1992a, 1992b)
concluded from the study of helium and deuterium abundances that
$\omega_{_{\rm BD}}>380$ when $N_\nu=3$ (see also Damour and Gundlach,
1991, and Serna {\em et al.}, 1992) and $\omega_{_{\rm BD}}>50$ when
$N_\nu=2$.

Kim and Lee (1995) calculated the allowed value for the gravitational
constant, electron chemical potential and entropy consistent with
observations up to lithium-7 and argued that beryllium-9 and bore-11
abundances are very sensitive to a change in $G$. Kim {\em et al.}
(1998) further included neutrino degeneracy. The degeneracy of the
electron-neutrino not only increases the radiation density but also
influences the weak interaction rates so that it cannot be absorbed in
a variation of $G$. It was shown that a higher gravitational constant
can be balanced by a higher electron-neutrino degeneracy so that the
range of (electron chemical potential, $G$) was wider.

Damour and Pichon (1999) extended these investigations by considering
a two-parameter family of scalar-tensor theories of gravitation
involving a non-linear scalar field-matter coupling function. They
concluded that even in the cases where before BBN the scalar-tensor
theory was far from general relativity, BBN enables to set quite small
constraints on the observable deviations from general relativity
today.

Let us also note the work by Carroll and Kaplinghat (2001) in
which they tried to constrain the expansion history of our
universe in a model-independent way during nucleosynthesis. They
assumed changes in the gravitational dynamics and not in the
particle physics processes. For that purpose the expansion rate at
the time of nucleosynthesis is approximated as $H(T)=(T/1\,{\rm
MeV})^\alpha H_1$ in order to infer the constraints on
$(\alpha,H_1)$. This a simple way to compare an alternative to
cosmology with data.

\section{Other constants}\label{sec_5}

Up to now, we have detailed the results concerning the two most
studied constants, $\ag$ and $\aem$. But, as we emphasized,
if $\aem$ is varying one also expects a variation of other constants
such as $\as$ and $\aw$. There are many theoretical reasons for
that. First, in Kaluza-Klein or string inspired models, all constants
are varying due either to the dilaton or the extra-dimensions (see
Section~\ref{sec_7} for details).

Another argument lies in the fact that if we believe in grand unified
theories, there exists an energy scale $\Lambda_{_{\rm GUT}}$ at which
all the (non-gravitational) couplings unify,
\begin{equation}\label{gut}
\aem(\Lambda_{_{\rm GUT}})=\aw(\Lambda_{_{\rm
GUT}})=\as(\Lambda_{_{\rm GUT}})\equiv \alpha_{_{\rm GUT}}.
\end{equation}
The value of the coupling constants at any energy scale smaller than
$\Lambda_{_{\rm GUT}}$ is obtained from the renormalization group
equations. It follows that a time variation of $\aem$ induces a time
variation of $\alpha_{_{\rm GUT}}$ and thus of $\aw$ and $\as$. In
such a framework, the varying parameters would then be $\alpha_{_{\rm
GUT}}$, $\Lambda_{_{\rm GUT}}/M_4$ and the Yukawa couplings.

The strong coupling at an energy scale $E$ is
related to the QCD scale $\Lambda_{_{\rm QCD}}$ by
\begin{equation}\label{transmu}
\as(E)=-\frac{2\pi}{\beta_0\ln(E/\Lambda_{_{\rm QCD}})}
\end{equation}
with $\beta_0=-11+2n_{\rm f}/3$, $n_{\rm f}$ being the number of quark
flavors. It follows that
\begin{equation}\label{rg0}
\frac{\Delta\Lambda_{_{\rm QCD}}}{\Lambda_{_{\rm
QCD}}}=\ln\left(\frac{E}{\Lambda_{_{\rm QCD}}}\right) \frac{\Delta\as}{\as}.
\end{equation}
The time variation of $\as$ is thus not the same at all
energy scales.  In the chiral limit, in which the quarks are massless,
the proton mass is proportional to the QCD energy scale,
$m_{\rm p}\propto\Lambda_{_{\rm QCD}}$, so that a change in $\as$ (or in
$\alpha_{_{\rm GUT}}$) induces a change in $\mu$  and we have
\begin{equation}
{\Delta m_{\rm p}}/{m_{\rm p}}={\Delta\Lambda_{_{\rm QCD}}}/{\Lambda_{_{\rm
QCD}}}.
\end{equation}
The energy-scale evolution of the three coupling constants in a 1-loop
approximation takes the form
\begin{equation}\label{rg1}
\alpha_i^{-1}(E)=\alpha_{_{\rm GUT}}^{-1}-\frac{b_i}{2\pi}\ln\left(
\frac{E}{\Lambda_{_{\rm GUT}}}\right)
\end{equation}
where the numerical coefficients depend on the choice of the
considered gauge group. For instance $b_i=(41/10,-19/16,-7)$ in
the standard model (SM) and $b_i=(33/5,1,-3)$ in its minimal
supersymmetric extension. In the case of
supersymmetric models (SUSY), Eq.~(\ref{rg1}) has to be replaced
by
\begin{eqnarray}\label{rg2}
\alpha_i^{-1}(E)&=&\left[\alpha_{_{\rm GUT}}^{-1}-\frac{b_i^{^{\rm SUSY}}}{2\pi}\ln\left(
\frac{E}{\Lambda_{_{\rm GUT}}}\right)\right]\Theta(E-\Lambda_{_{\rm
SUSY}})\nonumber\\
&+&\left[\alpha_i^{-1}(\Lambda_{_{\rm SUSY}})-\frac{b_i^{^{\rm SM}}}{2\pi}\ln\left(
\frac{E}{\Lambda_{_{\rm SUSY}}}\right)\right]\Theta(\Lambda_{_{\rm
SUSY}}-E).\nonumber
\end{eqnarray}

Using (\ref{rg1}), one can work out the variation of all
couplings once the grand unified group is chosen, assuming or not
supersymmetry.

In the string-inspired model by Damour and Polyakov (1994a,b),
Eq.~(\ref{fhad}) [obtained from Eq.~(\ref{qcd})] implies that
\begin{equation}\label{mag}
\Delta m_{\rm hadron}/m_{\rm hadron}\simeq 40 \Delta\aem/\aem,
\end{equation}
as was first pointed out by Taylor and Veneziano (1988).

Recently Calmet and Fritzsch (2001,2002), Dent and Fairbairn (2001)
and Langacker {\em et al.} (2002) tried to work out these
relationships in different models and confirmed the order of magnitude
(\ref{mag}).  Calmet and Fritzsch (2001) computed low energy effects
of a time varying fine structure constant within a GUT-like theory
with a constraint of the form (\ref{gut}) and focus their analysis on
the nucleon mass. Nevertheless, they assumed that the mechanisms of
electroweak and supersymmetry breaking as well as fermion mass
generation were left unchanged and thus that quarks, leptons, W and Z
masses do not vary. But, as seen from Eqs (\ref{weak1}-\ref{weak2})
below, one cannot vary $g_{_{\rm W}}$ with $M_{_{\rm W}}$ being fixed
without varying the Higgs vacuum expectation value which induces a
variation of the mass of the fermions. On this basis they concluded
that the result by Webb {\em et al.} (2001) on the fine structure
constant implies that $\Delta m_{\rm p}/m_{\rm
p}\simeq38\Delta\aem/\aem\simeq-4\times10^{-4}$ (keeping the Planck
mass constant) and that $\Delta
y/y\sim-121\Delta\aem/\aem\sim3\times10^{-4}$, which is above the
current observational constraints (see Section~\ref{subsec_5.3}).
Calmet and Fritzsch (2002) considered different scenarios: (i)
$\Lambda_{_{\rm GUT}}$ is constant and $\alpha_{_{\rm GUT}}$
time-dependent, (ii) only $\Lambda_{_{\rm GUT}}$ is time-dependent and
(iii) both are varying. They concluded that the most ``interesting''
situation, in view of the variation of $\aem$ and $\mu$, is the second
case. Langacker {\em et al.}  (2002) pointed out that changes in the
quark masses and in the Higgs vacuum expectation value were also
expected and they parameterized the effects of the variation of
$\alpha_{_{\rm GUT}}$ on the electroweak and Yukawa sector. They
assumed that $\alpha_{_{\rm GUT}}$ was the vacuum expectation value of
a slowly varying scalar field. They concluded that
$\Delta\Lambda_{_{\rm QCD}}/\Lambda_{_{\rm QCD}}\sim34\Delta\aem/\aem$
(with a precision of about 20\% on the numerical factor) and that a
variation of the fine structure of the magnitude of the one observed
by Webb {\em et al.} (2001) would imply $\Delta m_{\rm p}/m_{\rm
p}\simeq-2.5\times10^{-4}$.  They also argued that $\Delta
x/x\sim-32\Delta\aem/\aem\sim 8\times10^{-5}$, which is consistent
with current bounds, if one assumes the variation of the proton
gyromagnetic factor to be negligible. Earlier, Sisterna and Vucetich
(1990) tried to determine the compatibility of all the bounds by
restricting their study to ($\Lambda_{_{\rm QCD}},\aem,\gfermi,G$) and
then included the $u$, $d$ and $s$ quark masses (Sisterna and
Vucetich, 1991).

Since we do not have the theories of electroweak and supersymmetry
breaking as well as the ones for the generation of fermion masses, the
correlations between different low-energy observables remain
model-dependent. But, in this unification picture, one is abale to
derive stronger constraints. For instance Olive {\em et al.} (2002)
expressed the constraints from $\alpha$-, $\beta$-decays and Oklo in
fonction of $|\Delta\Lambda_{_{\rm QCD}}/\Lambda_{_{\rm QCD}}-\Delta
v/v|\sim50\Delta\aem/\aem$ to give respectively the tighter
constraints $<10^{-7}$, $<10^{-9}$ and $<10^{-10}$.  The goal of this
section is to discuss the constraints on some of these constantss which
are of importance while checking for consistency.

\subsection{Weak interaction}\label{subsec_5.1}

Most of the studies on the variation either of $\gfermi$ or $\aw$
concern BBN, Oklo, CMB and geochemichal dating.

The Fermi constant can be expressed in terms of $g_{_{\rm W}}$ and
of the mass of the boson W, $M_{_{\rm W}}$, as
\begin{equation}\label{weak1}
\gfermi=\frac{g^2_{_{\rm W}}}{8M_{_{\rm W}}^2}.
\end{equation}
In the standard model, $M_{_{\rm W}}^2$ is simply the product of $g_{_{\rm
W}}^2/4$ by the Higgs vacuum expectation value
$v^2\equiv\left<\phi\right>^2$, so that
\begin{equation}\label{weak2}
\gfermi={1}/{2v^2}.
\end{equation}
Thus, at tree level, $\gfermi$ is actually independent of the $SU(2)$
coupling and is a direct measurement of the magnitude of the
electroweak symmetry breaking. Note that a change in $v$ is
related to a change in the Yukawa couplings.

Kolb {\em et al.} (1985) considered the effect of the variations of
the fundamental constants on nucleosynthesis. As detailed in
Section~\ref{subsec_4.4}, they found the dependence of the helium
abundance on $G$, $\gfermi$ and $Q$, the variation of which were
related to the variation of $\aem$ (then related to the size of
extra-dimensions).  Kolb {\em et al.} (1985) did not considered
changes in $\gfermi$ due to the variation in $M_{_{\rm W}}$ and assumed
that $\delta\gfermi\propto\delta g_{_{\rm W}}$. Since $G$, $\aem$ and
$\gfermi$ where related to the volume of the extra-space, this study
gives no bound on the variation of $\gfermi$.

Dixit and Sher (1988) pointed out that the relation between $\gfermi$
and $g_{_{\rm W}}$ in the work by Kolb {\em et al.} (1985) was
ill-motivated and that the only way to vary $\gfermi$ was to vary
$v$. Changing $v$ has four effects on BBN: it changes (1) all weak
interaction rates, (2) $m_{\rm e}$, (3) the quark masses and hence $Q$ and
(4) the pion mass which affects the strong nuclear force and the
binding of the deuteron. Using results on the dependence of $m_{\rm e}$ and
$m_{\rm p}$ on $\aem$ and $v$ (Gasser and Leutwyler, 1982) they got
\begin{equation}
\frac{Q}{1\,{\rm MeV}}=1.293-0.9\frac{\Delta\aem}{\aem}+2.193
\frac{\Delta v}{v}.
\end{equation}
Besides, a change of 1\% of the quark masses changes the pion mass by
0.5\% which implies that the deuteron binding energy changes also by
0.5\% (Davies, 1972). They concluded that the helium abundance was given by
\begin{equation}
Y_{\rm p}=0.240-0.31{\Delta v}/{v}+0.38{\Delta\aem}/{\aem}
\end{equation}
and deduced that $\Delta v/v<0.032$ if $\aem$ is fixed and
$\Delta\aem/\aem<0.026$ if $v$ is fixed. They also noted that the
changes in $Q$, $m_{\rm e}$ and $\gfermi$ induced by $v$ tend to cancel
making the change in $\aem$, appearing only in $Q$, larger.

Scherrer and Spergel (1993) followed the same way and focused on two
cases: (1) that the Yukawa couplings are fixed so that both $\gfermi$
and the fermion masses vary in parallel and (2) that the Yukawa
couplings vary so that $\gfermi$ changes while the fermion masses are
kept constant. Considering the abundances of helium they deduced, assuming
$\aem$ fixed, that
\begin{equation}
-0.22<{\Delta\gfermi}/{\gfermi}<0.01
\end{equation}
in the first case and
\begin{equation}
-0.01<{\Delta\gfermi}/{\gfermi}<0.09
\end{equation}
in the second case.

To finish with cosmological constraints, a change in $\gfermi$ induces
a change in $m_{\rm e}$ which can be constrained by the CMB. The
electron mass appears in the expression of the Thomson cross section
(\ref{cmb3}) and on the binding energy of hydrogen (\ref{cmb5}) which
induces a change in the ionization fraction. Kujat and Scherrer (2000)
implemented these changes as in Section~\ref{subsec_4.3} and showed
that the upper limit on $\Delta m_{\rm e}/m_{\rm e}$ is of order
$10^{-2}-10^{-3}$ (keeping the Planck mass constant) for a maximum
multipole of $\ell\sim 500-2500$ if $\aem$ is assumed constant. The
degeneracy with $\aem$ is broken at high multipole so that one can
hope to detect a 1\% variation with a maximum multipole of
$\ell>1500$.

From Oklo data, Shlyakhter (1976) argued that the weak interaction
contribution to the total energy of the nucleus is of order
$10^{-5}(m_\pi/m_{\rm p})^2$ so that $\Delta g_{_{\rm W}}/g_{_{\rm W}}\sim
5\times10^6\Delta g_{_{\rm S}}/g_{_{\rm S}}$ to conclude that
\begin{equation}
\left|\Delta\aw/\aw\right|<4\times10^{-3}.
\end{equation}
But in fact, the change in $\aw$ is much more difficult to model than
the change in $\aem$. Damour and Dyson (1996) used the estimate by
Haugan and Will (1976) for the weak interaction contribution to the
nuclear ground state energy of samarium $E({}^{150}{\rm
Sm})-E({}^{149}{\rm Sm})\simeq5.6$~eV to conclude that, if no subtle
cancellation appears,
\begin{equation}
\left|{\Delta\aw}/{\aw}\right|<0.02.
\end{equation}

Concerning geolochemical data (see Section~\ref{subsec_5.1}),
Dyson (1972) pointed out that all $\beta$-decay rates are
proportional to $\aw^2$ so that all constraints are in fact
dependent on the combination $\aem^s\aw^2$, $s$ being defined in
Eq.~(\ref{s}). The degeneracy can be lifted by comparing different
nuclei, e.g. ${}^{187}_{75}{\rm Re}$ ($s_{\rm Re}=-18000$) and
${}^{40}_{19}{\rm K}$ ($s_{\rm K}=-30$). The constancy of the
decay rates of these two nuclei have approximatively the same
accuracy. From the constancy of the ratio
$$
\Delta\frac{\lambda_{\rm Re}}{\lambda_{\rm K}}=\left(s_{\rm Re}-s_{\rm K}
\right)\frac{\lambda_{\rm Re}}{\lambda_{\rm K}}\frac{\Delta\aem}{\aem}
$$
within a few parts in $10^{10}$ per year, one can deduce that,
independently of any assumption on $\aw$,
\begin{equation}
\left|\Delta\aem/\aem\right|<2\times10^{-5}
\end{equation}
and thus that
\begin{equation}
\left|\Delta\aw/\aw\right|<10^{-1}
\end{equation}
during the last $10^9$ years.

Wilkinson (1958) studied the variation of $\aw$ by using
pleochroic haloes, that is spheres formed by $\alpha$-ray tracks
around specks of uranium-bearing mineral in mica. The intensities
of the haloes of different radii give a picture of the natural
radioactive series integrated over geological time from which one
can deduce the proportion of different daughter-activities in the
decay chain from uranium to lead. This series contain elements
undergoing both $\alpha$- and $\beta$-decay. For instance Ac
branches 1.2\% by $\alpha$-decay and the rest by $\beta$-decay.
From $10^9$ years old samples, Wilkinson (1958) deduced that
\begin{equation}
\left|\Delta\aw/\aw\right|<10.
\end{equation}

Let us also point out some works (Agrawal {\em et al.}, 1998a,b)
in which the mass scale of the standard model and the scale of
electroweak symmetry breaking are constrained by mean of the anthropic
principle. Passarino (2001) investigated the effects of the time
variation of the Higgs vacuum expectation value and showed that the
classical equation of motion for the Higgs field in the standard model
accepts time dependent solution.

\subsection{Strong interaction}\label{subsec_5.15}

There is a very small number of works addressing this issue. Due to
the strong energy dependence of $\as$, it makes more sense to
constraint the variation of $\Lambda_{_{\rm QCD}}$. It has a lot of
implications on the stability properties of nuclei and it follows that
most of the constraints arise from nuclear considerations. Let us
remind that in the chiral limit, all dimensional parameters are
proportional to $\Lambda_{_{\rm QCD}}$ so that all dimensionless
ratios will be, in this limit, pure numbers and thus insensitive to a
change of the strong interaction. But, quark masses will play an
important role in the variation of dimensionless ratios and have to be
taken into account.

A change in the strong interaction affects the {light elements} and
(1) the most weakly bound nucleus, namely the deuteron, can be unbind
if it is weaker(2) there may exist stable dineutron and diproton if it
is stronger [and hydrogen would have been burned catastrophically at
the beginning of the universe, (Dyson, 1971)], (3) the rate of the
proton capture ($p+n\rightarrow D+\gamma$) is altered and (4) the
neutron lifetime changes. All these effects influence the
nucleosynthesis (Barrow, 1987). It will also be a catastrophe if the
deuteron was not stable (by affecting the the hydrogen burning
properties in stars).

Most of the early studies considered these stability properties by
modelling the nuclear force by a Yukawa approximation of the form
$V(r)\sim g_{_{\rm S}}^2\exp(-m_\pi r)$.  In the following of this
section, the cited bounds refer to suh a definition of $\as$. Davies
(1972) studied the stability of two-nucleon systems in terms of $\aem$
and $\as$ assuming that $\aw$ remains fixed and concluded that the
diproton is not bounded if
${\Delta\as}/{\as}-{\Delta\aem}/{\aem}<0.034$.  Rozental (1980)
assumed that the depth of the potential well in the deuteron is
proportional to $\as$, to state that a decrease of $\as$ of 10-15\%
would make it unstable. An increase of $\as$ would render the diproton
stable so that $\left|{\Delta\as}/{\as}\right|<10^{-1}$ at
nucleosynthesis. A previous and more detailed analysis by Davies
(1972) yield $\left|{\Delta\as}/{\as}\right|<4\times10^{-2}$ and
Pochet {\em et al.} (1991) concluded that
$\left|{\Delta\as}/{\as}\right|<4\times10^{-2}$ for the deuteron to be
stable and $\left|{\Delta\as}/{\as}\right|<6\times10^{-1}$ for the
diproton to be unstable.

Concerning high-$Z$ nuclei, Broulik an Trefil (1971) used the liquid
drop model of the nucleus and the observed half lives and abundances
of transuramium elements to constraint the variation of $\as/\aem$. In
this model, the stability of a nucleus can be discussed by comparing
the Coulomb repulsion between protons to the strong interaction
attraction modeled by a surface tension $T$ proportional to $\as$.
With increasing atomic weight, the individual nucleons become
progressively more weakly bound as the Coulomb force dominates.  A
nucleus is stable against spontaneous fission if
\begin{equation}\label{trefil}
  \frac{Z^2}{A}<\frac{40\pi}{3}\frac{r_0^2}{e^2}T.
\end{equation}
If $\as/\aem$ was larger in the past some unstable nuclei would have
been stable. The idea is thus to find unstable nuclei with long
half-life which do not occur naturally. The variation of $\as/\aem$
would make them stable in the past but this must have occured roughly
more than about ten times their lifetime since otherwise they will be
in detectable abundances. Assuming that $\aem$ is fixed, they
concluded from data on ${}^{244}_{94}{\rm Pu}$ that
$\left|{\Delta\as}/{\as}\right|<1.7\times10^{-3}$ on a time scale of
about $7.6\times10^8$~yr. Unfortunately, four month later it was
reported that ${}^{244}_{94}{\rm Pu}$ occurs naturally on Earth
(Hoffmann {\em et al.}, 1971) hence making the argument invalid.
Davies (1972) argued that the binding energy is expected to vary as
$\as^2$ (contrary to the ansatz by Broulik and Trefil, 1971) so that
the previous bound becomes
$\left|{\Delta\as}/{\as}\right|<8.5\times10^{-4}$.

Barrow (1987) studied the effect of the change of $\as$ on the BBN
predictions in Kaluza-Klein and superstring theories in which all the
couplings depends on the compactification radius. Assuming that the
deuteron binding energy, probably the most sensitive parameter of BBN,
scales as $\as$, he concluded that
\begin{equation}
t_{\rm N}/\tau_{\rm n}\propto G^{-1/2}\as^2\gfermi^2
\end{equation}
which affects the helium abundances from Eqs.~(\ref{n1}-\ref{n2}).
Flambaum and Shuryak (2001) discussed the effects of the variation of
$\as$ and took the quarks masses into account. They expressed their
results in terms of the parameter $\delta_\pi\equiv\delta\ln(
m_\pi/\Lambda_{_{\rm QCD}})=\delta\ln(\sqrt{(m_{\rm u}+m_{\rm
d})/\Lambda_{_{\rm QCD}}})$ where the pion mass $m_\pi$ determines the
range of the nuclear force.  They concluded that $|\delta_\pi|<0.005$
between BBN and today.

As detailed in Section~\ref{oklo}, Shlyakhter (1976) argued that
the change in the energy of the resonance is related to a change
in $g_{_{\rm S}}$ by
\begin{equation}
{\Delta g_{_{\rm S}}}/{g_{_{\rm S}}}\sim{\Delta E_r}/{V_0}
\end{equation}
and deduced that $|\Delta g_{_{\rm S}/}g_{_{\rm S}}|<1.9\times10^{-9}$.
Clearly, this analysis is not very reliable. Flambaum and Shuryak
(2001) estimated the variation of the resonance energy due to a
variation of the pion mass and concluded that $\Delta E_r/E_r\sim
3\times10^8|\delta_\pi|$ so that $|\delta_\pi|<7\times10^{-10}$.

Flambaum and Shuryak (2001) also argued that in the worst case all
strong interaction phenomena depend on $\Lambda_{_{\rm QCD}}+Km_{_{\rm
S}}$ where $K$ is some universal constant and $m_{_{\rm S}}$ the
strange quark mass but a real study of the effect of $m_{_{\rm S}}$ on
all hadronic masses remains to be done. It also follows that the
proton gyromagnetic factor can be time dependent and constrained by
observations such as those presented in Section~\ref{subsec_5.3}.

\subsection{Electron to proton mass ratio}\label{subsec_5.2}

An early limit on the variation of \footnote{In the literature $\mu$
refers either to $m_{\rm e}/m_{\rm p}$ or to its inverse. In the
present work we choose the first definition and we harmonize the
results of the different articles.} $\mu$ was derived by Yahil (1975)
who compared the concordance of K-Ar and Rb-Sr geochemical ages and
deduced that $|\Delta\mu/\mu|<1.2$ over the past $10^{10}\,{\rm yr}$.

As first pointed out by Thomson (1975) molecular absorption lines
can provide a test of the variation of $\mu$. The energy
difference between two adjacent rotational levels in a diatomic
molecule is proportional to $M r^{-2}$, $r$ being the bond length
and $M$ the reduced mass, and  that the vibrational transition of
the same molecule has, in first approximation, a $\sqrt{M}$
dependence. For molecular hydrogen $M=m_{\rm p}/2$ so that comparison of
an observed vibro-rotational spectrum with its present analog will
thus give information on the variation of $m_{\rm p}$ and $m_{\rm n}$.
Comparing pure rotational transitions with electronic transitions
gives a measurement of $\mu$.

Following Thompson (1975), the fre\-qu\-ency of
vi\-bra\-tion-ro\-ta\-tion transitions is, in the Born-Oppenheimer
approximation, of the form
\begin{equation}\label{mu1}
\nu\simeq E_I\left(c_{_{\rm elec}} +c_{_{\rm vib}}/\sqrt{\mu}
+c_{_{\rm rot}}/\mu\right)
\end{equation}
where $c_{_{\rm elec}}$, $c_{_{\rm vib}}$ and $c_{_{\rm rot}}$ are some
numerical coefficients. Comparing the ratio of wavelengths of various
electronic-vibration-rotational lines in quasar spectrum and in the
laboratory allow to trace the variation of $\mu$ since, at lowest
order, Eq.~(\ref{mu1}) implies
\begin{equation}
\frac{\Delta E_{ij}(z)}{\Delta E_{ij}(0)}=1+K_{ij}\frac{\Delta\mu}{\mu}
+{\cal O}\left(\frac{\Delta\mu^2}{\mu^2}\right).
\end{equation}
Varshalovich and Levshakov (1993) used the observations of a
damped Lyman-$\alpha$ system associated with the quasar PKS
0528-250 (which is believed to have molecular hydrogen in its
spectrum) of redshift $z=2.811$ and deduced that
\begin{equation}
|\Delta\mu/\mu|<4\times 10^{-3}.
\end{equation}
A similar analysis was first
tried by Foltz {\em et al.} (1988) but their analysis did not take
into account the wavelength-to-mass sensitivity and their result
hence seems not very reliable. Nevertheless, they concluded that
\begin{equation}
|\Delta\mu/\mu|<2\times10^{-4}
\end{equation}
at $z=2.811$. Cowie and Songaila (1995) observed the same quasar and
deduced that
\begin{equation}
\Delta\mu/\mu=(0.75\pm6.25)\times10^{-4}
\end{equation}
at 95\% C.L.  from the data on 19 absorption
lines. Varshalovich and Potekhin (1995) calculated the coefficient
$K_{ij}$ with a better precision and deduced that
\begin{equation}
|\Delta\mu/\mu|<2\times10^{-4}
\end{equation}
at 95\% C.L.. Lanzetta {\em et al.} (1995) and Varshalovich {\em et
al.} (1996b) used 59 transitions for H$_2$ rotational levels in PKS
0528-250 and got
\begin{equation}
\Delta\mu/\mu=(-8.3_{-6.6}^{+5.5})\times10^{-5}
\end{equation}
at 1.6$\sigma$ level and
\begin{equation}
\Delta\mu/\mu=(-1\pm1.2)\times10^{-4}
\end{equation}
at 2$\sigma$ level. These results were confirmed by Potekhin {\em et
al.}  (1998) using 83 absorption lines to get
\begin{equation}
\Delta\mu/\mu=(-7.5\pm9.5)\times10^{-5}
\end{equation}
at a $2\sigma$ level.

More recently, Ivanchik {\em et al.} (2001) measured, with the
VLT, the vibro-rotational lines of molecular hydrogen for two
quasars with damped Lyman-$\alpha$ systems respectively at
$z=2.3377$ and $3.0249$ and also argued for the detection of a
time variation of $\mu$. Their most conservative result is (the
observational data were compared to two experimental data sets)
\begin{equation}
\Delta\mu/\mu=(-5.7\pm3.8)\times10^{-5}
\end{equation}
at 1.5$\sigma$ and the authors cautiously point out that
additional measurements are necessary to ascertain this
conclusion. 1.5$\sigma$ is not really significant and this may not
survive further extended analysis. The result is also dependent on
the laboratory dataset used for the comparison since it gave
$\Delta\mu/\mu=(-12.2\pm7.3)\times10^{-5}$ with another dataset.

As in the case of Webb {\em et al.} (1999, 2001), this measurement is
very important in the sense that it is a non-zero detection that will
have to be compared with other bounds.  The measurements by Ivanchik
{\em et al.} (2001) is indeed much larger than one would expect from
the electromagnetic contributions. As seen above, the change in any
unified theory, the changes in the masses are expected to be larger
than the change in $\aem$. Typically, we expect
$\Delta\mu/\mu\sim\Delta\Lambda_{_{\rm QCD}}/\Lambda_{_{\rm QCD}}
-\Delta v/v\sim(30-40)\Delta\aem/\aem$, so that it seems that the
detection by Webb {\em et al.} (2001) is too large by a factor of
order 10 to be compatible with it.

Wiklind and Combes (1997) observed the quasar PKS~1413+135 with
redshift $z=0.247$ and used different transitions from the same
molecule to constrain the variation of $\mu$. They compared
different lines of HCO$^+$, HCN, CO and showed that the redshift
difference are likely to be dominated by the velocity difference
between the two species which limits the precision of the
measurements to $\Delta\mu/\mu\sim10^{-5}$ at $3\sigma$ level. In
one source (B3~1504+377) they observed a discrepancy of
$\Delta\mu/\mu\sim10^{-4}$

Pagel (1977, 1983) used another method to constrain $\mu$ based on
the measurement of the mass shift in the spectral lines of heavy
elements. In that case the mass of the nucleus can be considered
as infinite contrary to the case of hydrogen. A variation of $\mu$
will thus influence the redshift determined from hydrogen [see
Eq.~(\ref{51})]. He compared the redshifts obtained from spectrum
of hydrogen atom and metal lines for quasars of redshift ranging
from 2.1 to 2.7. Since
\begin{equation}
\Delta z\equiv z_{_{\rm H}}-z_{_{\rm
metal}}=(1+z)\frac{\Delta\mu}{1-\mu_0},
\end{equation}
he obtained that
\begin{equation}
|\Delta\mu/\mu|<4\times10^{-1}
\end{equation}
at $3\sigma$ level. This result is unfortunately not conclusive
because usually heavy elements and hydrogen belong to different
interstellar clouds with different radial velocity.

\subsection{Proton gyromagnetic factor}\label{subsec_5.3}

As seen in Section~\ref{subsec_4.2}, the hyperfine structure induces a
splitting dependent on $g_{\rm p}\mu\aem^2$.  The ratio between the
frequency $\nu_{21}$ of the hyperfine 21~cm absorption transition an
optical resonance transition of frequency $\nu_{_{\rm opt}}$ mainly
depends on
\begin{equation}\label{deltax}
{\nu_{21}}/{\nu_{_{\rm opt}}}\propto \aem^2 g_{\rm p}\mu\equiv x.
\end{equation}
By comparing the redshift of the same object determined from
optical data and the 21~cm transition, one deduces that
\begin{equation}
\Delta z=z_{_{\rm opt}}-z_{21}=(1+z){\Delta x}/{x}
\end{equation}

Savedoff (1956) used the spectrum of Cygnus A
and deduced that
\begin{equation}
\Delta x/x=(3\pm7)\times10^{-4}
\end{equation}
at $z\sim0.057$.
Wolfe {\em et al.} (1976) discovered a BL Lac object (AO
0235+164) having the same redshift determined either by the 21~cm
absorption line or by the ultraviolet doublet of ${\rm Mg}^+$. Using
that
\begin{equation}
{\nu_{_{\rm H}}}/{\nu_{_{\rm Mg}}}\propto g_{\rm p}\mu\aem^2
\left(1-3\mu+\ldots\right)
\end{equation}
they concluded that
\begin{equation}
\Delta x/x=(5\pm10)\times10^{-5}
\end{equation}
at redshift of $z=0.5$. They also got a constraint on the variation of
$g_{\rm p}\mu$ by comparing the separation of Mg~II doublet to
hydrogen to get $|\Delta g_{\rm p}\mu /g_{\rm
p}\mu|<6\times10^{-2}$. Wolfe and Davis (1979) used the 21~cm
absorption lines of neutral hydrogen in front of the quasar
QSO~1331+170 at a redshift $z\sim2.081$. They determined that the
cloud was at redshift $z\sim1.755$. The agreement between the 21~cm
and optical redshifts is limited by the error in the determination of
the optical redshift. They concluded that
\begin{equation}
|\Delta x/x|\leq2\times10^{-4}
\end{equation}
at a redshift $z\sim1.755$ and from another
absorber at redshift $z\sim0.524$ around the quasar AO~0235+164
gives
\begin{equation}
|\Delta x/x|\leq 2.8\times10^{-4}.
\end{equation}

Tubbs and Wolfe (1980) used a set of four quasars among which
MC3~1331+17 for which $z_{21}=1.77642\pm2\times10^{-5}$ is known with
very high precision and deduced that
\begin{equation}
|\Delta x/x|<2\times10^{-4}.
\end{equation}
Cowie and Songaila (1995) used the observations of ${\rm C}^0$
absorption and fine structure to get the better optical redshift
$z_{_{\rm opt}}=1.77644\pm2\times10^{-5}$ which enables them to
improve the constraint to
\begin{equation}
\Delta x/x=(7\pm11)\times10^{-6}.
\end{equation}
Besides the uncertainty in the determination of the optical
redshift, since the 21~cm optical depth depends sensitively on
spin temperature while resonance-line optical depths do not, the
two regions of absorption need not coincide. This induces an
uncertainty $\Delta z=\pm(1+z)(\Delta v_{_{\rm opt}}/c)$ into
Eq.~(\ref{deltax}) [see e.g. Wolfe and Davis (1979) for a
discussion].

Drinkwater {\em et al.} (1998)
compared the hydrogen hyperfine structure to molecular absorption for
three systems at redshift $z=0.24,0.67$ and 0.68 and used CO
absorption lines. This allows to constrain $y\equiv g_{\rm p}\aem^2$ and
they got
\begin{equation}
|\Delta y/y|<5\times10^{-6}.
\end{equation}
Assuming that the change in $g_{\rm p}$ and $\aem$ are not correlated
they deduced that $|\Delta g_{\rm p}/g_{\rm p}|<5\times10^{-6}$ and
$|\Delta\aem/\aem|<2.5\times10^{-6}$.  Varshalovich and Potekhin
(1996) used the CO and hyperfine hydrogen redshift toward PKS~1413+135
($z=0.247$) to get
\begin{equation}
\Delta y/y=(-4\pm6)\times10^{-5}
\end{equation}
and PKS~1157+0.14 ($z=1.944$)
\begin{equation}
\Delta y/y=(7\pm10)\times10^{-5}
\end{equation}
Murphy {\em et al.} (2001c) improved
the precision of this measurement by fitting Voigt profiles to the H 21
cm profile instead of using published redshifts and got
\begin{equation}
\Delta y/y=(-0.2\pm0.44)\times10^{-5}
\end{equation}
at $z=0.25$ and
\begin{equation}
\Delta y/y=(-0.16\pm0.54)\times10^{-5}
\end{equation}
at $z=0.68$. With the same systems Carrilli {\em et al.} (2001) found
\begin{equation}
|\Delta y/y|<1.7\times10^{-5}
\end{equation}
both at $z=0.25$ and $z=0.68$. Murphy {\em et al.} (2001c) argued that
one can estimate the velocity to $1.2\,{\rm km\cdot s}^{-1}$ instead
of the $10\,{\rm km\cdot s}^{-1}$ assumed by Carrilli {\em et al.}
(2001) so that their results in fact lead to $\Delta
y/y=(1\pm0.03)\times10^{-5}$ at $z=0.25$ and $\Delta y
/y=(1.29\pm0.08)\times10^{-5}$ at $z=0.68$.

\subsection{The particular case of the cosmological constant}
\label{sec_5.4}

The cosmological constant has also been loosing its status of
constant. In this section, we briefly review the observations
backing up this fact and then describe the theoretical models in
favor of a time dependent cosmological constant and some links
with the variation of other fundamental constants.

The combination of recent astrophysical and cosmological observations
[among which the luminosity distance-redshift relation up to $z \sim
1$ from type Ia supernovae (Riess {\em et al.}, 1998; Perlmutter {\em
et al.}, 1998), the cosmic microwave background temperature
anisotropies (de Bernardis {\em et al.}, 2000) and gravitational
lensing (Mellier, 1999)] seems to indicate that the universe is
accelerating and that about 70\% of the energy density of the
universe is made of a matter with a negative pressure (i.e. having an
equation of state $w \equiv P / \rho < 0$).

There are many different candidates to account for this exotic type of
matter. The most simple solution would be a cosmological
constant (for which $w = - 1$) but one will then have to face the
well known {\it cosmological constant problem} (Weinberg, 1989), i.e.
the fact that the value of this cosmological constant inferred from
the cosmological observations is extremely small --- about 120 order of
magnitude --- compared with the energy scales of high energy physics
(Planck, GUT and even electroweak scales). Another way is to
argue that there exists a (yet unknown) mechanism which makes the
cosmological constant strictly vanish and to find another matter
candidate (referred to as ``dark energy'') able to explain the
cosmological observations.

Among all the proposals (see e.g.  Bin\'etruy 2000 and Carroll, 2000
for a review) quintessence seems to be a promising mechanism. In these
models, a scalar field is rolling down a runaway potential decreasing
to zero at infinity hence acting as a fluid with an effective equation
of state in the range $- 1 \leq w \leq 1$ if the field is
minimally coupled. Runaway potentials such as exponential potential
and inverse power law potentials
\begin{equation}
\label{01}
V (\phi) = {M^{4 + \alpha}}/{\phi^\alpha} ,
\end{equation}
with $\alpha > 0$ and $M$ a mass scale, arise in models where
supersymmetry is dynamically broken (Bin\'etruy, 1999) and in which
flat directions are lifted by non-perturbative effects.

One of the underlying motivation to replace the cosmological
constant by a scalar field comes from superstring models in which
any dimensionful parameter is expressed in terms of the string
mass scale and the vacuum expectation value of a scalar field.
However, the requirement of slow roll (mandatory to have a
negative pressure) and the fact that the quintessence field
dominates today imply, if the minimum of the potential is zero,
that (i) it is very light, roughly of order $\sim 10^{- 33}\,{\rm
eV}$ (Carroll, 1998) and that (ii) the vacuum expectation value of
the quintessence field today is of order of the Planck mass. It
follows that coupling of this quintessence field leads to
observable long-range forces and time dependence of the constant
of nature.

Carroll (1998) considered the effect of the coupling of this very
light quintessence field to ordinary matter via an interaction of the
form $\beta_i(\phi/M){\cal L}_i$ and to the electromagnetic field as
$\phi F^{\mu\nu}\widetilde F_{\mu\nu}$. Chiba and Kohri (2001) also
argued that an ultra-light quintessence field induces a time variation
of the coupling constant if it is coupled to ordinary matter and
studied a coupling of the form $\phi F^{\mu\nu}F_{\mu\nu}$. Dvali and
Zaldarriaga (2002) showed that it will be either detectable as a
quintessence field or by tests of the equivalence principle, as also
concluded by Wetterich (2002).

It was proposed that the quintessence field is also the dilaton (Uzan,
1999; Banerjee and Pavon, 2001; Esposito-Far\`ese and Polarski, 2001;
Riazuelo and Uzan, 2000; Gasperini {\em et al.}, 2002). The same
scalar field drives the time variation of the cosmological constant
and of the gravitational constant and it has the property to also have
tracking solutions (Uzan, 1999).

Another motivation for considering the link between a dynamical
cosmological constant and the time variation of fundamental constants
comes from the origin of the inverse power law potential. As shown
by Bin\'etruy (1999), it can arise from supersymmetry breaking by non
perturbative effects such as gaugino condensation. The same kind of
potential was also considered by Vayonakis (1988) while discussing the
variation of the fundamental couplings in the framework of
10-dimensional supergravity.

The variation of fundamental constants has also other implications on
the measurement of the cosmological constant. Riazuelo and Uzan (2002)
considered the effect of the variation of the gravitational constant
on supernovae data. Besides changing the luminosity distance-redshift
relation, the variation of $G$ changes the standard picture, according
to which type Ia supernovae are standard candles, in two ways. First
the thermonuclear energy release proportional to the synthetized
nickel mass is changing (and hence the maximum of the light curve);
second the time scale of the supernovae explosion and thus the width
of the light curve is also changed. Riazuelo and Uzan (2002) derived
the modified magnitude-redshift relation to include the effect of the
variation of $G$, using a one-zone analytical model for the supernovae
and was confirmed by numerical simulations (Gazta\~naga {\em et al.},
2002).

Barrow and Magueijo (2001) considered the effect of a time dependent
fine structure constant on the interpretation of the supernovae
data. Their study was restricted to a class of varying speed of
light theories (see Section~\ref{subsec_7.15}) which have
cosmological solutions very similar to quintessence. But, only the
effect on the Hubble diagram was studied and the influence of the
change of the fine structure constant on the thermonuclear burst
of the supernovae, and hence on its light curve, was not
considered at all.

Up to now there is no observational evidence of a time variation of
the cosmological constant. The measurement of the equation of state of
the dark energy can be hoped to be possible very soon, the best
candidate method being the use of large-scale structure growth and
weak gravitational lensing (Benabed and Bernardeau, 2001). But, it
seems that the variation of constants and the dark energy are somehow
related (Dvali and Zaldarriaga, 2002; Chiba and Khori, 2001; Banks
{\em et al.}, 2002; Wetterich, 2002; Fujii, 2002), at least they share
the properties to be very light and to appear in many models with a
runaway potential.

\subsection{Attempts to constrain the variation of
dimensionful constants}
\label{sec_5.5}

As emphasized in Section~\ref{sec_1}, considering the variation of
dimensionful constants is doubtful and seems meaningless but such
attempts have nevertheless been performed. We briefly review and
comment them.  These investigations were mainly motivated by the
construction of cosmological models alternative to the big bang
scenario and in which the redshift needs to have another
interpretation.

Bahcall and Salpeter (1965) proposed to look for a time variation
of the Planck constant by comparing the light emitted by two
quasars.  Their idea is based on the remark that a prism is
sensitive to the energy $E$ of the photon and a diffraction
grating to its wavelength $\lambda$ so that any difference in the
comparison of the wavelengths of a particular spectral line could
be attributed to a change in $\hbar$. Their study led to a null
result in terms of experimental errors.

Noerdlinger (1973) [and later Blake (1977a)] tried to measure
$E\lambda$. His argument was that the intensity of the
Rayleigh-Jeans tail of the Planck spectrum of the CMB photon
determines $k_{_{\rm B}}T$ whereas the turnover point of the
spectrum determines $h\nu/k_{_{\rm B}}T$. It follows that one can
determine the value of $hc$ at the time of recombination, leading
to the constraint $|\Delta\ln hc|<0.3$.

Further works were performed by Solheim {\em et al.} (1973) and Baum
and Florentin-Nielsen (1976) who compared the light of nearby and
distant galaxies in order to test the constancy of
$E\lambda$. Bekenstein (1979) demonstrated that these experiments were
meaningless since the constancy of $E\lambda$ was interpreted as the
constancy of $\hbar c$ but that this latter fact was implicitely
assumed in the two experiments since the wave vector and momentum of
the photon were both parallely propagated. This is only possible if
their proportionality factor $\hbar c$ is constant, hence ensuring the
null result of the two experiments.

\section{Theoretical motivations}\label{sec_7}

One general feature of extra-dimensional theories, such as
Kaluza-Klein and string theories, is that the ``true'' constants
of nature are defined in the full higher dimensional theory so
that the effective 4-dimensional constants depends, among other
things, on the structure and sizes of the extra-dimensions. Any
evolution of these sizes either in time or space, would lead to a
spacetime dependence of the effective 4-dimensional constants.

We present in Sections~\ref{subsec_7.0} and \ref{subsec_7.1} some
results concerning Kaluza-Klein theories and string theories.  We
end in Section~\ref{subsec_7.15} by describing some phenomenological
approaches initiated by Bekenstein (1982).

\subsection{Kaluza-Klein theories}\label{subsec_7.0}

The aim of the early model by Kaluza (1921) and Klein (1926) to
consider a 5-dimensional spacetime with one spatial
extra-dimension $S^1$ (assumed to be of radius $R_{_{\rm KK}}$)
was to unify electromagnetism and gravity (for a review see e.g.
Overduin and Wesson, 1997).  Starting from the 5-dimensional
Einstein-Hilbert Lagrangian
\begin{equation}\label{kkaction}
S_5=\frac{1}{2}\int\dd^5{\bf x}\sqrt{-g_5}M_5^3R_5,
\end{equation}
we decompose the 5-dimensional metric as
\begin{equation}\label{kkvac}
\dd s^2_5=g_{\mu\nu}\dd x^\mu\dd x^\nu+ \hbox{e}^{2\sigma}\left(
A_\mu\dd x^\mu+\dd y\right)^2.
\end{equation}
This form still allows 4-dimensional reparametrizations of the
form $y'=y+\lambda(x^\mu)$ provided that
$A'_\mu=A_\mu-\partial_\mu\lambda$ so that gauge transformations
arise from the higher dimensional coordinate transformations
group. Any field $\phi$ can be decomposed as
\begin{equation}\label{kkmodes}
\phi(x^\mu,y)=\sum_{n\in Z}\phi^{(n)}(x^\mu)\hbox{e}^{iny/R_{_{\rm KK}}}.
\end{equation}
The 5-dimensional Klein-Gordon equation for a massless field becomes
\begin{equation}
\nabla_\mu\nabla^\mu\phi^{(n)}=(n/R_{_{\rm KK}})^2\phi^{(n)}
\end{equation}
so that $\phi^{(n)}$ has a mass $m_n=n/R_{_{\rm KK}}$. At energies
small with respect to $m_{_{\rm KK}}=R_{_{\rm KK}}^{-1}$, only
$y$-independent fields remain and the physics is 4-dimensional. The
effective action for the massless fields is obtained from the relation
$R_5=R_4-2\hbox{e}^{-\sigma}\Delta
\hbox{e}^{\sigma}-\hbox{e}^{2\sigma}F^2/4$ with
$F_{\mu\nu}=\partial_\mu A_\nu-\partial_\nu A_\mu$ so that
\begin{eqnarray}\label{l4}
S_4&=&\pi\int\dd^4\bx\sqrt{-g}\hbox{e}^{2\sigma} R_{_{\rm KK}}M_5^3
\left[R_4-\partial_\mu\sigma\partial^\mu\sigma\right.\nonumber\\
&&\left.-\frac{1}{4}
\hbox{e}^{2\sigma}F_{\mu\nu}F^{\mu\nu}\right].
\end{eqnarray}
The field equations do not determine the compactification radius and
only the invariant radius $\rho=R_{_{\rm KK}}\exp(\sigma)$
distinguishes non-equivalent solutions (one can set $R_{_{\rm
KK}}$ to unity without loss of generality).

Setting $A_\mu=R_{_{\rm KK}}\widetilde A_\mu$, the covariant derivative is
$\partial_\mu+ip_yA_\mu=\partial_\mu+in\widetilde A_\mu$ so that the
charges are integers. The 4-dimensional Yang-Mills coupling,
identified as the coefficient $-1/4g^2_{_{\rm YM}}$ of $\widetilde
F^2$, and gravitational constant are given by
\begin{equation}
M_4^2=2\pi\rho M_5^3,\quad
4g^{-2}_{_{\rm YM}}=M_4^2\rho^2/2,
\end{equation}
$2\pi\rho$ being the volume of the extra-space. Note that as long as
one considers vacuum as in Eq.~(\ref{kkvac}), there is a conformal
undeterminacy that has to be lifted when adding matter fields. This
generalizes to the case of $D$ extra-dimensions (see e.g. Cremmer and
Scherk, 1977 and Forg\'acs and Horv\'ath, 1979 for the case of two
extra-dimensions) to
\begin{equation}\label{kkDdim}
G\propto \rho^{-D},\quad
\alpha_i(m_{_{\rm KK}})=K_i(D)G\rho^2
\end{equation}
where the constants $K_i$ depends only on the dimension and
topology of the compact space (Weinberg, 1983b) so that the only
fundamental constant of the theory is $M_{4+D}$. A theory on
${\cal M}_4\times {\cal M}_D$ where ${\cal M}_D$ is a
$D$-dimensional compact space generates a low-energy quantum field
theory of the Yang-Mills type related to the isometries of ${\cal
M}_D$ [for instance Witten (1981) showed that for $D=7$, it can
accommodate the Yang-Mills group $SU(3)\times SU(2)\times U(1)$].
Indeed the two main problems of these theories is that one cannot
construct chiral fermions in four dimensions by compactification
on a smooth manifold with such a procedure and that gauge theories
in five dimensions or more are not renormalisable.

The expression for the structure constants at lower
energy are obtained by the renormalisation group (Marciano, 1987; Wu and
Wang, 1986)
\begin{eqnarray}\label{kk2}
\alpha_i^{-1}(mc^2)&=&\alpha_i^{-1}(m_{_{\rm KK}}c^2)
-\frac{1}{\pi}\sum_j C_{ij}\left[\ln\frac{m_{_{\rm KK}}}{m_j}\right.
\nonumber\\
&&\qquad\qquad\qquad\left.
-\theta(m-m_j)\ln\frac{m_j}{m}\right]
\end{eqnarray}
where the sum is over all leptons, quarks, gluons... and the
$C_{ij}$ are constants that depend on the spin and group
representation (Georgi {\em et al.}, 1974). Note however that this
relation is obtained by considering the renormalization group in
four dimensions and does not take into account the contribution of
the Kaluza-Klein modes in loops.

Chodos and Detweiler (1980) illustrated the effect of the fifth
dimension by considering a 5-dimensional vacuum solution of the
Kasner form
\begin{equation}
\dd s^2=-\dd t^2+\sum_{i=1..4}\left(\frac{t}{t_0}\right)^{2p_i}
\left(\dd x^i\right)^2
\end{equation}
with $\sum p_i=\sum p_i^2=1$ and assuming compact spatial sections
$0\leq x_i<L$. In order to ensure local isotropy and homogeneity, they
choose the solution $p_1=p_2=p_3=1/2$ and $p_4=-1/2$ so that the
universe has four macroscopic spatial dimensions at the time $t_0$ and
looks spatially 3-dimensional at a time $t\gg t_0$ with a small
compact dimension of radius $(T_0/t)^{1/2}L$. Considering $A_\mu$ as a small
metric perturbation, they deduced that
\begin{equation}
{\aem}/{\ag}={t}/{t_0}
\end{equation}
hence offering a realization of Dirac large number hypothesis.  Freund
(1982) studied $(4+D)$ Kaluza-Klein cosmologies starting both in a
$(4+D)$-dimensional Einstein gravity or a $(4+D)$-dimensional
Brans-Dicke gravity.

Using the expressions (\ref{kkDdim}-\ref{kk2}), Marciano (1984) related the
time dependence of the different couplings and restricted his discussion
to the cases where $\dot K_i$ and $\dot m_j$ vanish. In the case where
$\dot\alpha_i(m_{_{\rm KK}})=0$ (as studied in Chodos and Detweiler, 1980) one
can relate the time variation of the gravitational and fine structure
constant as
\begin{equation}
\frac{\dot\aem}{\aem}=-\frac{\aem}{2\pi}\sum_j\left(\frac{5}{3}C_{1j}
+C_{2j}\right)\frac{\dot G}{G}.
\end{equation}
In the case where $\dot\alpha_i(m_{_{\rm KK}})\not=0$ (as studied in
Freund, 1982), it was shown that the time variation of $\as$ is
enhanced at low energy so that constraints on the time variation of
$m_{\rm e}/m_{\rm p}$ provide a sensitive test. It is also claimed
that in the case of an oscillating $m_{_{\rm KK}}$ the amplitude of
the oscillations will be damped by radiation in our 3-dimensional
spacetime due to oscillating charges and that experimental bounds can
be circumvented.

Kolb {\em et al.} (1985) used the variation (\ref{kkDdim}) to
constrain the time variation of the radius of the extra-dimensions
during primordial nucleosynthesis (see section~\ref{subsec_4.4}) but
their ansatz concerning the variation of $\gfermi$ was
ill-motivated. They deduced $|\Delta R_{_{\rm KK}}/R_{_{\rm
KK}}|<1\%$. Barrow (1987) took the effects of the variation of
$\as\propto R_{_{\rm KK}}^{-2}$ (see Section~\ref{subsec_5.15}) and
deduced from the helium abundances that for $|\Delta R_{_{\rm
KK}}/R_{_{\rm KK}}|<0.7\%$ and $|\Delta R_{_{\rm KK}}/R_{_{\rm
KK}}|<1.1\%$ respectively for $D=2$ and $D=7$ Kaluza-Klein theory and
that $|\Delta R_{_{\rm KK}}/R_{_{\rm KK}}|<3.4\times10^{-10}$ from the
Oklo data.

It follows that the radius of the extra-dimensions has to be
stabilized but no satisfactory and complete mechanism has yet been
found. Li and Gott (1998) considered a 5-dimensional Kaluza-Klein
inflationary scenario which is static in the internal dimension
and expanding in the other dimensions and solve the 5-dimensional
semi-classical Einstein equations including the Casimir effect. In
particular, it was deduced that the effective 4-dimensional
cosmological constant is related to the fine structure constant by
$G\Lambda_{_{\rm eff}}=(15g_*/2048\pi^7)\aem^2$.

\subsection{Superstring theories}\label{subsec_7.1}

There exist five anomaly free, supersymmetric perturbative string
theories respectively known as type I, type IIA, type IIB, SO(32)
heterotic and $E_8\times E_8$ heterotic theories (see e.g. in
Polchinski, 1997). One of the definitive predictions of these theories
is the existence of a scalar field, the dilaton, that couples directly
to matter (Taylor and Veneziano, 1988) and whose vacuum expectation
value determines the string coupling constant (Witten, 1984). There
are two other excitations that are common to all perturbative string
theories, a rank two symmetric tensor (the graviton) $g_{\mu\nu}$ and
a rank two antisymmetric tensor $B_{\mu\nu}$. The field content then
differs from one theory to another. It follows that the 4-dimensional
couplings are determined in terms of a string scale and various
dynamical fields (dilaton, volume of compact space, \ldots). When the
dilaton is massless, we expect {\it three} effects: (i) a scalar
admixture of a scalar component inducing deviations from general
relativity in gravitaional effects, (ii) a variation of the couplings
and (iii) a violation of the eak equivalence principle. Our purpose is
to show how the 4-dimensional couplings are related to the string mass
scale, to the dilaton and the structure of the extra-dimensions mainly
on the example of heterotic theories.

The two {\it heterotic theories} originate from the fact that left-
and right-moving modes of a closed string are independent. This
reduces the number of supersymmetry to $N=1$ and the quantization of
the left-moving modes imposes that the gauge group is either $SO(32)$
or $E_8\times E_8$ depending on the fermionic boundary conditions. The
effective tree-level action is
\begin{eqnarray}\label{het}
S_{H}&=&\int\dd^{10}{\bf x}\sqrt{-g_{10}}\hbox{e}^{-2\Phi}
         \left[M_{_{H}}^8\left\lbrace R_{10}+4\Box\Phi-4(\nabla\Phi)^2
         \right\rbrace\right.\nonumber\\
     &&\left.-\frac{M_{_{H}}^6}{4}F_{AB}F^{AB}
         +\ldots\right].
\end{eqnarray}
When compactified on a 6-dimensional Calabi-Yau space, the effective
4-dimensional action takes the form
\begin{eqnarray}\label{het4}
S_{H}&=&\int\dd^{4}{\bf x}\sqrt{-g_{4}}\phi
\left[M_{_{H}}^8\left\lbrace R_{4}+\left(\frac{\nabla\phi}{\phi}\right)^2
-\frac{1}{6}\left(\frac{\nabla V_6}{V_6}\right)^2\right\rbrace\right.
\nonumber\\
&&-\left.\frac{M_{_{H}}^6}{4}F^2\right]+\ldots
\end{eqnarray}
where $\phi\equiv V_6\hbox{e}^{-2\Phi}$ couples identically to the
Einstein and Yang-Mills terms. It follows that
\begin{equation}
M_4^2=M_{_{H}}^8\phi,\qquad
g^{-2}_{_{\rm YM}}=M_{_{H}}^6\phi
\end{equation}
at tree-level. Note that to reach this conclusion, one has to
assume that the matter fields (in the `dots' of Eq.~(\ref{het4})
are minimally coupled to $g_4$, see e.g. the discussion by Maeda,
1988).

The strongly coupled SO(32) heterotic string theory is equivalent to
the weakly coupled type I string theory.  {\it Type I superstring}
admits open strings, the boundary conditions of which divide the
number of supersymmetries by two. It follows that the tree-level
effective bosonic action is $N=1$, $D=10$ supergravity which takes the
form, in the string frame,
\begin{eqnarray}
S_{I}&=&\int\dd^{10}{\bf x}\sqrt{-g_{10}}M_{_{I}}^6\hbox{e}^{-\Phi}
        \left[\hbox{e}^{-\Phi}
        M_{_{I}}^2R_{10}\right.\nonumber\\
     &&\left.\qquad-\frac{F^2}{4}+\ldots\right]
\end{eqnarray}
where the dots contains terms describing the dynamics of the dilaton,
fermions and other form fields. At variance with (\ref{het}), the
field $\Phi$ couples differently to the gravitational and Yang-Mills
terms because the graviton and Yang-Mills fields are respectively
excitation of close and open strings. It follows that $M_I$ can be
lowered even to the weak scale by simply having $\exp\Phi$ small
enough. Type I theories require $D9$-branes for consistancy. When
$V_6$ is small, one can use T-duality (to render $V_6$ large, which
allows to use a quantum field theory approach) and turn the $D9$-brane
into a $D3$-brane so that
\begin{eqnarray}
S_{I}&=&\int\dd^{10}{\bf x}\sqrt{-g_{10}}
\hbox{e}^{-2\Phi}M_{_{I}}^8R_{10}\nonumber\\
&&-\int\dd^{4}{\bf x}\sqrt{-g_{4}}\hbox{e}^{-\Phi}
\frac{1}{4}F^2+\ldots
\end{eqnarray}
where the second term describes the Yang-Mills fields localized on the
$D3$-brane. It follows that
\begin{equation}
M_4^2=\hbox{e}^{-2\Phi}V_6M_{_{I}}^8,\qquad
g^{-2}_{_{\rm YM}}=\hbox{e}^{-\Phi}
\end{equation}
at tree-level. If one compactifies the $D9$-brane on a 6-dimensional
orbifold instead of a 6-torus, and if the brane is localized at an
orbifold fixed point, then gauge fields couple to fields $M_i$ living
only at these orbifold fixed points with a (calculable) tree-level
coupling $c_i$ so that
\begin{equation}
M_4^2=\hbox{e}^{-2\Phi}V_6M_{_{I}}^8,\qquad
g^{-2}_{_{\rm YM}}=\hbox{e}^{-\Phi}+c_iM_i.
\end{equation}
The coupling to the field $c_i$ is a priori non universal.  At
strong coupling, the 10-dimensional $E_8\times E_8$ heterotic
theory becomes M-theory on $R^{10}\times S^1/Z_2$ (Ho\v{r}ava and
Witten, 1996). The gravitational field propagates in the
11-dimensional space while the gauge fields are localized on two
10-dimensional branes.

At one-loop, one can derive the couplings by including Kaluza-Klein
excitations to get (see e.g. Dudas, 2000)
\begin{equation}
g^{-2}_{_{\rm YM}}=M_{_{H}}^6\phi-\frac{b_a}{2}(RM_{_H})^2+\ldots
\end{equation}
when the volume is large compared to the mass scale and in that case
the coupling is no more universal. Otherwise, one would get a more
complicated function. Obviously, the 4-dimensional effective
gravitational and Yang-Mills couplings depend on the considered
superstring theory, on the compactification scheme but in any case
they depend on the dilaton.

Wu and Wang (1986) studied the cosmological behavior of the theory
(\ref{het}) assuming a 10-dimensional metric of the form ${\rm
diag}(-1, R_3(t)^2\tilde g_{ij}(x), R_6(t)^2\tilde g_{mn}(y))$ where
$R_3$ and $R_6$ are the scale factors of the external and internal
spaces. The rate of evolution of the size of the internal space was
related to the time variation of the gravitational constant. The
effect of a potential for the size of the internal space was also
studied.

Maeda (1988) considered the ($N=1, D=10$)-supergravity model
derived from the heterotic superstring theory in the low energy limit
and assumed that the 10-dimensional spacetime is compactified on a
6-torus of radius $R(x^\mu)$ so that the effective 4-dimensional
theory described by (\ref{het4}) is of the Brans-Dicke type with
$\omega=-1$.  Assuming that $\phi$ has a mass $\mu$,  and
couples to the matter fluid in the universe as $S_{_{\rm
matter}}=\int\dd^{10}{\bf x}\sqrt{-g_{10}}\exp(-2\Phi){\cal L}_{_{\rm
matter}}(g_{10})$, the reduced 4-dimensional matter action is
\begin{equation}
S_{_{\rm matter}}=\int\dd^{4}{\bf
x}\sqrt{-g}\phi{\cal L}_{_{\rm matter}}(g).
\end{equation}
The cosmological evolution of $\phi$ and $R$ can then be computed and
Maeda (1988) deduced that $\dot\aem/\aem\simeq10^{10}(\mu/1\,{\rm
eV})^{-2}\,{\rm yr}^{-1}$. In this approach, there is an ambiguity in
the way to introduce the matter fluid.

Vayonakis (1988) considered the same model but assumed that
supersymmetry is broken by non-perturbative effects such as
gaugino condensation. In this model, and contrary to the work by
Maeda (1988), $\phi$ is stabilized and the variation of the
constants arises mainly from the variation of $R$ in a runaway
potential.

Damour and Polyakov (1994a, 1994b) argued that the effective action
for the massless modes taking into account the full string loop
expansion is of the form
\begin{eqnarray}
S&=&\int\dd^4{\bf x}\sqrt{-\hat g}\left[M_s^2
\left\lbrace B_g(\Phi)\hat R+4B_\Phi(\Phi)\left[\hat \Box\Phi
-(\hat\nabla\Phi)^2\right]
\right\rbrace\right.\nonumber\\
&&\left.-B_F(\Phi)\frac{k}{4}\hat F^2-B_\psi(\Phi)\bar{\hat
\psi}\hat\dslash\hat\psi+\ldots\right]
\end{eqnarray}
in the string frame, $M_s$ being the string mass scale. The functions
$B_i$ are not known but can be expanded as
\begin{equation}\label{ans}
B_i(\Phi)=\hbox{e}^{-2\Phi}+c^{(i)}_0+c^{(i)}_1\hbox{e}^{2\Phi}+
c^{(i)}_2\hbox{e}^{4\Phi}
+\ldots
\end{equation}
in the limit $\Phi\rightarrow-\infty$, so that these functions can
exhibit a local maximum. After a conformal transformation
($g_{\mu\nu}=CB_g\hat g_{\mu\nu},
\psi=(CB_g)^{-3/4}B_\psi^{1/2}\hat\psi$), the action in Einstein frame
takes the form
\begin{eqnarray}
S&=&\int\frac{\dd^4{\bf x}}{16\pi G}\sqrt{-g}\left[
R-2(\nabla\phi)^2-\frac{k}{4}B_F(\phi)F^2\right.\nonumber\\
&&\left.\qquad-\bar
\psi\dslash\psi+\ldots\right]
\end{eqnarray}
from which it follows that the Yang-Mills coupling behaves as
$g^{-2}_{_{\rm YM}}=kB_F(\phi)$. This also implies that the QCD mass scale
is given by
\begin{equation}\label{qcd}
\Lambda_{_{\rm QCD}}\sim M_s(CB_g)^{-1/2}\hbox{e}^{-8\pi^2kB_F/b}
\end{equation}
where $b$ depends on the matter content. It follows that the mass of
any hadron, proportional to $\Lambda_{_{\rm QCD}}$ in first
approximation, depends on the dilaton, $m_A(B_g, B_F,\ldots)$. With
the anstaz (\ref{ans}), $m_A(\phi)$ can exhibit a minimum $\phi_m$
that is an attractor of the cosmological evolution that drives the
dilaton towards a regime where it decouples from matter. But, one
needs to assume for this mechanism to apply, and particularly to avoid
violation of the equivalence principle at an unacceptable level, that
all the minima are the same, which can be implemented by setting
$B_i=B$. Expanding $\ln B$ around its maximum $\phi_m$ as
$\ln B\propto-\kappa(\phi-\phi_m)^2/2$, Damour and Polyakov (1994a,
1994b) constrained the set of parameters $(\kappa,\phi_0-\phi_m)$
using the different observational bounds. This toy model allows to
address the unsolved problem of the dilaton stabilization and to study
all the experimental bounds together.

Damour, Piazza and Veneziano (2002a,b) extended this model to a case
where the coupling functions have a smooth finite limit for infinite
value of the bare string coupling, so that $B_i=C_i+{\cal O}({\rm
e}^{-\phi}$). The dilaton runs away toward its attractor at infinity
during a stage of inflation. The amplitude of residual dilaton
interaction is related to the amplitude of the primordial density
fluctuations and it can induce a variation of the fundamental
constants, provided it couples to dark matter or dark energy. It is
concluded that, in this framework, the largest allowed variation of
$\aem$ is of order $2\times10^{-6}$, which is reached for a violation
of the universality of free fall of of order $10^{-12}$.

Kolb {\em et al.} (1985) argued that in 10-dimensional superstring
models, $G\propto R^{-6}$ and $\aem\propto R^{-2}$ to deduce that
$|\Delta R/R|<0.5\%$. This was revised by Barrow (1987) who included
the effect of $\as$ to deduce that helium abundances impose $|\Delta
R/R|<0.2\%$. Recently Ichikawa and Kawasaki (2002) considered a model
in which all the couplings vary due to the dilaton dynamics and
constrain the variation of the dilaton field from nucleosynthesis as
$-1.5\times10^{-4}<\sqrt{16\pi G}\Delta\phi<6.0\times10^{-4}$. From
the Oklo data, Barrow (1987) concluded that $|\Delta
R/R|<1.5\times10^{-10}$.

To conclude, superstring theories offer a theoretical framework to
discuss the value of the fundamental constants since they become
expectation values of some fields. This is a first step towards
their understanding but yet, no complete and satisfactory
mechanism for the stabilization of the extra-dimension and dilaton
is known.

\subsection{Other investigations}\label{subsec_7.15}

Independently of string theory, Bekenstein (1982)
formulated a framework to incorporate a varying fine structure
constant. Working in units in which $\hbar$ and $c$ are
constant, he adopted a classical description of the electromagnetic
field and made a set of assumptions to obtain a reasonable
modification of Maxwell equations to take into account the effect of
the variation of the elementary charge [for instance to take into
account the problem of charge conservation which usually derived from
Maxwell equations]. His eight postulates are that (1) for a constant
$\aem$ electromagnetism is described by Maxwell theory and the
coupling of the potential vector $A_\nu$ to matter is minimal, (2) the
variation of $\aem$ results from dynamics, (3) the dynamics of
electromagnetism and $\aem$ can be obtained from an invariant action
that is (4) locally gauge invariant, (5) electromagnetism is causal
and (6) its action is time reversal invariant, (7) the shortest length
scale is the Planck length and (8) gravitation is described by a
metric theory which satisfies Einstein equations.

Assuming that the charges of all particles vary in the same way, one can
set $e=e_0\epsilon(x^\mu)$ where $\epsilon(x^\mu)$ is a dimensionless
universal field (it should be invariant under
$\epsilon\rightarrow\hbox{constant}\times\epsilon$ through a
redefinition of $e_0$). The electromagnetic tensor generalizes to
\begin{equation}\label{beken1}
F_{\mu\nu}=\epsilon^{-1}\nabla_{[\mu}\left(\epsilon A_{\nu]}\right)
\end{equation}
and the electromagnetic action is given by
\begin{equation}\label{beken2}
S_{_{\rm EM}}=\frac{-1}{16\pi}\int F_{\mu\nu}F^{\mu\nu}\sqrt{-g}\dd^4\bx.
\end{equation}
The dynamics of $\epsilon$ can be shown to derive from the action
\begin{equation}\label{beken3}
S_{_\epsilon}=\frac{-1}{2}\frac{\hbar c}{\ell^2} \int
\frac{\partial_\mu\epsilon\partial^\mu\epsilon}{\epsilon^{-2}}\sqrt{-g}\dd^4\bx
\end{equation}
where $\ell$ is length scale which needs to be small enough to be
compatible with the observed scale invariance of electromagnetism
($\ell_{_{\rm Pl}}<\ell<10^{-15}-10^{-16}$~cm around which
electromagnetism merges with the weak interaction). Finally, the matter
action for point particles of mass $m$ takes the form
$S_m=\sum\int[-mc^2+(e/c)u^\mu
A_\mu]\gamma^{-1}\delta^3(x^i-x^i(\tau))\dd^4\bx$ where $\gamma$ is the
Lorentz factor and $\tau$ the proper time.

Varying the total action gives the electromagnetic equation
\begin{equation}\label{beken4}
\nabla_\mu\left(\epsilon^{-1}F^{\mu\nu}\right)=4\pi j^\nu
\end{equation}
and the equation for the dynamics of $\epsilon$
\begin{equation}\label{beken5}
\Box\epsilon=\frac{\ell^2}{\hbar
c}\left[\epsilon\frac{\partial\sigma}{\partial\epsilon} -\frac{1}{8\pi}
F_{\mu\nu}F^{\mu\nu}\right]
\end{equation}
with $\sigma=\sum mc^2\gamma^{-1}\delta^3(x^i-x^i(\tau))/\sqrt{-g}$.
The Maxwell equation (\ref{beken4}) is the same as electromagnetism in
a material medium with dielectric constant $\epsilon^{-2}$ and
permeability $\epsilon^2$ [this was the original description proposed
by Fierz (1955) and Lichn\'erowicz (1955); see also Dicke (1964)].

On cosmological scales, it can be shown that the dynamical equation for
$\epsilon$ can be cast under the form
\begin{equation}\label{bevo}
\left(a^3\dot\epsilon/\epsilon\right)^.=-a^3\zeta\frac{\ell^2}{\hbar
c}\rho_mc^2
\end{equation}
where $\zeta={\cal O}(10^{-2})$ is a dimensionless (and
approximatively constant) measuring the fraction of mass in Coulomb
energy for an average nucleon compared with the free proton mass and
$\rho_m$ is the matter density. Since $\rho_m\propto a^{-3}$,
Eq. (\ref{bevo}) can be integrated to relate
$(\dot\epsilon/\epsilon)_0$ to $\ell/\ell_{_{\rm Pl}}$ and the
cosmological parameters. In order to integrate this equation,
Bekenstein assumed that $\zeta$ was constant, which was a reasonable
assumption at low redshift. Livio and Stiavelli (1998) extended this
analysis and got $\zeta=1.2\times10^{-2}(X+4/3Y)$ where $X$ and $Y$
are the mass fraction of hydrogen and helium.

Replacing the quantity in the brackets of the r.h.s. of
Eq.~(\ref{beken5}) by $\zeta\rho_mc^2$ with $\zeta={\cal O}(10^{-2})$,
the static form Eq.~(\ref{beken5}) is analogous to the standard Poisson
equation so that $\ln\epsilon$ is proportional to the gravitational
potential
\begin{equation}
\ln\epsilon=\frac{\zeta}{4\pi c^2}\frac{\ell}{\ell_{_{\rm Pl}}}\Phi
\end{equation}
from which it follows that a test body of mass $m$ and of
electromagnetic energy $E_{_{\rm EM}}$ experiences an acceleration of
$\vec a=-\nabla\Phi-M^{-1}(\partial E_{_{\rm
EM}}/\partial\epsilon)\nabla\epsilon$.

From the confrontation of the results of the spatial and cosmological
variation of $\epsilon$ Bekenstein (1982) concluced, given his
assumptions on the couplings, that $\aem$ ``{\it is a parameter, not a
dynamical variable}''. This problem was recently by passed by Olive and
Pospelov (2001) who generalized the model to allow additional coupling
of a scalar field $\epsilon^2=B_F(\phi)$ to non-baryonic dark matter
(as first proposed by Damour {\em et al.}, 1990) and cosmological
constant, arguing that in certain classes of dark matter models, and
particularly in supersymmetric ones, it is natural to expect that
$\phi$ would couple more strongly to dark matter than to baryon. For
instance, supersymmetrizing Bekenstein model, $\phi$ will get a
coupling to the kinetic term of the gaugino of the form
$M_*^{-1}\phi\bar\chi\partial\chi$ so that, assuming that the gaugino
is a large fraction of the stable lightest supersymmetric particle,
then the coupling to dark matter would be of order $10^3-10^4$ times
larger. Such a factor could almost reconcile the constraint arising
from the test of the universality of free fall with the order of
magnitude of the cosmological variation. This generalization of
Bekenstein model relies on an action of the form
\begin{eqnarray}\label{olive}
S&=&-\frac{1}{2}M_4^2\int R\sqrt{-g}\dd^4\bx\nonumber\\
 &&+\int\left[\frac{1}{2}M_*^2\partial_\mu\phi\partial^\mu\phi-\frac{1}{4}
   B_F(\phi)
    F_{\mu\nu}F^{\mu\nu}\right]\sqrt{-g}\dd^4\bx
    \nonumber\\
 &&+\int\left\lbrace\sum\bar
   N_i[i\dslash-m_iB_{N_i}(\phi)]N_i+\frac{1}{2}\bar\chi\partial\chi
   \right\rbrace\sqrt{-g}\dd^4\bx
    \nonumber\\
  &&-\int\left[M_4^2B_\Lambda(\phi)\Lambda+\frac{1}{2}M_\chi
    B_\chi(\phi)\chi{}^T\chi\right]\sqrt{-g}\dd^4\bx
\end{eqnarray}
where the sum is over proton
[$\dslash=\gamma^\mu(\partial_\mu-ie_0A_\mu)$] and neutron
[$\dslash=\gamma^\mu\partial_\mu$]. The functions $B$ can be expanded
(since one focuses on small variations of the fine structure constant
and thus of $\phi$) as $B_X=1+\zeta_X\phi+ \xi_X\phi^2/2$. It follows
that $\aem(\phi)={e_0^2}/{4\pi B_F(\phi)}$ so that
$\Delta\aem/\aem=\zeta_F\phi+(\xi_F-2\zeta_F^2)\phi^2/2$.  This
framework extends the analysis by Bekenstein (1982) to a 4-dimensional
parameter space ($M_*,\zeta_F,\zeta_m,\zeta_\Lambda$). It contains the
Bekenstein model ($\zeta_F=-2$, $\zeta_\Lambda=0$,
$\zeta_m\sim10^{-4}\xi_F$), a Jordan-Brans-Dicke model ($\zeta_F=0$,
$\zeta_\Lambda=-2\sqrt{2/2\omega+3}$, $\xi_m=-1/\sqrt{4\omega+6}$), a
string-like model ($\zeta_F=-\sqrt{2}$, $\zeta_\Lambda=\sqrt{2}$,
$\zeta_m=\sqrt{2}/2$) so that $\Delta/\aem/\aem=3$) and
supersymmetrized Bekenstein model ($\zeta_F=-2$, $\zeta_\chi=-2$,
$\zeta_m=\zeta_\chi$ so that $\Delta\aem/\aem\sim5/\omega$). In all
the models, the universality of free fall sets a strong constraint on
$\zeta_F/\sqrt{\omega}$ (with $\omega\equiv M_*/2M_4^2$) and the
authors showed that a small set of models was compatible with the
cosmological variation and the equivalence principle tests.

The constraint arising from the universality of free fall can be
fulfilled if one sets by hand $B_F-1\propto[\phi-\phi(0)]^2$ where
$\phi(0)$ is the value of the field today. It then follows that the
cosmological evolution will drive the system toward a state in which
$\phi$ is almost stabilized today but allowing for cosmological
variation of the constants of nature. In their two-parameter
extension, Livio and Stiavelli (1998) found that only variations of
$\Delta\aem/\aem$ of $8\times10^{-6}$ and $9\times 10^{-7}$
respectively for $z<5$ and $z<1.6$ were compatible with Solar system
experiments.

The formalism developed by Bekenstein (1982) was also applied to the
the strong interaction (Chamoun {\em et al.}, 2000, 2001) by simply
adding a term $f_{abc}A^b_\mu A^c_\nu$ to describe the gluon tensor
field $G_{\mu\nu}^a$, $f_{abc}$ being the structure constants of the
non-Abelian group.  It was also implemented in the braneworld context
(e.g. Youm, 2001) and Magueijo {\em et al.} (2001) studied the effect
of a varying fine structure constant on a complex scalar field
undergoing an electromagnetic $U(1)$ symmetry breaking in this
framework. Armend\'ariz-Pic\'on (2002) derived the most general low
energy action including a real scalar field that is local, invariant
under space inversion and time reversal, diffeomormism invariant and
with a U(1) gauge invariance. This form includes the previous form
(\ref{olive}) of Bekenstein's theory as well as scalar-tensor theories
and long wavelength limit of bimetric theories.

Recently Sandvik {\em et al.} (2001) claimed to have generalized
Bekenstein model by simply redefining $a_\mu\equiv \epsilon A_\mu$,
$f_{\mu\nu}\equiv \partial_{[\mu}a_{\nu]}$ and $\psi\equiv\ln\epsilon$
so that the covariant derivative becomes
$D_\mu\equiv\partial_\mu+ie_0a_\mu$. It follows that the total action
including the Einstein-Hilbert action for gravity the actions (\ref{beken2})
and (\ref{beken3}) for the modified electromagnetism and normal matter takes
the form
\begin{equation}
S=\int\sqrt{-g}\dd^4\bx\left( {\cal L}_{\rm grav}+{\cal L}_{\rm mat}+{\cal
L}_{\psi} +{\cal L}_{_{\rm EM}}\hbox{e}^{-2\psi} \right)
\end{equation}
with ${\cal L}_{\psi}=-(\omega/2)\partial_\mu\psi\partial^\mu\psi$ so
that the Einstein equation are the ``standard'' Einstein equations
with an additive stress-energy tensor for the scalar field
$\psi$. Indeed, Bekenstein (1982) did not take into account the effect
of $\epsilon$ (or $\psi$) in the Friedmann equation and studied only
the time variation of $\epsilon$ in a matter dominated universe. In
that sense Sandvik {\em et al.} (2002) extended the analysis by
Bekenstein (1982) by solving the coupled system of Friedmann and
Klein-Gordon equations. They studied numerically in function of
$\zeta/\omega_{\rm SBM}$ (with $\omega_{\rm SBM}=\ell_{_{\rm
Pl}}^2/\ell$) and showed that cosmological and astrophysical data can
be explained with $\omega_{\rm SBM}=1$ if $\zeta$ ranges between
0.02\% and 0.1\% (that is about one order of magnitude smaller than
Bekenstein's value based on the argument that dark matter has to be
taken into account). An extension of the discussion of the
cosmological scenarios was performed in Barrow {\em et al.} (2002c)
and it was shown that $\aem$ is constant during the radiation era,
then evolves logarithmically with the cosmic time during the matter
era and then tends toward a constant during a curvature or
cosmological constant era. The scalar-tensor case with both
varying $G$ and $\aem$ was considered by Barrow {\em et al.}
(2002a,b).

Sandvik {\em et al.} (2002), following Barrow and O'Toole (2001),
estimated the spatial variations to be of order
$\Delta\ln\aem\sim4.8\times10^{-4}GM/c^2r$ (Magueijo, 2001) to
conclude that on cosmological scale $\Delta\ln\aem\sim 10^{-8}$ if
$GM/c^2r\sim10^{-4}$, as expected on cosmological scales if $(\delta
T/T)_{_{\rm CMB}}\sim GM/c^2r$. On the Earth orbit scale, this leads
to the rough estimate $|\nabla\ln\aem|\sim 10^{-23}-10^{-22}\,{\rm
cm}^{-1}$ which is about ten orders of magnitude higher than the
constraint arising from the test of the universality of free fall.
Nevertheless, Magueijo {\em et al.}  (2002) re-analyzed the violation
of the universality of free fall and claimed that the theory is still
compatible with equivalence principle tests provided that $\zeta_m\la
1$ for dark matter. This arises probably from the fact that only
$\aem$ is varying while other constants are fixed so that the dominant
factor in Eq.~(\ref{fhad}) is absent.\\

Wetterich (2002) considered the effect of the scalar field responsible
for the acceleration of the universe (the ``cosmon'') on the couplings
arising from the coupling of the cosmon to the kinetic term of the
gauge field as $Z_F(\phi)F^2/4$. Focusing on grand unified theory, so
that one gets a coupling of the form ${\cal L}=Z_F(\phi){\rm
Tr}(F^2)/4+iZ_\psi(\phi)\bar\psi\dslash\psi$ and assuming a runaway
expotenial potential, he related the variation of $\aem$, $\as$,
the nucleon masses to the arbitrary function $Z_F$ and to the
$\phi$-dependent electroweak scale so that the different bounds can be
discussed in the same framework.

Chacko {\em et al.} (2002) proposed that the variation of the fine
structure constant could be explained by a late second order phase
transition at $z\sim1-3$ (that is around $T\sim10^{-3}$~eV) inducing a
change in the vacuum expectation value of a scalar field. This can be
implemented for instance in supersymmetric theories with low energy
symmetry breaking scale. This will induce a variation of the masses of
electrically charge particle and. From the renormalization group
equation $\aem^{-1}=\aem^{-1}(\Lambda)+\sum_i(b_{i+1}/2\pi)
\ln(m_{i+1}/m_i)$ and assuming that $\aem^{-1}(\Lambda)$ was fixed,
one would require that $\sum\delta m_i/m_i={\cal O}(10^{-2})$ to
explain the observations by Webb {\em et al.} (2001), so that the
masses have to increase. Note that it will induce a time variation of
the Fermi constant. Such models can occur in a large class of
supersymmetric theories. Unfortunately, it is yet incomplete and its
viability depends on the exitence of an adjustment mechanism for the
cosmological constant. But, it offers new way of thinking the
variation of the constants at odd with the previous analysis involving
a rolling scalar field.\\

Motivated by resolving the standard cosmological puzzles (horizon,
flatness, cosmological constant, entropy, homogeneity problems)
without inflation, Albrecht and Magueijo (1998) introduced a
cosmological model in which the speed of light is varying.  Earlier
related attempts were investigated by Moffat (1993a,1993b).  Albrecht
and Magueijo (1998) postulated that the Friedmann equations are kept
unchanged from which it follows that the matter conservation has to be
changed and get a term proportional to $\dot c/c$.  The flatness and
horizon problems are not solved by a period of accelerated expansion
so that, contrary to inflation, it does not offer any explanation for
the initial perturbations (see however Harko and Mak, 1999).  Albrecht
and Magueijo (1998) considered an abrupt change in the velocity of
light as may happen during a phase transition. It was extended to
scenarios in which both $c$ and $G$ were proportional to some power of
the scale factor by Barrow (1999) (see also Barrow and Magueigo,
1999a,b). The link between this theory and Bekenstein theory was
investigated by Barrow and Magueijo (1998).  Magueijo {\em et al.},
(2002) investigated the test of universality of free fall. A
Lagrangian formulation would probably requires the introduction of an
``ether'' vector field to break local Lorentz invariance as was used
in e.g. Lubo {\em et al.}  (2002).

Clayton and Moffat (1999) implemented a varying speed of light
model by considering a bimetric theory of gravitation in which one
metric $g_{\mu\nu}$ describe the standard gravitational vacuum
whereas a second metric $g_{\mu\nu}+\beta\psi_\mu\psi_\nu$,
$\beta$ being a dimensionless constant and $\psi^\mu$ a dynamical
vector field, describes the geometry in which matter is
propagating (see also Bekenstein, 1993). When choosing
$\psi_\mu=\partial_\mu\phi$ this reduces to the models developed
by Clayton and Moffat (2000, 2001). Some cosmological implications
were discussed by Moffat (2001, 2002) but no study of the
constraints arising from Solar system experiments have been taken
into account. Note that Dirac (1979) also proposed that a varying
$G$ can be reconciled with Einstein theory of gravity if the space
metric was different from the ``atomic'' metric. Landau and
Vucetich (2000) investigated the constraints arising from the
violation of the charge conservation. Other realizations arise
from the brane world picture in which our universe is a
3-dimensional brane embedded in a higher dimensional spacetime.
Kiritsis (1999) showed that when a test brane is moving in a black
hole bulk spacetime (Kehagias and Kiritsis, 1999) the velocity of
light is varying as the distance between the brane and the black
hole. Alexander (2000) generalizes this model (see also Steer and
Parry, 2002) by including rotation and expansion of the bulk so
that the speed of light gets stabilized at late time. Carter {\em
et al.} (2001) nevertheless showed that even if a Newton-like
force is recovered on small scales such models are very
constrained at the post-Newtonian level. Brane models allowing for
the scalar field in the bulk naturally predicts a time variable
gravitational constant (see e.g. Brax and Davis, 2001).

\subsection{A new cosmological constant problem?}\label{subsec_7.29}

The question of the compatibility between an observed variation of the
fine structure constant and particle physics models was put forward by
Banks {\em et al.} (2002). As seen above, in the low energy limit, the
change of the fine structure constant can be implemented by coupling a
scalar field to the photon kinetic term $F^{\mu\nu}F_{\mu\nu}$, but
this implies that the vacuum energy computed in this low energy limit
must depend on $\aem$. Estimating that
\begin{equation}
\Delta \Lambda_{_{\rm vac}}\sim\Lambda^4\Delta\aem,
\end{equation}
leads to a variation of order $\Delta\Lambda_{_{\rm
vac}}\sim10^{28}\, ({\rm eV})^4$ for a variation $\Delta\aem\sim
{\cal O}(10^{-4})$ and for $\Lambda=\Lambda_{_{\rm QCD}}\sim
100\,{\rm MeV}$. Indeed, this contrasts with the average energy
density of the universe of about $10^4\,({\rm eV})^4$ during the
matter era so that the universe was dominated by the cosmological
constant at $z\sim3$, which is at odds with observations. It was
thus concluded that this imposes that
\begin{equation}
\left|\Delta\aem/\aem\right|<10^{-28}.
\end{equation}
Contrary to the standard cosmological constant problem, the vacuum
zero-point energy to be removed is time-dependent and one can only
remove it for a fixed value of $\aem$. Whereas the cosmological
constant problem involves the fine tuning of a parameter, this now
implies the fine tuning of a function!

It follows that a varying $\aem$ cannot be naturally explained in a
field theory approach. A possible way out would be to consider that
the field is in fact an axion (see Carroll, 1998; Choi, 2000; Banks
and Dine, 2001). Some possible links with Heisenberg relations and
quantum mechanics were also investigated by Ra\~nada (2002).  Besides,
the resolution of the cosmological constant problem may also provide
the missing elements to understand the variation of the
constants. Both preoblems can be hoped to be solved by string theory.

\section{Conclusions}\label{sec_8}

The experimental and observational constraints on the variation of the
fine structure, gravitational constants, of the electron to proton
mass ration and different combinations of the proton gyromagnetic
factor and the two previous constants, as well as bounds on $\aw$ are
summarized in tables \ref{table_3}, \ref{table_1} and \ref{table_4}.

The developments of high energy physics theories such as
multi-dimensional and string theories provide new motivations to
consider the time variation of the fundamental constants. The
observation of the variability of these constants constitutes one of
the very few hope to test directly the existence of extra-dimensions
and to test these high energy-physics models. In the long run, it may
help to discriminate between different effective potentials for the
dilaton and/or the dynamics of the internal space. But indeed,
independently of these motivations, the understanding of the value of
the fundamental constants of nature and the discussion of their status
of constant remains a central question of physics in general:
questioning the free parameters of a theory accounts to question the
theory itself.  It is a basic and direct test of the law of gravity.

As we have shown, proving that a fundamental constant has changed is
not an obvious task mainly because observations usually entangle a set
of constants and because the bounds presented in the literature often
assume the constancy of a set of parameters.  But, in GUT,
Kaluza-Klein and string inspired models, one expects all the couplings
to vary simultaneously.  Better analysis of the degeneracies as
started by Sisterna and Vucetich (1990, 1991) (see also Landau and
Vucetich, 2002) are really needed before drawing definitive
conclusions but such analysis are also dependent in the progresses in
our understanding of the fundamental interactions and particularly of
the QCD theory and on the generation of the fermion masses.

Other progresses require (model-dependent) investigations of the
compatibility of the different bounds. It has also to be remembered
that arguing about the non-existence of something to set constraints
(e.g. Broulik and Trefil, 1971) is very dangerous. On an observational
point of view, one needs to further study the systematics (and
remember some erroneous claims such as those by Van Flandern, 1975)
and to propose new experiments (see e.g. Karshenboim, 2000, 2001 who
proposed experiments based on the hyperfine structure of deuterium an
ytterbium-171 as well as atoms with magnetic moment; Braxmaier {\em et
al.}, 2001; Torgerson, 2000 who proposed to compare optical frequency
references; Sortais {\em et al.}, 2001 who improved the sensitivity of
frequency standards, the coming satellite experiments ACES, MICROSCOPE and
STEP\ldots). On local scales, the test of the universality of free
fall sets drastic constraints and one can hope to use similar methods
on cosmological scales from the measurements of weak gravitational
lensing (Uzan and Bernardeau, 2001) or from structure formation
(Martins {\em et al.}, 2002). The complementarity between local
experiments and geo-astrophysical observations is necessary since
these methods test different time-scales and are mainly sensitive
either to rapid oscillations or a slow drift of the constants.

The recent astrophysical observations of quasars tend to show that
both the fine structure constant and the electron to proton mass ratio
have evolved. These two measurements are non-zero detections and thus
very different in consequences compared with other bounds.  They draw
the questions of their compatibility with the bounds obtained from
other physical systems such as e.g. the test of the universality of
free fall and Oklo but also on a more theoretical aspect of the
understanding of such a late time variation which does not seem to be
natural from a field theory point of view. Theoretically, one expects
{\it all} constants to vary and the level of their variation is also
worth investigating.  One would need to study the implication of these
measurements for the other experiments and try to determine their
expected level of detection. Both results arise from the observation
of quasar absorption spectra; it is of importance to ensure that all
systematics are taken into account and are confirmed by independent
teams, using e.g. the VLT which offers a better signal to noise and
spectral resolution.

The step from the standard model+general relativity to string theory
allows for dynamical constants and thus starts to address the question
of why the constants have the value they have. Unfortunately, no
complete and satisfactory stabilization mechanism is known yet and we
have to understand why, if confirmed, the constants are still varying
and whether such a variation induces a new cosmological constant
problem.

The study of the variation of the constants offers a new link between
astrophysics, cosmology and high-energy physics complementary to
primordial cosmology. It is deeply related to the test of the law of
gravitation, both of the deviations from general relativity and the
violation of the weak equivalence principle. But yet much work is
needed both to disentangle the observations and to relate them to
theoretical models.

\section*{Acknowledgments}

It is a pleasure to thank Robert Brandenberger, Michel Cass\'e,
Thibault Damour, Emilian Dudas, Nathalie Deruelle, Gilles
Esposito-Far\`ese, Patrick Peter and Patrick Petitjean for their
numerous comments and suggestions to improve this text.  I want also
to thank Pierre Bin\'etruy, Francis Bernardeau, Philippe Brax, Brandon
Carter, Christos Charmousis, C\'edric Deffayet, Ruth Durrer, Gia
Dvali, Bernard Fort, Eric Gourgoulhon, Christophe Grojean, Joseph
Katz, David Langlois, Roland Lehoucq, J\'er\^ome Martin, Yannick
Mellier, Jihad Mourad, Kenneth Nordtvedt, Keith Olive, Simon Prunet,
Alain Riazuelo, Christophe Ringeval, Christophe Salomon, Aur\'elien
Thion, Gabriele Veneziano, Filippo Vernizzi for discussions on the
subject. This work was initially motivated by the questions of Ren\'e
Cuillierier and the monday morning discussions of the Orsay cosmology
group. I want to thank Patricia Flad for her help in gathering the
literature.

\begin{figure}
\begin{center}
\psfig{file=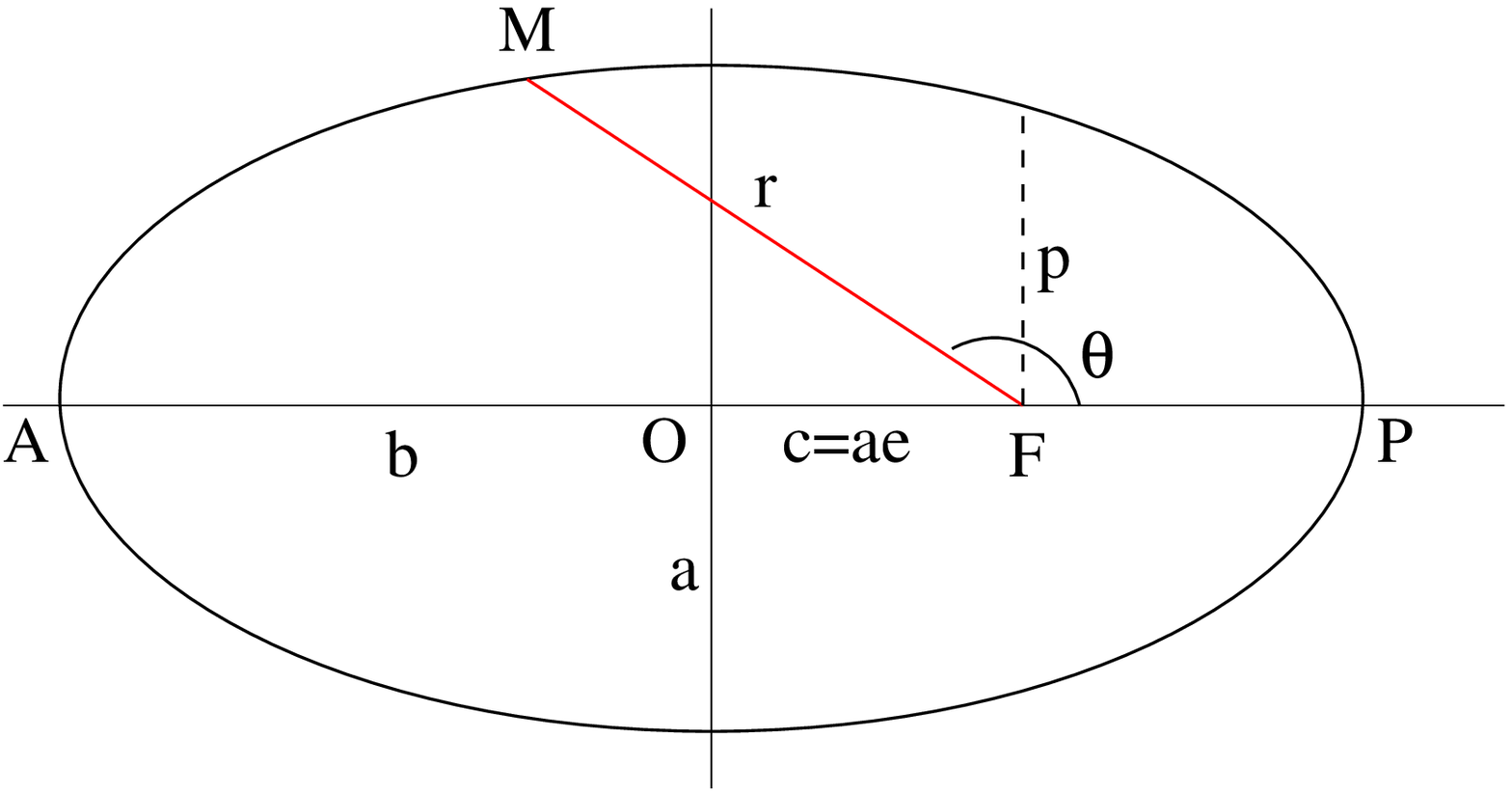,width=7cm} \caption{The standard
orbital parameters. $a$ and $b$ is the semi-major and semi-minor
axis, $c=ae$ the focal distance, $p$ the semi-latus rectum,
$\theta$ the true anomaly. $F$ is the focus, $A$ and $B$ the
periastron and apoastron (see e.g. Murray and Dermott, 2000). It
is easy to check that $b^2=a^2(1-e^2)$ and that $p=a(1-e^2)$ and
one defines the frequency or mean motion as $n=2\pi/P$ where $P$
is the period.} \label{fig4}
\end{center}
\end{figure}

\begin{figure}
\begin{center}
\epsfig{file=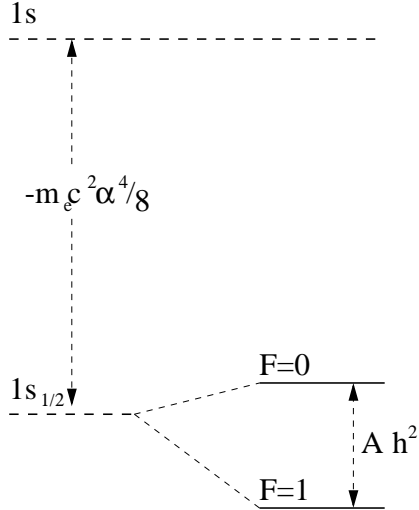,width=5.5cm}
\caption{Hyperfine structure of the $n=1$ level of the hydrogen atom.
The fine structure Hamiltonian induces a shift of $-m_{\rm e}c^2\aem^4/8$ of
the level $1s$. $J$ can only take the value $+1/2$. The hyperfine
Hamiltonian (\ref{hfH}) induces a splitting of the level $1s_{1/2}$ into
the two hyperfine levels $F=0$ and $F=+1$. The transition between these
two levels corresponds to the 21~cm ray with
$Ah^2=1,420,405,751.768\pm0.001$~Hz and is of first importance in
astronomy.}
\label{fig1}
\end{center}
\end{figure}

\begin{figure}
\begin{center}
\psfig{file=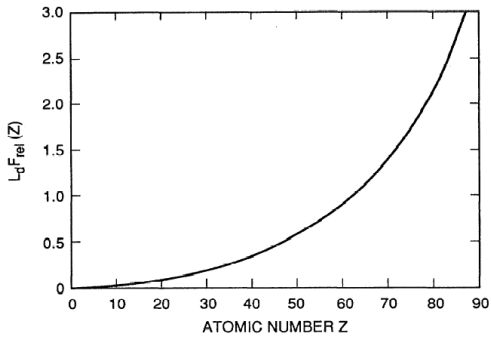,width=7cm}
\caption{The correction function $F_{_{\rm rel}}$. [From Prestage {\em
et al.}  (1995)].}
\label{figfrel}
\end{center}
\end{figure}

\begin{figure}
\begin{center}
\epsfig{file=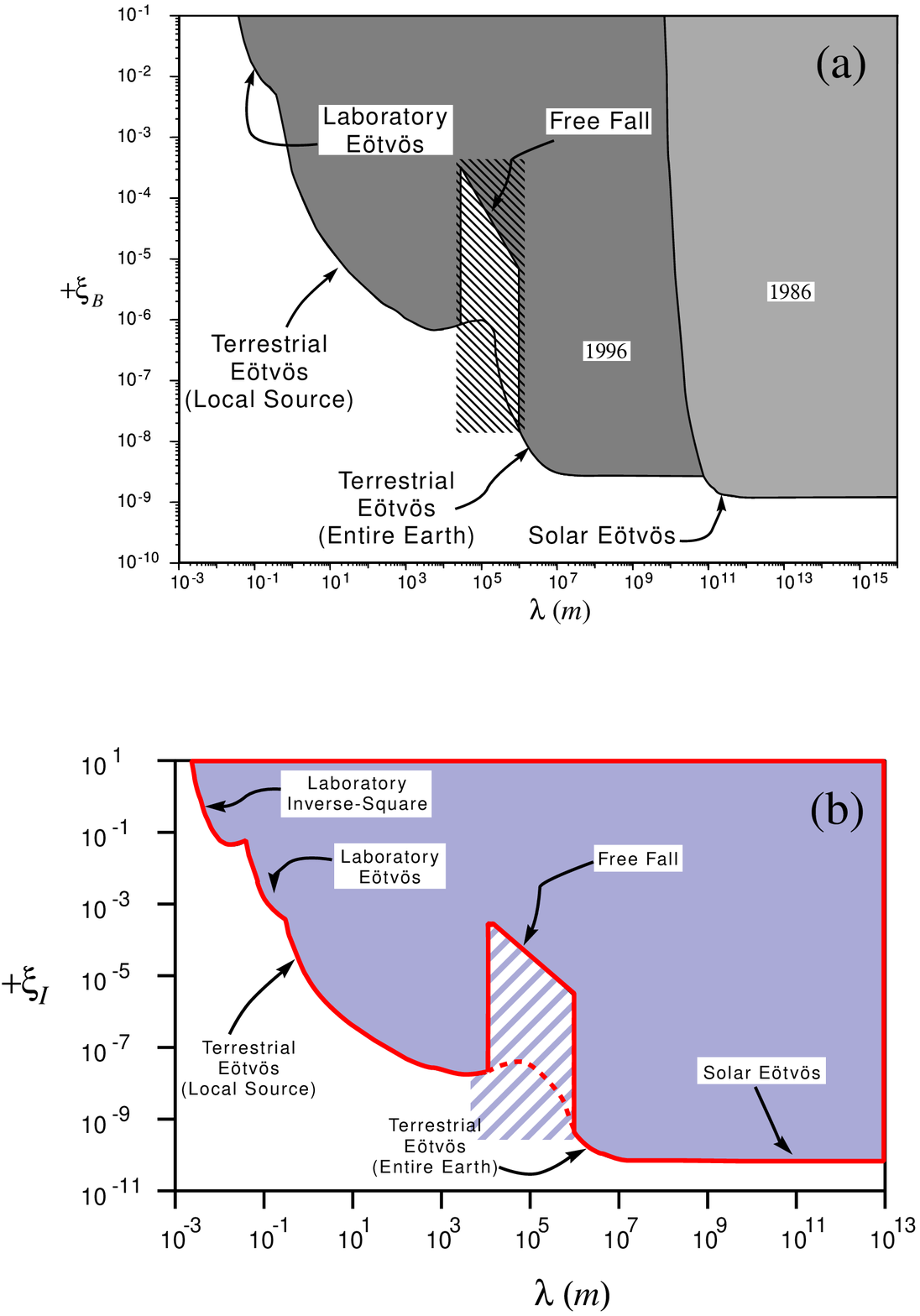,width=8cm}
\caption{Constraints on the coupling $\xi_B$ (upper panel) and $\xi_I$
(lower panel) respectively to $N+Z$ and $N-Z$ as a function of the
length scale $\lambda$. The shaded regions are excluded at
$2\sigma$. [From Fischbach and Talmadge (1997)].}
\label{figwep}
\end{center}
\end{figure}

\begin{figure}
\begin{center}
\psfig{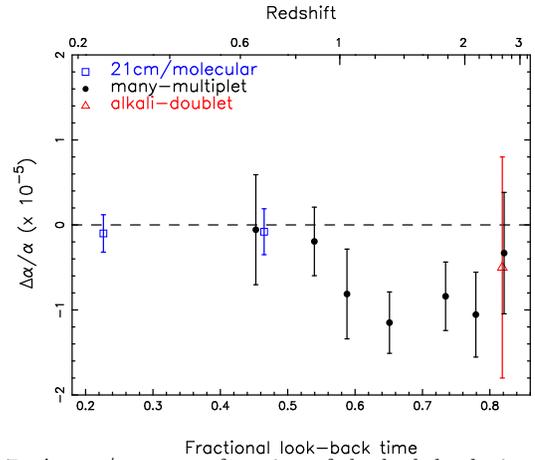}
\caption{ $\Delta\aem/\aem$ as a function of the look-back time
computed with the cosmological parameters
$(\Omega_m,\Omega_\Lambda)=(0.3,0.7)$ and $h=0.68$. Squares refer to
the data by (Murphy {\em et al.} (2001c) assuming $g_{\rm p}$ constant;
triangles refer to Si~IV systems by Murphy {\em et al.} (2001d); dots
correspond to Mg~II and Fe~II systems for redshidts smaller than 1.6
(Webb {\em et al.}, 2001) and higher redshifts come from Murphy {\em et
al.} (2001a). [From Webb {\em et al.} (2001)].}
\label{figwebb}
\end{center}
\end{figure}



\appendix
\onecolumn

\begin{table}
\caption{Comparison of absorption lines and the combinations of
the fundamental constants that can be constrained.} \label{table0}
\begin{tabular}{|lc|}
Comparison & constant \\
\hline
fine structure doublet                       & $\aem$ \\
hyperfine H vs optical                       & $g_{\rm p}\mu\aem^2$ \\
hyperfine H vs fine-structure                & $g_{\rm p}\mu\aem$ \\
rotational vs vibrational modes of molecules & $\mu$ \\
rotational modes vs hyperfine H              & $g_{\rm p}\aem^2$\\
fine structure doublet vs hyperfine H        & $g_{\rm p}$
\end{tabular}
\end{table}


\begin{table}
\caption{Summary of the constraints on the variation
         of the fine structure constant $\Delta\aem/\aem$.
         } \label{table_3}
\begin{tabular}{|lrlll|}
Reference   &  Constraint  &  Redshift & Time ($10^9\,{\rm yr}$)
      & Method \\
\hline
(Savedoff, 1956)&
       $(1.8\pm1.6)\times10^{-3}$&
       0.057 & &
       Cygnus A (N~II, Ne~III)\\
(Wilkinson, 1958)&
       $(0\pm8)\times10^{-3}$&
        & 3-4 &
       $\alpha$-decay\\
(Bahcall {\em et al.}, 1967)&
       $(-2\pm5)\times10^{-2}$&
       1.95 & &
       QSO (Si~II, Si~IV)\\
(Bahcall and Schmidt, 1967)&
       $(1\pm2)\times10^{-3}$&
       0.2 & &
       radio galaxies (O~III)\\
(Dyson, 1967)&
       $(0\pm9)\times10^{-4}$&
        & 3 &
       Re/Os\\
(Gold, 1968)&
       $(0\pm4.66)\times10^{-4}$&
       & 2 &
       Fission\\
(Chitre and Pal, 1968)&
       $(0\pm3_{-2}^{+2})\times10^{-4}$&
       & 1 &
       Fission\\
(Dyson, 1972)&
       $(0\pm4)\times10^{-4}$&
        & 2 &
       $\alpha$-decay\\
(Dyson, 1972)&
       $(0\pm1)\times10^{-3}$&
        & 2 &
        Fission\\
(Dyson, 1972)&
       $(0\pm5)\times10^{-6}$&
       &  1 &
       Re/Os\\
(Shlyakhter, 1976)&
       $(0\pm1.8)\times10^{-8}$&
       & 1.8 &
       Oklo \\
(Wolfe {\em et al.}, 1976)&
       $(0\pm3)\times10^{-2}$&
       0.524 & &
       QSO (Mg~I)\\
(Irvine, 1983a)&
       $(0\pm9)\times10^{-8}$&
       & 1.8 &
       Oklo \\
(Lindner {\em et al.}, 1986)&
       $(-4.5\pm9)\times10^{-6}$&
        &  4.5 &
       Re/Os \\
(Kolb {\em et al.}, 1986)&
       $(0\pm1)\times10^{-4}$&
        $10^8$& &
        BBN \\
(Potekhin and Varshalovich, 1994)&
       $(2.1\pm2.3)\times10^{-3}$ &
       3.2 & &
       QSO (C~IV, Si~IV,\ldots)\\
(Varshalovich and Potekhin, 1994)&
       $(0\pm1.5)\times10^{-3}$ &
       3.2 & &
       QSO (C~IV, Si~IV,\ldots)\\
(Cowie and Songaila, 1995)&
       $(-0.3\pm1.9)\times10^{-4}$ &
       $2.785-3.191$ &&
       QSO\\
(Prestage {\em et al.}, 1995)&
       $(0\pm1.42)\times10^{-14}$ &
       0 & 140\,{\rm days}&
       Atomic cloks \\
(Damour and Dyson, 1996) &
       $(0.15\pm1.05)\times10^{-7}$&
       &1.8&
       Oklo \\
(Varshalovich {\em et al.}, 1996a)&
       $(2\pm7)\times10^{-5}$ &
       $2.8-3.1$ &&
       QSO (Si~IV)\\
(Bergstr\"om {\em et al.}, 1999)&
       $(0\pm2)\times10^{-2}$ &
       $10^8$ &&
       BBN \\
(Webb {\em et al.}, 1999)&
      $(-0.17\pm0.39)\times10^{-5} $&
      $0.6-1$&&
      QSO (Mg~II, Fe~II)\\
(Webb {\em et al.}, 1999)&
      $(-1.88\pm0.53)\times10^{-5} $&
      $1-1.6$&&
      QSO (Mg~II, Fe~II) \\
(Ivanchik {\em et al.}, 1999)&
      $(-3.3\pm6.5\pm8)\times10^{-5} $&
      $2-3.5$&&
      QSO (Si~IV)\\
(Fujii {\em et al.}, 2000)&
       $(-0.36\pm1.44)\times10^{-7} $ &
       & 1.8&
       Oklo\\
(Varshalovich {\em et al.}, 2000a)&
      $(-4.5\pm4.3\pm1.4)\times10^{-5} $&
      $2-4$&&
      QSO (Si~IV)\\
(Avelino {\em et al.}, 2001)&
      $(-3.5\pm5.5)\times10^{-2} $&
      $10^3$&&
      CMB\\
(Landau {\em et al.}, 2001)&
      $(-5.5\pm8.5)\times10^{-2} $&
      $10^3$&&
      CMB\\
(Webb {\em et al.}, 2001)&
      $(-0.7\pm0.23)\times10^{-5} $&
      $0.5-1.8$&&
      QSO (Fe~II, Mg~II)\\
(Webb {\em et al.}, 2001)&
      $(-0.76\pm0.28)\times10^{-5} $&
      $1.8-3.5$&&
      QSO (Ni~II, Cr~II, Zn~II)\\
(Webb {\em et al.}, 2001)&
      $(-0.5\pm1.3)\times10^{-5} $&
      $2-3$&&
      QSO (Si~IV)\\
(Murphy {\em et al.}, 2001a)&
      $(-0.2\pm0.3)\times10^{-5} $&
      $0.5-1$&&
      QSO (Mg~I, Mg~II,\ldots)\\
(Murphy {\em et al.}, 2001a)&
      $(-1.2\pm0.3)\times10^{-5} $&
      $1-1.8$&& QSO (Mg~I, Mg~II,\ldots)\\
(Murphy {\em et al.}, 2001a)&
      $(-0.7\pm0.23)\times10^{-5} $&
      $0.5-1.8$&& QSO (Mg~I, Mg~II,\ldots)\\
(Murphy {\em et al.}, 2001d)&
      $(-0.5\pm1.3)\times10^{-5} $&
      $2-3$&&
      QSO (Si~IV)\\
(Sortais {\em et al.}, 2001)&
      $(8.4\pm13.8)\times10^{-15} $&
      & 24 months&
      Atomic clock\\
(Nollet and Lopez, 2002)&
      $ (3\pm7)\times 10^{-2}$&
      $10^8$&&
      BBN\\
(Ichikawa and Kawasaki, 2002)&
      $(-2.24\pm3.75)\times10^{-4} $&
      $10^8$&&
      BBN\\
(Olive {\em et al.}, 2002)&
      $(0\pm1)\times10^{-7} $&
      &1.8&
      Oklo\\
(Olive {\em et al.}, 2002)&
      $(0\pm3)\times10^{-7} $&$\sim0.45$
      &4.6&
      Re/Os\\
(Olive {\em et al.}, 2002)&
      $(0\pm1)\times10^{-5} $&
      & &
      $\alpha$-decay
\end{tabular}
\end{table}


\begin{table}
\caption{The different atomic clock experiments. We recall the transitions
which are compared and the constraint on the time variation
obtained. SCO refers to superconductor cavity oscillator and the
reference to (Breakiron, 1993) is cited in Prestage {\em et al.}
(1995). fs and hfs refer respectively to fine structure and hyperfine
structure.}
\label{table9}
\begin{tabular}{|llccc|}
Reference & Experiment & Constant & Duration & Limit (${\rm yr}^{-1}$) \\
\hline
(Turneature and Stein, 1974) & hfs of Cs vs SCO & $g_{\rm p}\mu\aem^3$
                            & 12 days & $<1.5\times10^{-12}$ \\
(Godone {\em et al.}, 1983) & hfs of Cs vs fs of Mg & $g_{\rm p}\mu$
                            & 1 year & $<2.5\times10^{-13}$ \\
(Demidov, 1992)             & hfs of Cs vs hfs of H & $\aem g_{\rm p}/g_I$
                            & 1 year  & $<5.5\times10^{-14}$ \\
(Breakiron, 1993)           & hfs of Cs vs hfs of H & $\aem g_{\rm p}/g_I$
                            &         & $<5\times10^{-14}$ \\
(Prestage {\em et al.}, 1995) & hfs of HG$^{+}$ vs hfs of H & $\aem g_{\rm p}
                                   /g_I$
                            & 140 days & $<2.7\times10^{-14}$ \\
(Sortais {\em et al.}, 2001)&hfs of Cs vs hfs of Rb&
                            & 24 months        & $(4.2\pm6.9)\times10^{-15}$

\end{tabular}
\end{table}


\begin{table}
\caption{Summary of the constraints on the time variation of the
Newton constant $G$. The constraints labelled by $^*$ refer to bounds
on the rate of decrease of $G$ (that is $-\dot G/G<\ldots$).}
\label{table_1}
\begin{tabular}{|lrl|}
Reference  & Constraint (yr$^{-1}$)  & Method \\
\hline
(Teller, 1948) &
      $(0\pm2.5)\times10^{-11}$  &
      Earth temperature \\
(Shapiro {\em at al.}, 1971) &
      $(0\pm4)\times10^{-10}$  &
      Planetary ranging \\
(Morison, 1973) &
      $(0\pm2)\times10^{-11}$  &
      Lunar occultations \\
(Dearborn and Schramm, 1974) &
      $^*<4\times10^{-11}$  &
      Clusters of galaxies \\
(van Flandern, 1975) &
      $(-8\pm5)\times10^{-11}$  &
      Lunar occultations \\
(Heintzmann and Hillebrandt, 1975) &
      $(0\pm1)\times10^{-10}$  &
      Pulsar spin-down \\
(Reasenberg and Shapiro, 1976) &
      $(0\pm1.5)\times10^{-10}$  &
      Planetary ranging\\
(Mansfield, 1976) &
      $^*<(-5.8\pm1)\times10^{-11}$  &
      Pulsar spin-down\\
(Williams {\em et al.}, 1996)&
      $(0\pm3)\times10^{-11} $  &
      Planetary ranging       \\
(Blake, 1977b) &
      $(-0.5\pm2)\times10^{-11}$  &
      Earth radius \\
(Muller, 1978) &
      $(2.6\pm1.5)\times10^{-11}$  &
      Solar eclipses \\
(McElhinny {\em et al.}, 1978) &
      ${}^*<8\times10^{-12}$  &
      Planetary radii\\
(Barrow, 1978) &
      $(2\pm9.3)h\times10^{-12}$  &
      BBN\\
(Reasenberg {\em et al.}, 1979)&
      $^*<10^{-12} $  &
      Viking ranging     \\
(van Flandern, 1981) &
      $(3.2\pm1.1)\times10^{-11}$  &
      Lunar occultation \\
(Rothman and Matzner, 1981)&
      $(0\pm1.7)\times10^{-13} $  &
      BBN     \\
(Hellings {\em et al.}, 1983) &
      $(2\pm4)\times10^{-12}$  &
      Viking ranging\\
(Reasenberg, 1983) &
      $(0\pm3)\times10^{-11}$  &
      Viking ranging\\
(Damour {\em et al.}, 1988) &
      $(1.0\pm2.3)\times10^{-11}$  &
      PSR 1913+16\\
(Shapiro, 1990) &
      $(-2\pm10)\times10^{-12}$  &
      Planetary ranging \\
(Goldman, 1990) &
      $^*<(3.85\pm{1.65})\times10^{-11}$  &
      PSR 0655+64 \\
(Accetta {\em et al.}, 1990)&
      $(0\pm9)\times10^{-13}$  &
      BBN \\
(M\"uller {\em et al.}, 1991)&
      $(0\pm1.04)\times10^{-11}$  &
      Lunar laser ranging \\
(Anderson {\em et al.}, 1991)&
      $(0.0\pm2.0)\times10^{-12}$  &
      Planetary ranging \\
(Damour and Taylor, 1991) &
      $(1.10\pm1.07)\times10^{-11}$  &
      PSR 1913+16\\
(Chandler, 1993) &
      $(0\pm1)\times10^{-11}$  &
      Viking ranging\\
(Dickey {\em et al.}, 1994)&
      $(0\pm6)\times10^{-12} $  &
      Lunar laser ranging       \\
(Kaspi {\em et al.}, 1994)&
      $(4\pm5)\times10^{-12} $  &
       PSR B1913+16     \\
(Kaspi {\em et al.}, 1994)&
      $(-9\pm18)\times10^{-12} $  &
       PSR B1855+09     \\
(Demarque {\em et al.}, 1994)&
      $(0\pm2)\times10^{-11} $  &
       Heliosismology     \\
(Guenther {\em et al.}, 1995)&
      $(0\pm4.5)\times10^{-12} $  &
       Heliosismology     \\
(Garcia-Berro {\em et al.}, 1995)&
      $^*<(3_{-1}^{+3})\times10^{-11} $  &
       White dwarf  \\
(Williams {\em et al.}, 1996)&
      $(0\pm8)\times10^{-12} $  &
      Lunar laser ranging       \\
(Thorsett, 1996)&
      $(-0.6\pm4.2)\times10^{-12} $  &
      Pulsar statistics     \\
(Del'Innocenti {\em et al.}, 1996)&
      $(-1.4\pm2.1)\times10^{-11} $  &
      Globular clusters    \\
(Guenther {\em et al.}, 1998)&
      $(0\pm1.6)\times10^{-12} $  &
       Heliosismology     \\
\end{tabular}
\end{table}


\begin{table}
\caption{Summary of the constraints on the variation of the constant
$k$. We use the notation $\mu\equiv m_{\rm e}/m_{\rm p}$, $x\equiv
\aem^2g_{\rm p}\mu$ and $y\equiv \aem^2g_{\rm p}$.}
\label{table_4}
\begin{tabular}{|lcrrrl|}
Reference&
      Constant &  Constraint&
      redshift & Time ($10^9\,{\rm yr}$) &
      Method \\
 \hline
 (Yahil, 1975)
     & $\mu$ &
     $(0\pm1.2)$
     & & 10 &
     Rb-Sr, K-Ar\\
(Pagel, 1977)
     & $\mu$ &
     $(0\pm4)\times10^{-1}$
     & 2.1-2.7 & &
     QSO\\
(Foltz {\em et al.}, 1988)
     & $\mu$ &
     $(0\pm2)\times10^{-4}$
     & 2.811 & &
     QSO\\
(Varshalovich and Levshakov 1993)&
     $\mu$      &
     $(0\pm4)\times10^{-3}$
     & 2.811&
     & QSO\\
(Cowie and Songaila, 1995)&
     $\mu$
     &  $(0.75\pm6.25)\times10^{-4}$
     & 2.811 &
     & QSO\\
(Varshalovich and Potekhin, 1995)&
     $\mu$
     &  $(0\pm2)\times10^{-4}$
     & 2.811 &
     & QSO\\
(Varshalovich {\em et al.}, 1996a)&
     $\mu$      &
     $(0\pm2)\times10^{-4}$
     & 2.811 &
     & QSO\\
(Varshalovich {\em et al.}, 1996b)&
     $\mu$      &
     $(-1\pm1.2)\times10^{-4}$
     & 2.811 &
     & QSO\\
(Potekhin {\em et al.}, 1988)
     & $\mu$ &
     $(-7.5\pm9.5)\times10^{-5}$
     & 2.811 & &
     QSO\\
(Ivanchik {\em et al.}, 2001)&
     $\mu$      &
     $(-5.7\pm3.8)\times10^{-5}$
     & $2.3-3$ &
     & QSO\\
\hline
(Savedoff, 1956)&
     $x$ &
     $(3\pm7)\times10^{-4}$
     & $0.057$ &
     & Cygnus A \\
(Wolfe {\em et al.}, 1976)&
     $x$ &
     $(5\pm10)\times10^{-5}$
     & $\sim 0.5$ & &QSO (Mg~I)\\
(Wolfe and Davis, 1979)&
     $x$
     & $(0\pm2)\times10^{-4}$
     & 1.755& &
     QSO\\
(Wolfe and Davis, 1979)&
     $x$ &
     $(0\pm2.8)\times10^{-4}$
     & 0.524& &
     QSO\\
(Tubbs and Wolfe, 1980)&
     $x$ &
     $(0\pm1)\times10^{-4}$
     & 1.776 &
     & QSO\\
(Cowie and Songaila, 1995)&
     $x$ &
     $(7\pm11)\times10^{-6}$
     &1.776 &
     & QSO\\
\hline
(Varshalovich and Potekhin, 1996)&
     $y$ &
     $(-4\pm6)\times10^{-5}$
     & 0.247 &
     &QSO\\
(Varshalovich and Potekhin, 1996)&
     $y$ &
     $(-7\pm10)\times10^{-5}$
     & 1.94 &
     &QSO\\
(Drinkwater {\em et al.}, 1998)&
     $y$ &
     $(0\pm5)\times10^{-6}$
     &0.25, 0.68 &
     &QSO\\
(Carrilli {\em et al.}, 2001)&
     $y$ &
     $(0\pm3.4)\times10^{-5}$
     &0.25, 0.68&
     &QSO\\
(Murphy {\em et al.}, 2001c)&
     $y$ &
     $(-0.2\pm0.44)\times10^{-5}$
     &0.25 &
     &QSO\\
(Murphy {\em et al.}, 2001c)&
     $y$ &
     $(-0.16\pm0.54)\times10^{-5}$
     &0.68 &
     &QSO\\
\hline
(Wolfe {\em et al.}, 1976)&
     $g_{\rm p}\mu$ &
     $(0\pm0.68)\times10^{-2}$
     &0.524 &
     &QSO\\
 (Turneature and Stein, 1994)&
     $g_{\rm p}\mu\aem^3$ &
     $ (0\pm9.3)\times10^{-16}$
     & & 12 days
     & Atomic clocks\\
(Godone {\em et al.}, 1993)&
     $g_{\rm p}\mu$ &
     $(0\pm5.4)\times10^{-13}$
     & & 1 year
     & Atomic clocks\\
 \hline
(Wilkinson, 1958)&
     $\aw$ & $ (0\pm1)\times10^{1}$
     & & 1
     & Fission\\
(Dyson, 1972)&
     $\aw$ & $ (0\pm1)\times10^{-1}$
     & & 1
     & $\beta$-decay\\
(Shlyakhter, 1976)&
     $\aw$ & $ (0\pm4)\times10^{-3}$
     & & 1.8
     & Oklo\\
(Damour and Dyson, 1996)&
     $\aw$ & $(0\pm2)\times10^{-2} $
     & & 1.8
     & Oklo
\end{tabular}
\end{table}

\end{document}